\documentclass[fleqn,usenatbib]{mnras}

\usepackage{newtxtext,newtxmath}

\usepackage[T1]{fontenc}
\usepackage{ae,aecompl}
\usepackage{hyperref}
\usepackage{tabularx}
\usepackage{subcaption}
\usepackage{enumerate}

\newcommand{\kepler}{\textit{Kepler}}

\usepackage{graphicx}	
\usepackage{amsmath}	
\usepackage{comment}
\usepackage{tabularx}
\usepackage{xcolor}

\title[Single-Moon Modeling for Multi-Moon Systems]{On the Impact and Utility of Single-Exomoon Modeling for Multi-Moon Systems}

\author[Teachey \& Agarwal]{Alex Teachey$^{1}$ \& Garvit Agarwal$^{1,2}$ \thanks{E-mail: amteachey@asiaa.sinica.edu.tw}
\\
$^{1}$Academia Sinica Institute of Astronomy and Astrophysics \\ 
11F of AS/NTU Astronomy-Mathematics Building, No.1, Sec. 4, Roosevelt Rd, Taipei 10617, Taiwan, R.O.C.\\
$^{2}$Indian Institute of Science Education and Research, Pune 
}

\date{Accepted 2024 February 23. Received 2024 February 20; in original form 2024 January 12}

\pubyear{2023}

\begin{document}

\label{firstpage}
\pagerange{\pageref{firstpage}--\pageref{lastpage}}
\maketitle

\begin{abstract}
The search for exomoons in time-domain photometric data has to-date generally consisted of fitting transit models that are comprised of a planet hosting a single moon. This simple model has its advantages, but it may not be particularly representative, as most of the major moons in our Solar System are found in multi-moon satellite systems. It is critical that we investigate, then, the impact of applying a single-moon model to systems containing multiple moons, as there is the possibility that utilizing an inaccurate or incomplete model could lead to erroneous conclusions about the system. To that end, in this work we produce a variety of realistic multi-moon light curves, perform standard single-moon model selection, and analyze the impacts that this model choice may have on the search for exomoons. We find that the number of moons in a system fit with a single-moon model generally has little impact on whether we find evidence for a moon in that system, and other system attributes are individually not especially predictive. However, the model parameter solutions for the moon frequently do not match any real moon in the system, instead painting a picture of a ``phantom'' moon. We find no evidence that multi-moon systems yield corresponding multi-modal posteriors. We also find a systematic tendency to overestimate planetary impact parameter and eccentricity, to derive unphysical moon densities, and to infer potentially unphysical limb darkening coefficients. These results will be important to keep in mind in future exomoon search programs.

\end{abstract}




\section{Introduction}

Despite a search that has lasted now more than a decade, a clear, unambiguous exomoon discovery remains elusive. To date there are arguably only two transiting exomoon candidates in the literature that have survived a battery of tests to remain viable \citep[Kepler-1625 b-i and Kepler-1708 b-i;][]{K1625, Loose_Ends, K1708, Kipping:2024}, and yet even these detections are not without their skeptics \citep[e.g.][]{Kreidberg:2019, Heller:2019, Heller:2023}. And so the question naturally arises, what is taking so long?

There are a number of key challenges in the detection of exomoons using time-domain photometry. Moons, which are generally expected to be small, and as a rule will be less massive than their host planet, are prone to having their transits buried in the noise. On top of this, their appearance alternately before, after, and during the planet transits can confound simple periodic signal search algorithms such as a box-least squares, and defy easy phase-folding -- though methods have been developed to address these issues \citep[e.g.][]{QATS, transit_origami}. In addition, exomoon transit signals may be mimicked by high-frequency, low-amplitude stellar activity \citep[e.g.][]{HEKV}, or by planets passing in front of star spots \citep[e.g.][]{Beky:2014}. Exomoon transit signals are also vulnerable to removal by detrending algorithms. And to make matters worse, there are allowable and plausible satellite system architectures with inclinations that effectively hide the moons, as the moons can pass above or below the stellar disk for a sizeable fraction of transit epochs and produce no transit signal at all \citep[][]{Martin:2019}.

The search for exomoons in time-domain photometry typically consists of fitting at least two competing transit models -- a planet-only model, and a planet+moon model -- to see which model better explains the data. The model that includes the moon carries several additional free parameters, and so this model is typically penalized for its extra complexity, such that only a significant improvement over the planet-only model will be sufficient to claim evidence of a moon. Photodynamical moon models, produced by algorithms such as \textsc{Luna} \citep{LUNA}, \textsc{Pandora} \citep{Pandora}, and \textsc{Gefera} \citep{Gefera} accurately reproduce the planet and moon transit signals of varying duration, and produce the associated transit timing and duration variations, as well.

These codes support modeling transits of a planet hosting a single moon. This is entirely sufficient if we are looking for analogues of the Earth-Moon system, but if the Solar System is any guide, we can expect exomoons may often be found in multi-moon systems. Multiple-moon systems are certainly anticipated in other planetary systems, as well \citep[e.g.][]{Inderbitzi:2020, Cilibrasi:2021}. Even so, single-moon modeling has been the standard for exomoon hunting\footnote{See however \citealt{Heller:2016}, who examined the patterns of TTVs and TDVs for systems hosting $N\geq1$ moons.}, because unfortunately, modeling multi-moon system transits pose some additional challenges. For instance, the system can no longer be approximated as a nested 2-body problem; the moons will interact with one another gravitationally, and many possible architectures will be unstable over the long term. As a result, system architectures would need to be vetted through a stability analysis incorporating computationally expensive $N$-body simulations. While these simulations can be performed after a transit fit, utilizing the joint posteriors to test which solutions are long-term stable, we would ideally prefer the stability analysis to be performed during the fitting process, so that unstable solutions would be rejected at runtime. But this will increase the computation times enormously. 

And with regard to model fitting, each additional moon adds greater complexity to the model, with 7+ free parameters apiece, which can severely impact the computational demand. This is also somewhat of a predicament, because even though from a mathematical standpoint the model is increasingly contrived with each additional moon, and has to be penalized for the additional complexity to guard against over-fitting, there is nothing particularly contrived about a multiple moon system existing in nature. We therefore may run some risk of biasing ourselves against finding systems with multiple moons.

The choice thus far has therefore simply been to model every system with a single moon, with the implicit assumption that it should work just as well if more than one moon is present. It is far simpler to build, implement, and compute, and just as searching a multi-planet system for a single planet should yield results, so too should we be able to recover one of the several moons. But how well does a single-moon transit model really work for systems containing multiple moons? This important question has been left largely unexplored until now.

There are four major possible outcomes from the modeling of a multi-moon system with a single moon: (1) the single-moon model fit correctly identifies one of the system's multiple moons, perhaps the largest one, or the one most responsible for the planet's observed transit timing and duration variations; (2) the single-moon model correctly identifies that a moon is present, but the properties it derives for this moon are spurious, not representing any real object in the system (what we might call a ``phantom'' moon); (3) the single-moon model returns multi-modal posteriors, with each of the modes corresponding to one of the real moons in the system; and (4) the single-moon model is so grossly at odds with the data in hand that a solution cannot be found that significantly improves on the planet-only model (perhaps because of inconsistent transit transit timings or durations), and as a result the evidence comes out firmly against the presence of moons.

Possibility (1) is a very favorable outcome, as it means that as we are searching for one moon, we have emerged with one moon, accurately modeled, and the rest of the moons can be found at a later date (The system is now, presumably, an attractive target for follow-up). Possibility (2) is a bit more worrying, as an inaccurate model solution may fail to hold up to additional scrutiny on subsequent analysis / observation, and the moon hypothesis may subsequently be rejected, erroneously. Possibility (3) meanwhile presents a very exciting opportunity, and is perhaps the best-case scenario, as multiple moons could potentially be recovered all at once while utilizing the straightforward and computationally efficient single-moon model (though it might be challenging to disentangle these posteriors). And finally, possibility (4) is the most worrying: if indeed exomoons are residing in multiple moon systems to any significant degree, we might have dismissed these systems entirely, and they are unlikely to get a second look any time soon.

There are still other potential observational impacts of fitting the data with an inaccurate or incomplete model, some of which might provide indications that something is anomalous. We might see, for example, unphysical or otherwise peculiar limb darkening solutions, planet and moon solutions with exceptionally high or low densities, erroneous eccentricity or impact parameter solutions. Of course, any such predictors would be especially valuable to identify as we carry the exomoon search forward. Some of these parameters may be independently observable, which might in turn indirectly suggest the presence of an undetected moon skewing the model results.

In this work we investigate  the impact of modeling multiple-exomoon systems using a single-moon model. We begin by describing the production of our artificial dataset (section \ref{sec:sims}). We go on to describe the model fitting (section \ref{sec:model_fitting}), our results (section \ref{sec:results}) and their implications.

\label{sec:introduction}


\section{Artificial Dataset}
\label{sec:sims}

\begin{figure*}
    \begin{subfigure}{.23\textwidth}
    \centering
    \includegraphics[width=1\linewidth]{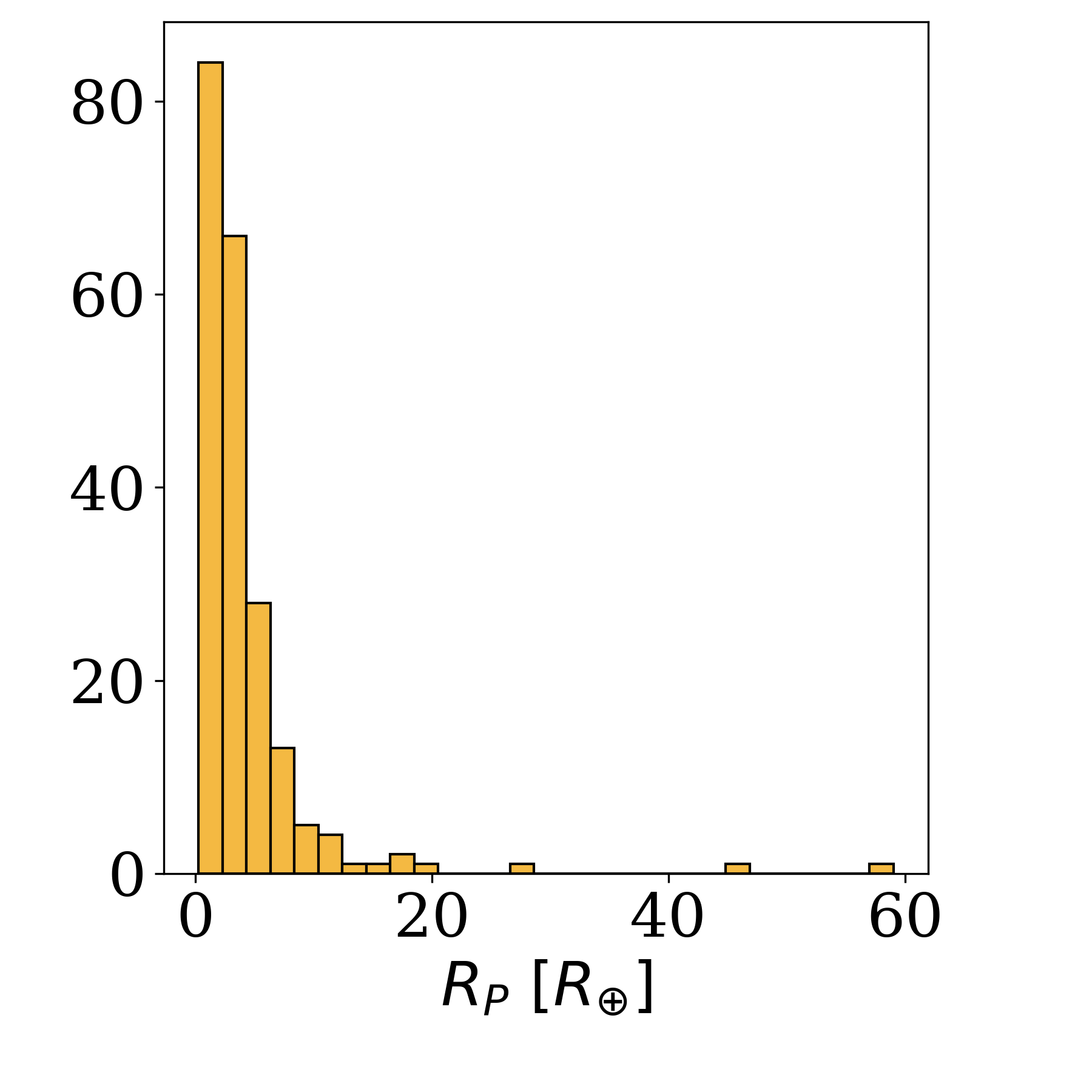}
    \end{subfigure}
    \begin{subfigure}{.23\textwidth}
    \centering
    \includegraphics[width=1\linewidth]{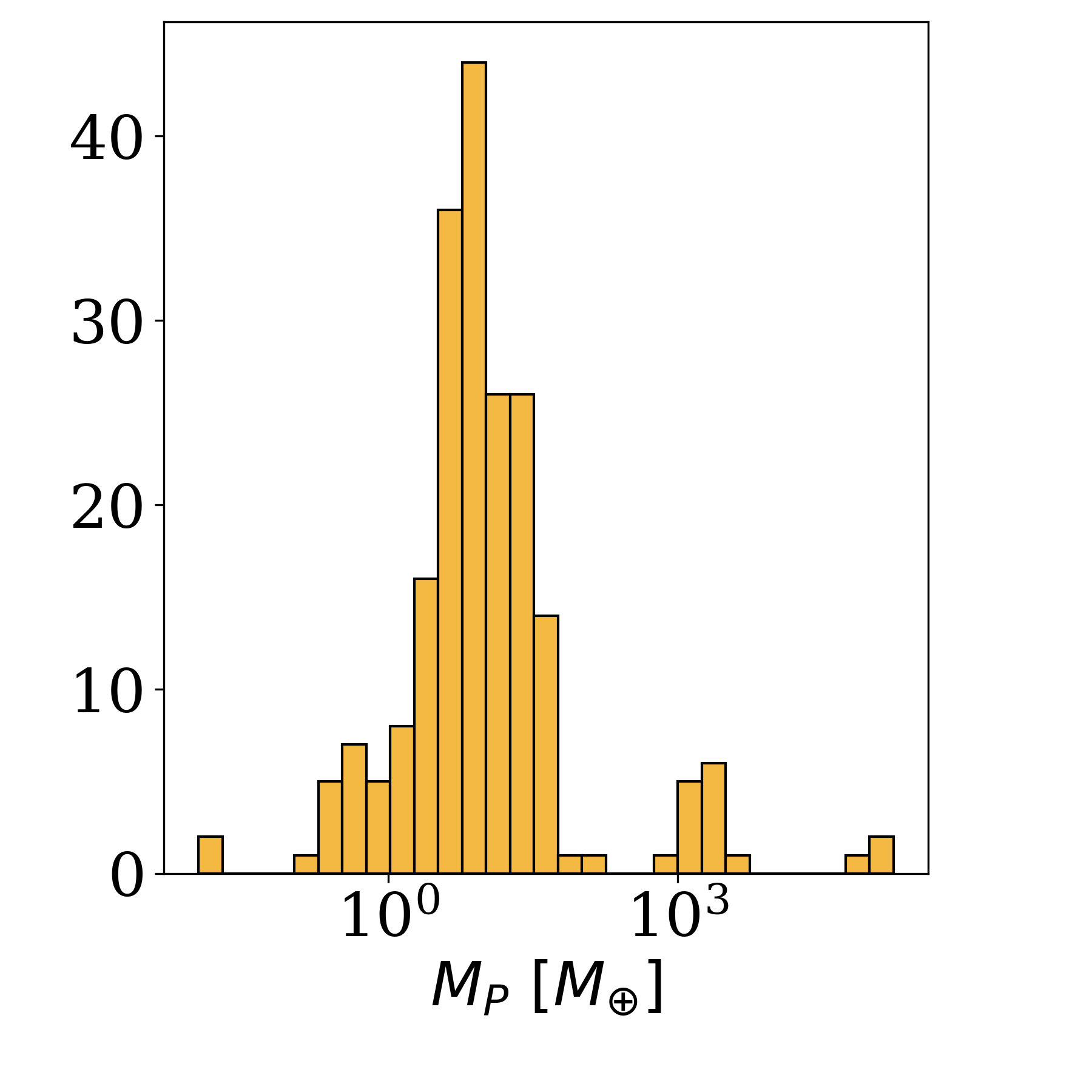}
    \end{subfigure}
        \begin{subfigure}{.23\textwidth}    
        \centering
    \includegraphics[width=1\linewidth]{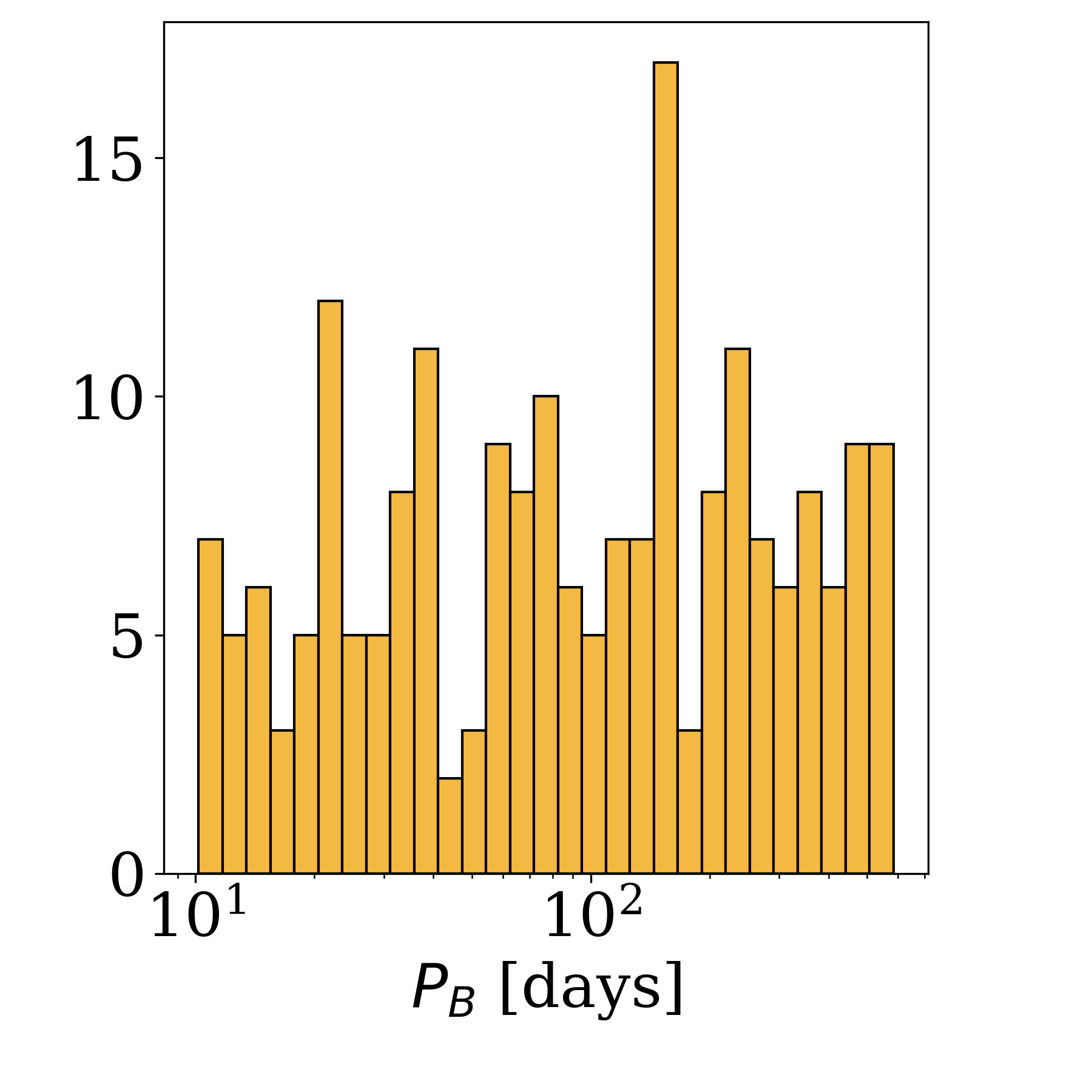}
    \end{subfigure}
    \begin{subfigure}{.23\textwidth}
    \centering
    \includegraphics[width=1\linewidth]{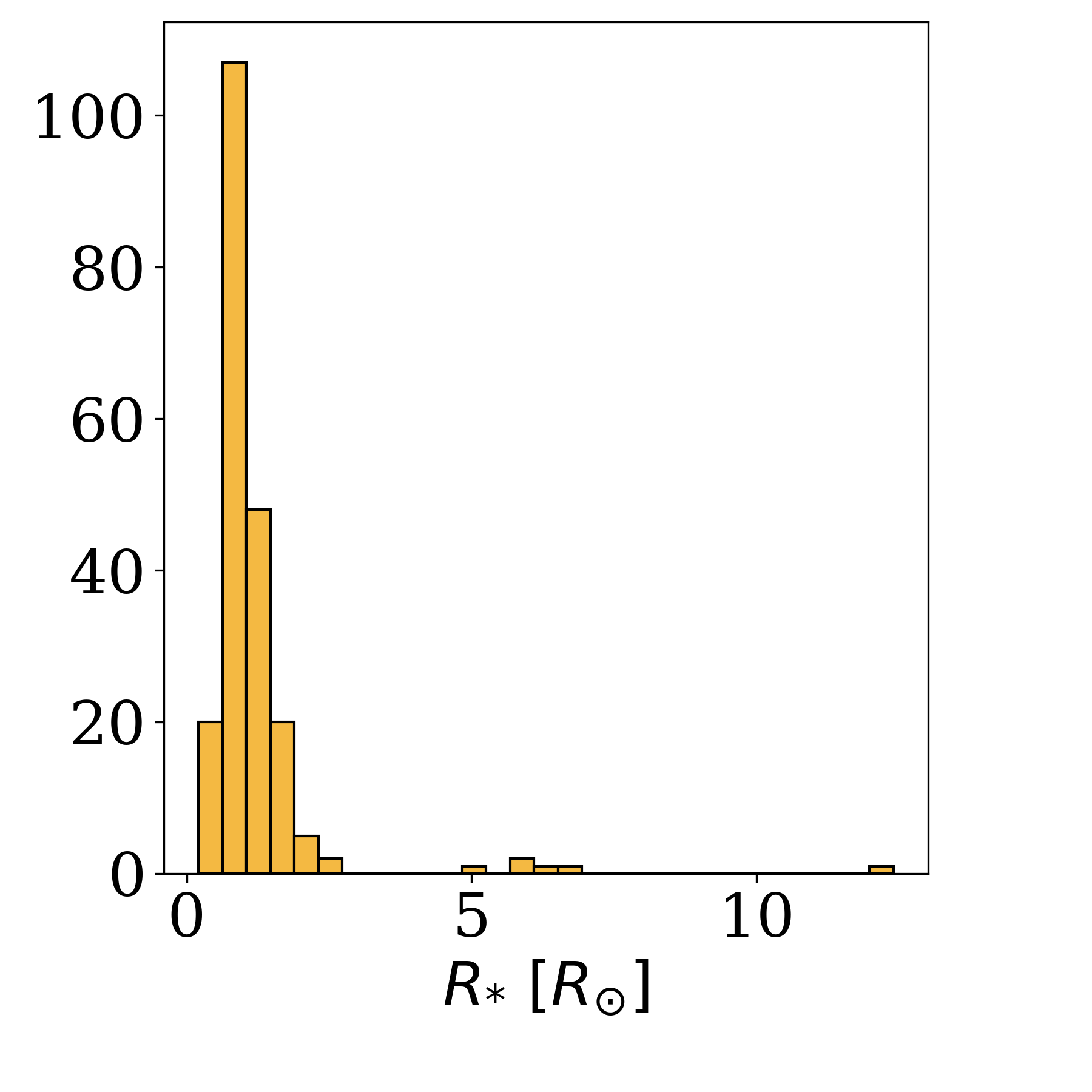}
    \end{subfigure}
    \begin{subfigure}{.23\textwidth}
    \centering
    \includegraphics[width=1\linewidth]{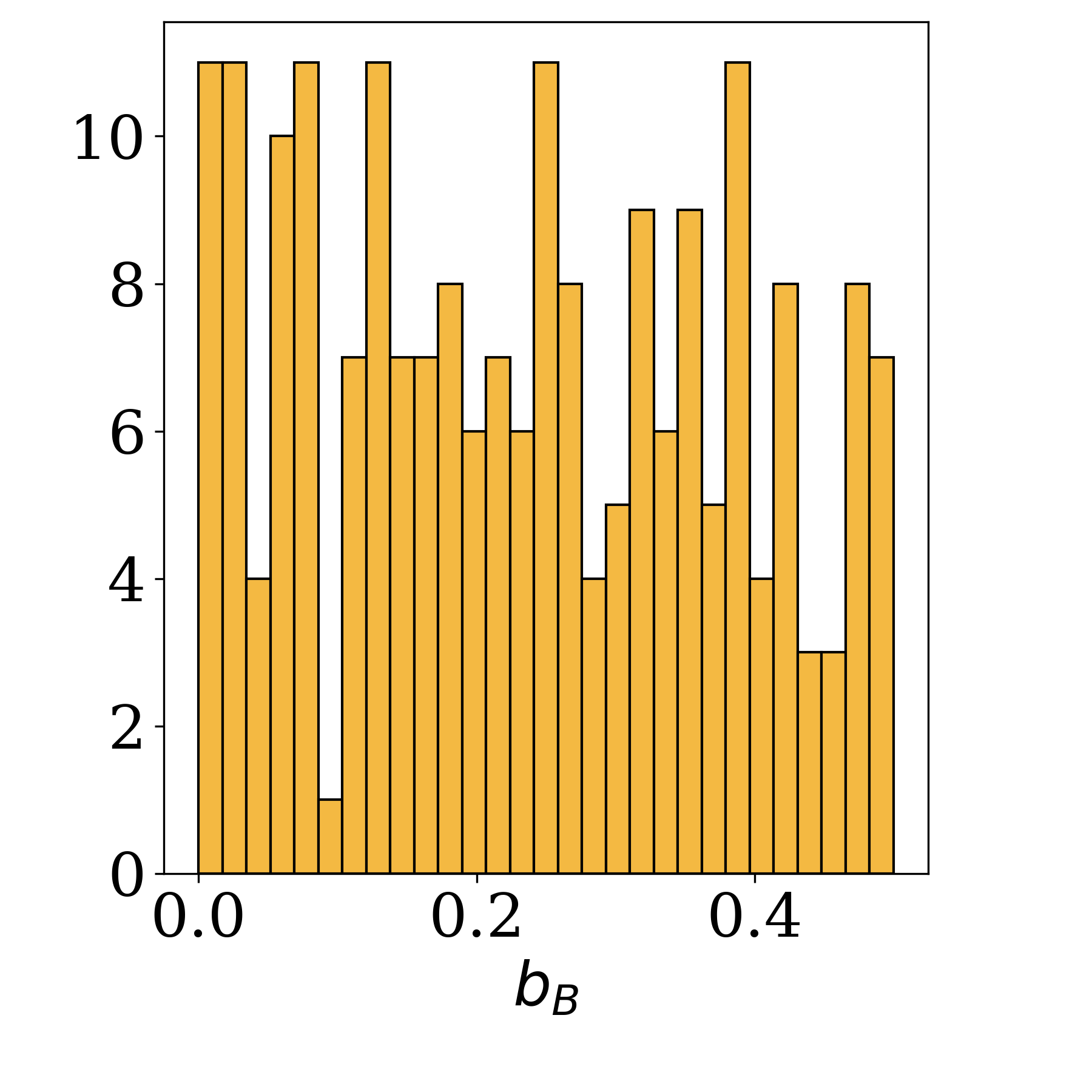}
    \end{subfigure}
    \begin{subfigure}{.23\textwidth}
    \centering
    \includegraphics[width=1\linewidth]{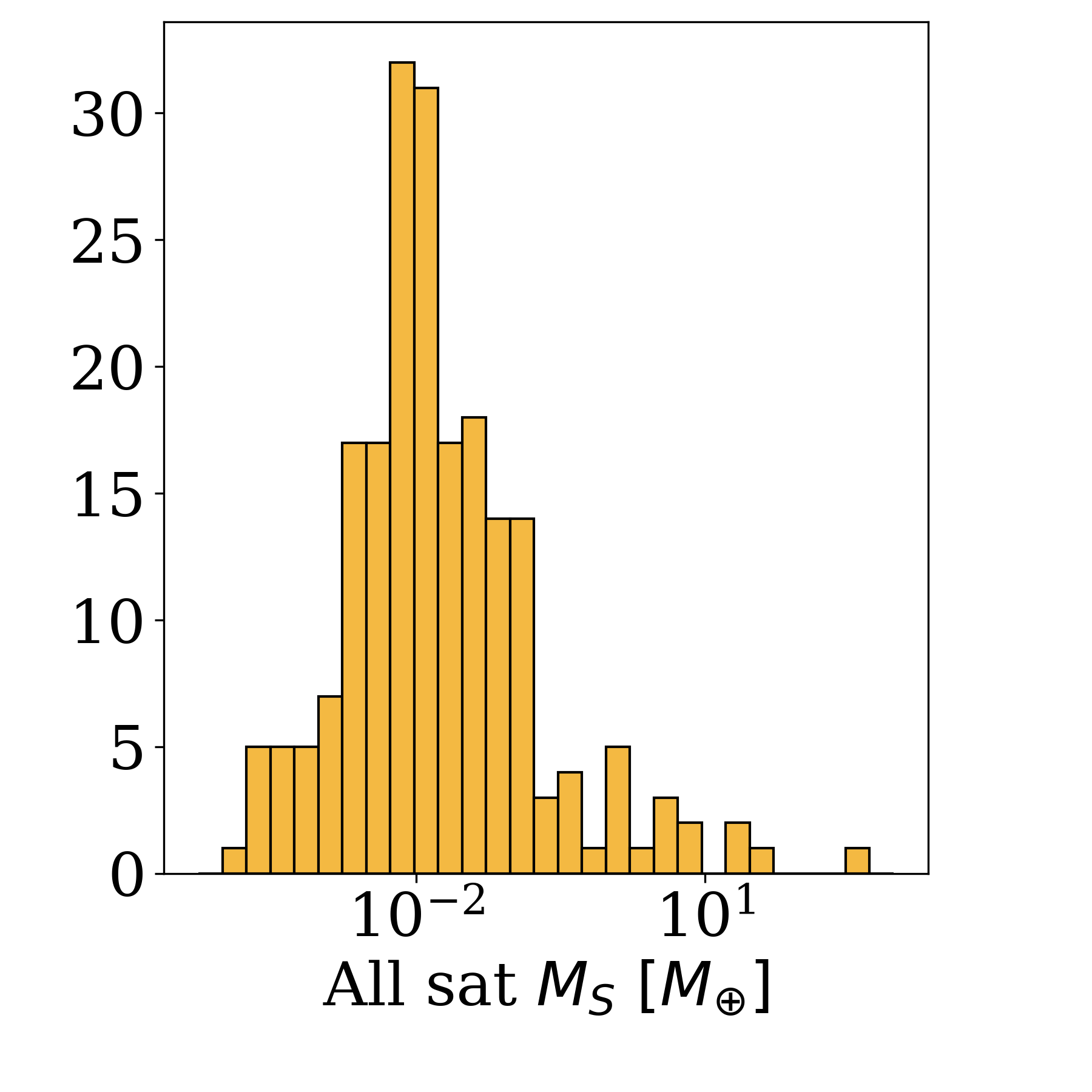}
    \end{subfigure}
    \begin{subfigure}{.23\textwidth}
    \centering
    \includegraphics[width=1\linewidth]{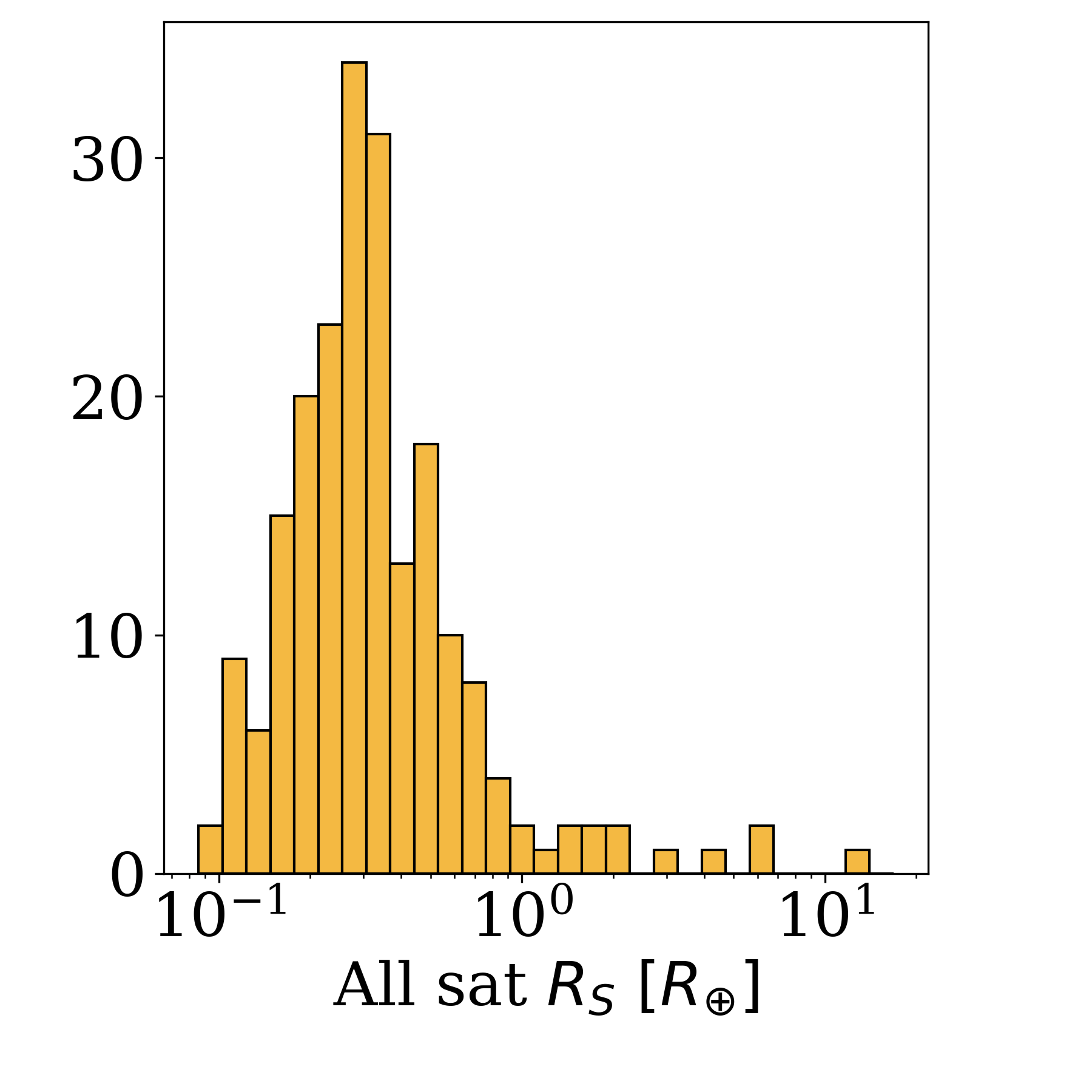}
    \end{subfigure}
    \begin{subfigure}{.23\textwidth}
    \centering
    \includegraphics[width=1\linewidth]{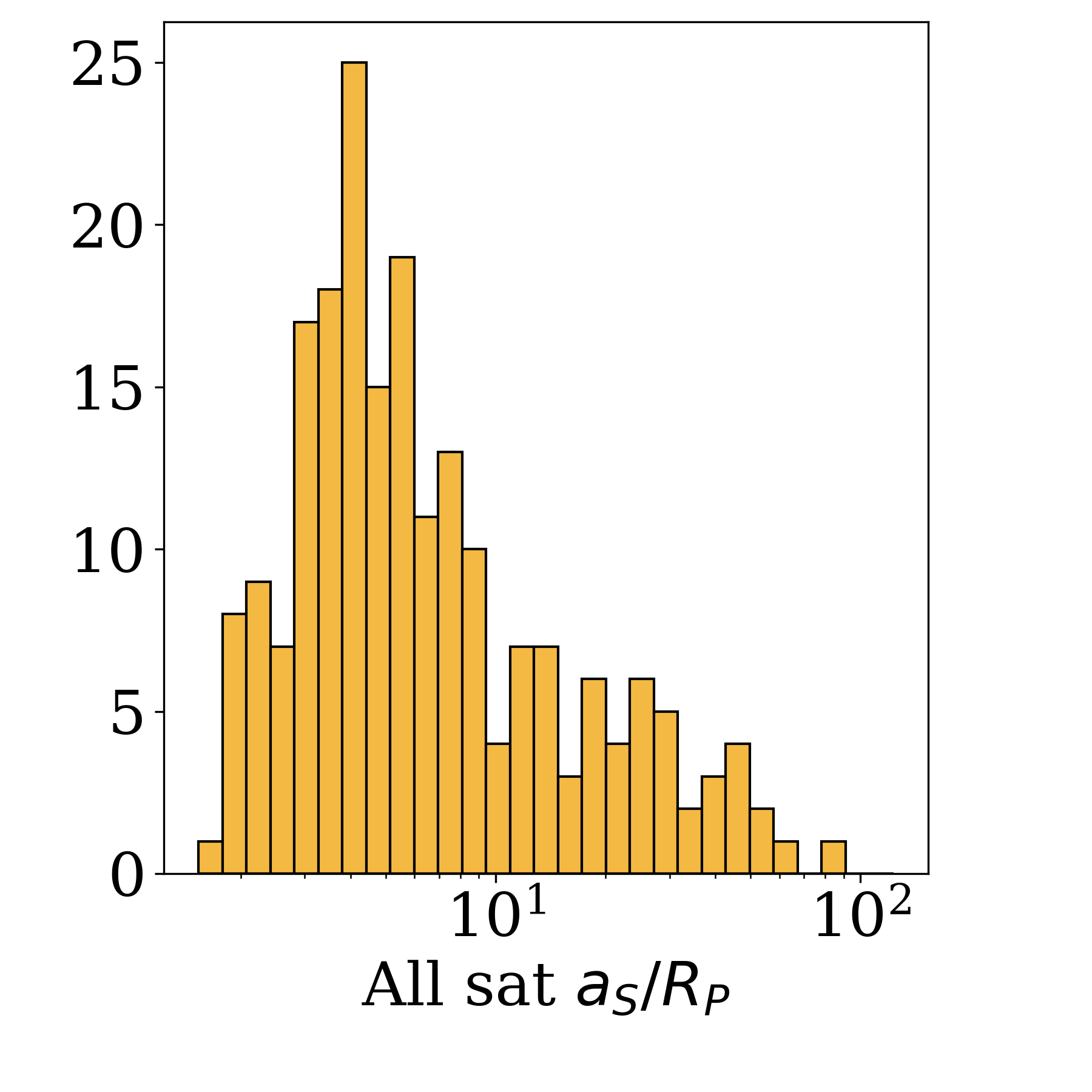}
    \end{subfigure}
    \begin{subfigure}{.23\textwidth}
    \centering
    \includegraphics[width=1\linewidth]{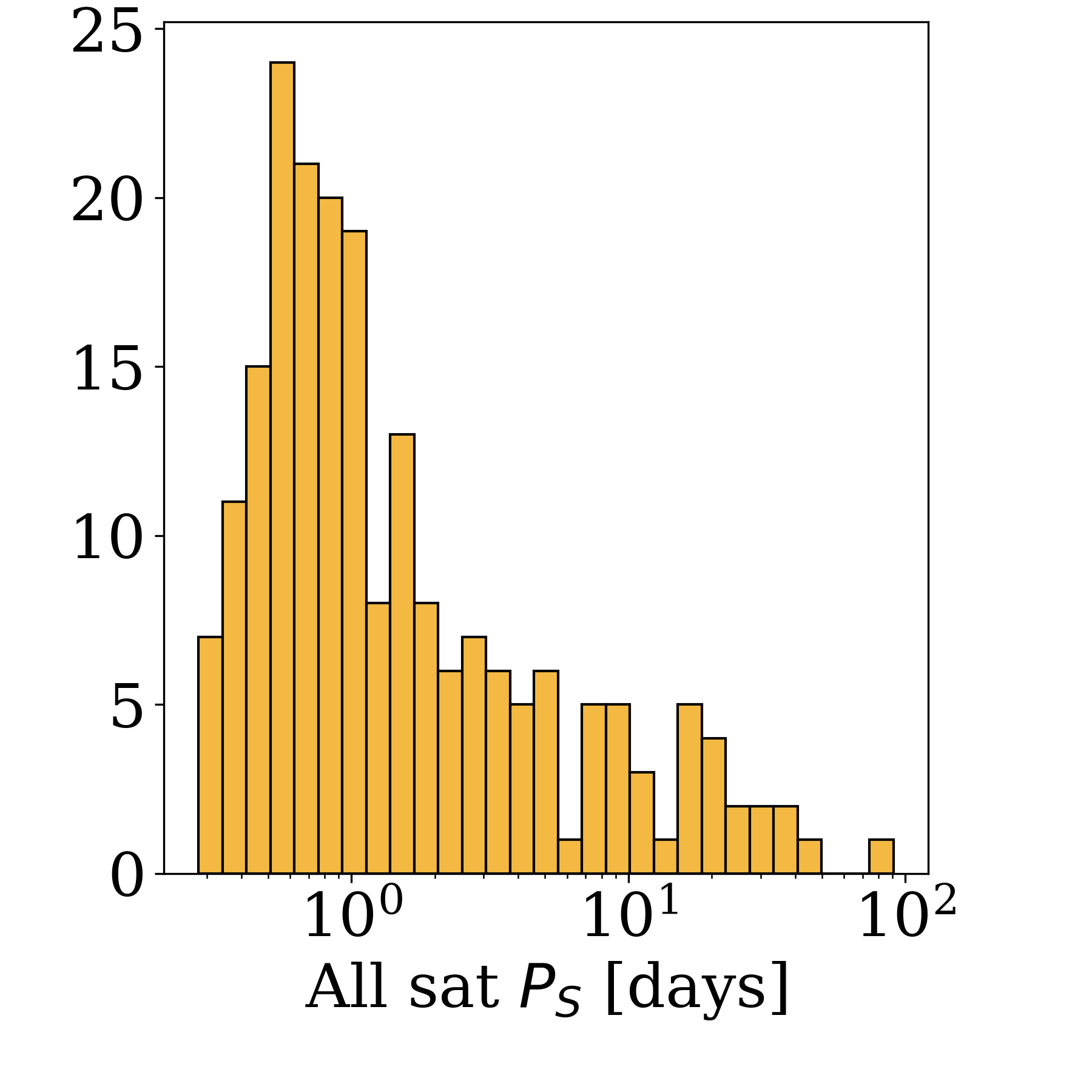}
    \end{subfigure} 
    \begin{subfigure}{.23\textwidth}
    \centering
    \includegraphics[width=1\linewidth]{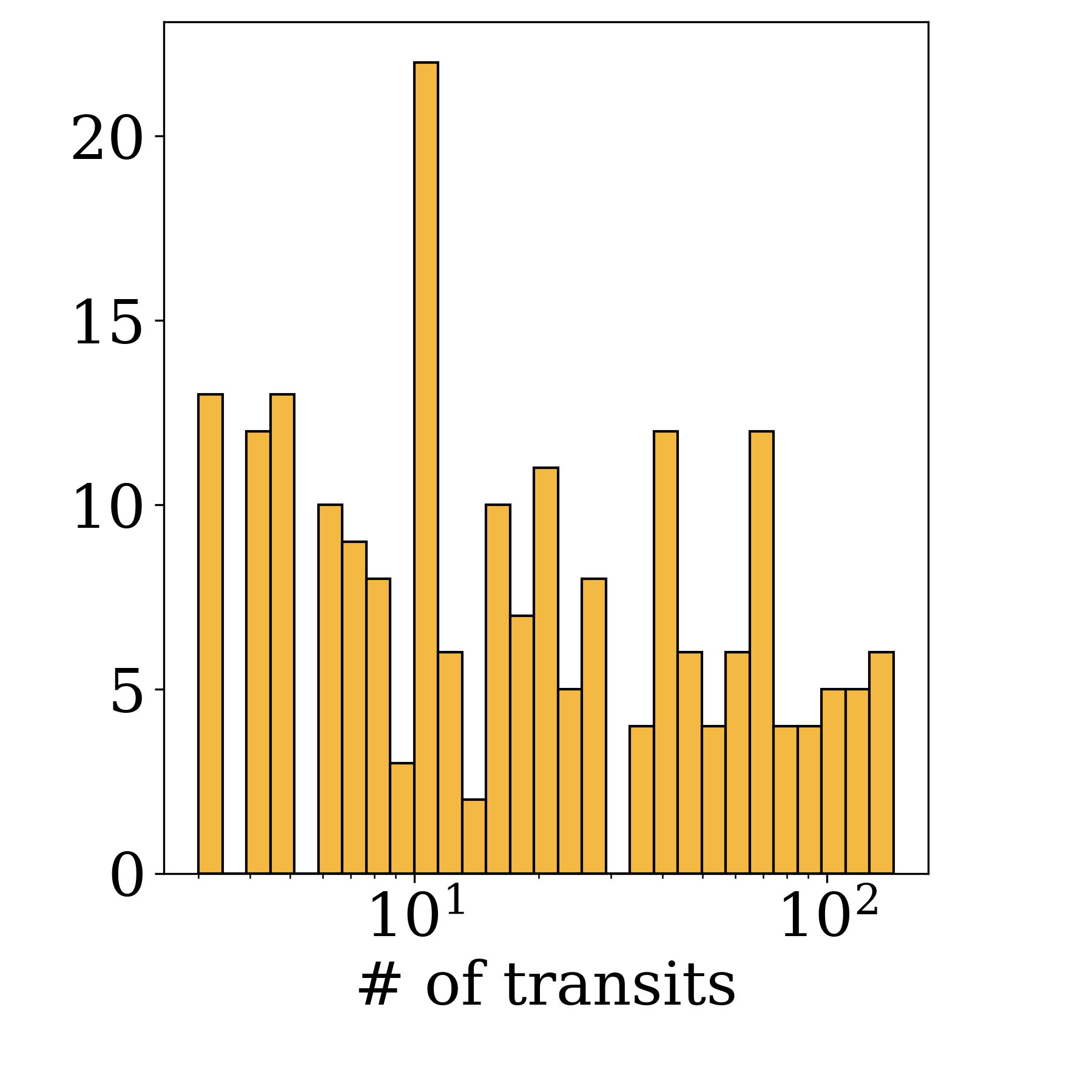}
    \end{subfigure} 
    \begin{subfigure}{.23\textwidth}
    \centering
    \includegraphics[width=1\linewidth]{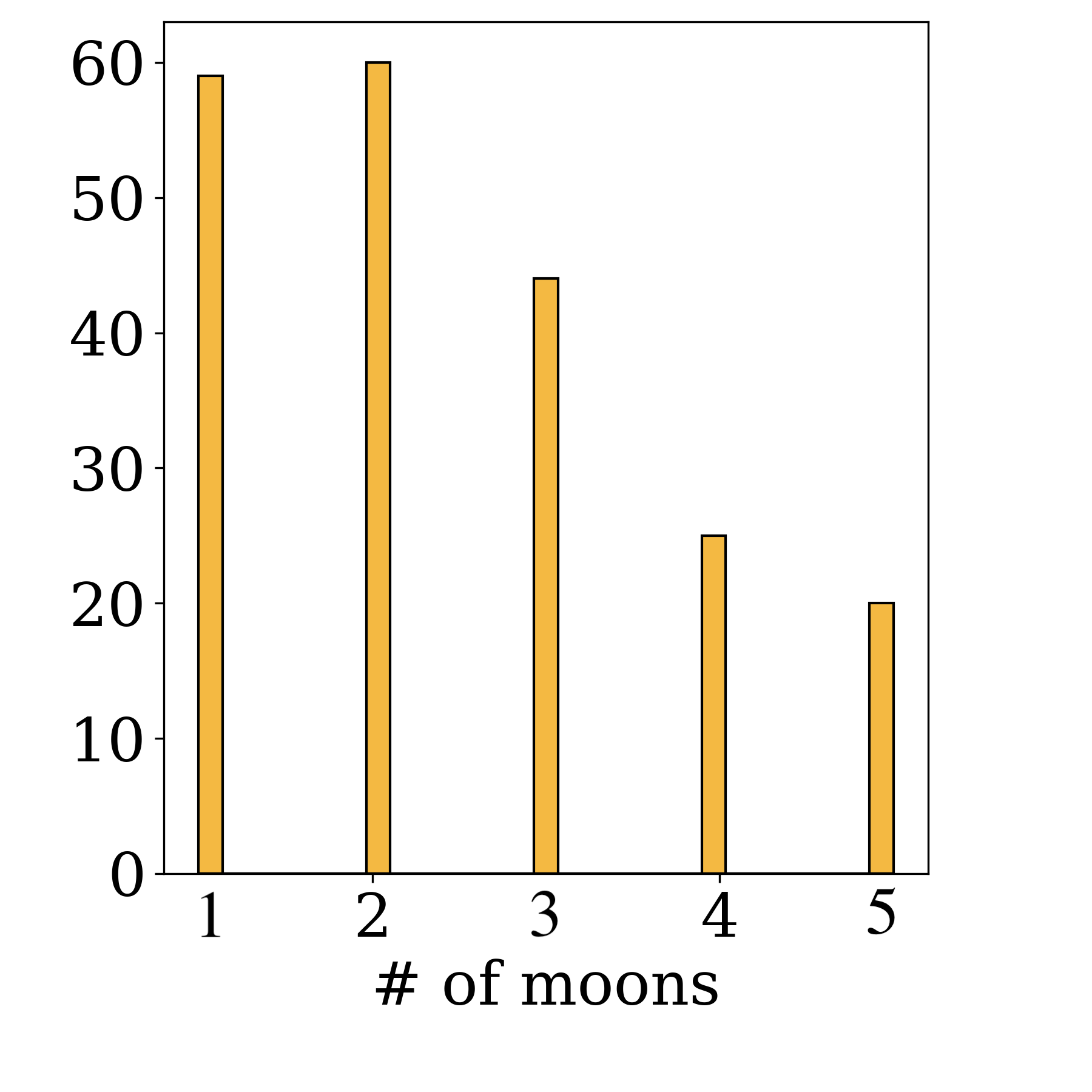}
    \end{subfigure} 
    \begin{subfigure}{.23\textwidth}
    \centering
    \includegraphics[width=1\linewidth]{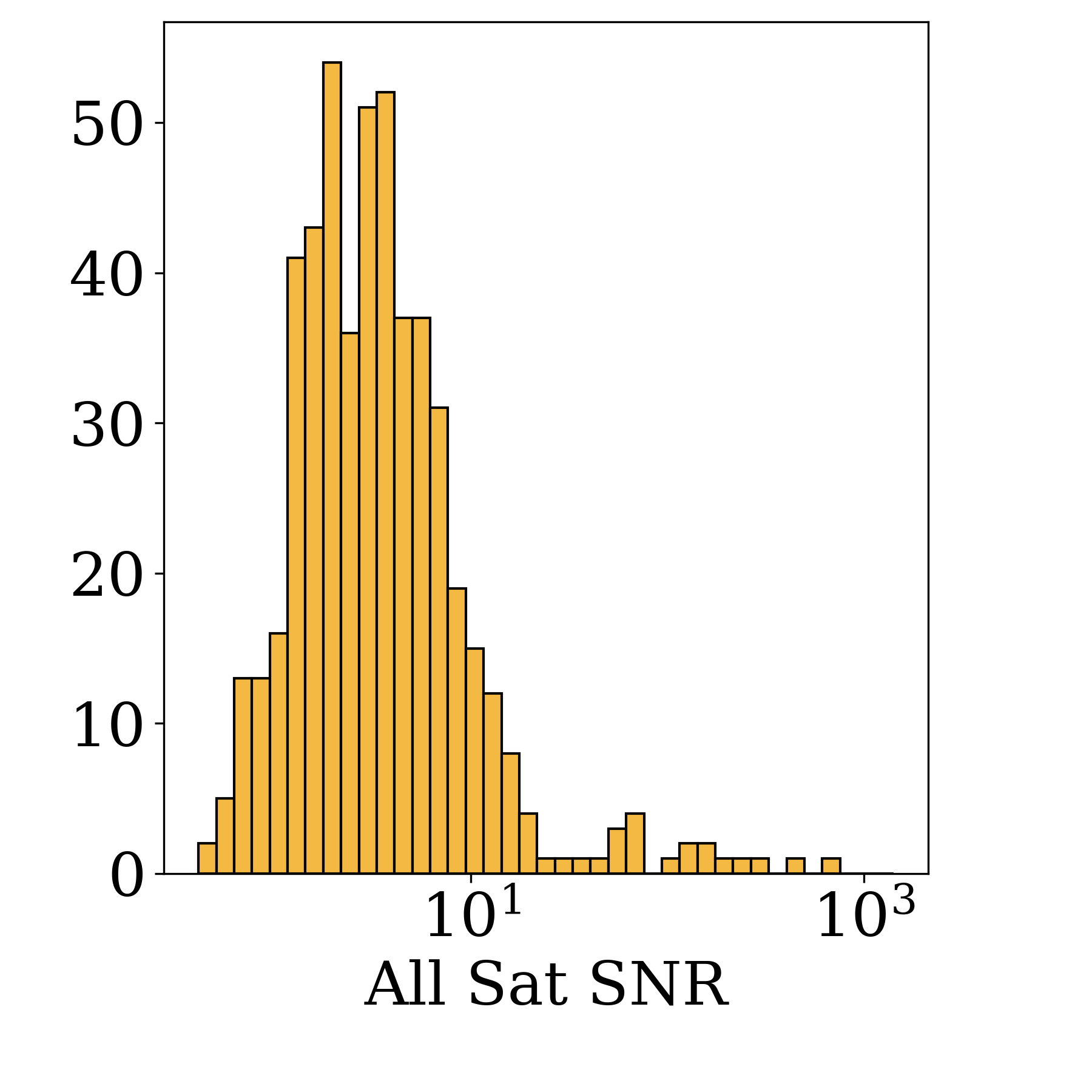}
    \end{subfigure} 
    \caption{Distributions of various ground truth features of the artificial light curves generated in this work. \textit{Top row:} planetary radii $R_P$, in Earth radii; planetary masses $M_P$, in Earth masses; orbital periods of the planetary system barycenter $P_B$, in days; and radius of the star $R_{*}$, in Solar radii. \textit{Middle row:} impact parameter of the system barycenter $b_B$; masses of all the satellites $M_S$, in Earth masses; radii of all the satellites $R_S$, in Earth radii; and the semimajor axes of the moons in units of planetary radii ($a_S / R_P$). \textit{Bottom row:} orbital periods of the moons $P_S$, in days; the number of transits for each planet; the number of moons in each system; and the single transit signal-to-noise ratios SNR for each moon.}
\label{fig:simulation_demographics}
\end{figure*}

\subsection{\textit{N}-body simulations}
Single-moon modeling has typically utilized a nested two-body assumption, treating the star-planet and planet-moon systems in isolation for the purpose of locating the objects in space relative to one another \citep[e.g.][]{LUNA}. This simplifies the calculations considerably, as the positions of each object pair can be computed analytically (no numerical integrations required). Of course, perturbations of the star will produce long-term changes in the system, but we expect those changes to be negligible on the short timescale of our observations; any rapidly evolving systems could very well be unstable, or solutions suggesting such a system might be rejected on these grounds. To first order, we can assume stability of a single moon if it is comfortably within the planet's Hill sphere, or some fraction of it\footnote{Prograde moons can be stable out to about 40 to 50 percent of the Hill radius, while retrograde moons can be stable almost all the way to the edge of the Hill sphere \citep[][]{Domingos:2006, Dobos:2021}}.

In the present context simulating multi-moon systems, we may also ignore the star's effect on the moons' orbits, owing to the long timescales of any associated changes. But unlike the single-moon case, we must now also ensure that system architectures are long-term stable; the moons will interact with one another, and some architectures simply will not be found in nature. We of course do not want to include these. To this end, we began by generating systems containing 1 - 5 moons using \textsc{Rebound} \citep{Rebound}, and analyzed their long-term stability following the methodology outlined in \citealt[][]{Teachey:2021}, requiring a long-term stability probabilities of 90\% or higher. Consistent with that work, we found that systems containing more moons generally required a lower total satellite system mass in order to maintain stability.

To generate realistic systems, we pulled as much as possible from the catalogue of known planetary systems, primarily using the NASA Exoplanet Archive \citep[][]{NEA}.  Each simulated star took its attributes (mass, radius, effective temperature, and surface gravity) from a real \textit{Kepler} planet host star. Quadratic limb darkening coefficients for the \textit{Kepler} bandpass were sourced by finding the closest matching effective temperature and surface gravity in the \citealt{Claret:2011} catalogue of limb darkening coefficients. 

Planet radii were also drawn at random from the NASA Exoplanet Archive (but separately from the stellar radius draw), and an appropriate mass was computed from the radius using \textsc{Forecaster} \citep[][]{forecaster}. The planet's orbital period was drawn from a log-uniform distribution between 10 and 1500 days. The planet's impact parameter was chosen from a uniform distribution between 0 and 0.5, in order to prevent grazing transits that would adversely affect the detectability of the moons. Every planet's eccentricity was set to zero, so as to be consistent with minimal perturbation of the system by the host star. This is also consistent with the general expectation that detectable exomoons will be found in dynamically cold systems.

The total satellite system mass was drawn from a uniform distribution between $10^{-4}$ and $10^{-2}$ times the mass of the planet -- roughly the range between the mass ratio for Jupiter's regular moons and the mass ratio of the Earth-Moon system. Following the methodology of \citealt{Teachey:2021}, this total satellite mass was divided randomly between the generated moons, resulting in some systems having roughly-equal mass moons while others having one or two moons dominating the mass budget. Half of the systems were initialized with integer-ratio resonances between the moon periods, and half were initialized at random periods. In both cases, the first moons' initial semi-major axes was chosen at random, half of these from a uniform distribution and the other half from a log-uniform distribution; each subsequent moon was placed farther out than the last, either in or out of resonance as the case may be.   As in \citealt{Teachey:2021}, the resonant systems had period ratios for neighboring moons $b:a$ selected randomly such that $a \in \{ 1,2,3,4,5 \} $ and $b \in \{ a+1, a+2, ..., 4a \}$. As a result, there are pile-ups at the reducible ratios; for example, there will be more 2:1 resonances, as the algorithm also produces 4:2, 6:3, 8:4 and 10:5 resonances; likewise, the 3:2 resonance also includes 6:4 draws. Additional integer ratio resonances will also exist between non-neighboring moons: for example, if we have a 3:2 resonance between moons I and II, and the resonance between moons II and III is 7:3, moons I and III will be in 7:2 resonance. Once again, we allow the stability of the system to dictate whether such a system is plausible, rather than concerning ourselves with whether a system might have formed in nature. 

Radii for the moons were also computed using the mass-radius relationship in \textsc{Forecaster.} The moons' orbits were co-planar with one another, but together could be inclined by up to 10 degrees from the plane of the planet's orbit.  This choice is in keeping with the near-coplanarity (generally $< 0.5^{\circ}$) of the major moons of Jupiter, Saturn, and Uranus, and helped to mitigate a tendency towards significant inclination changes during the simulations which would ultimately produce many more unstable systems; it comes at the expense of removing an additional, potentially complicating degree of freedom that could be present in real systems. The moon's longitude of ascending node was selected from a uniform distribution between 0 and 2$\pi$. Moon eccentricities were also initialized at zero, in view of anticipated tidal circularization on short timescales, though the moons could end the simulations with small non-zero eccentricities due to their mutual interactions.  A total of 1000 resonant chain systems and 1000 non-resonant chain systems were generated, which were then ranked by  by median transit depth of the satellites. A subset of the highest ranking systems  in each architecture were then selected to have artificial light curves generated.

\subsection{Light curve production}
Initially we experimented with using unique donor light curves taken from the \textit{Kepler} catalogue for each model injection, drawn from a list of light curves with no known planet transit signals, in order to simulate the randomness of a real sample (including a range of astrophysical trends, differing levels of photometric noise, and varying time baselines and footprints), as well as the complications associated with handling real data (including cleaning outliers and detrending). However, we eventually found that this added too many variables and needless complications to our experiment, obscuring the role of the system architecture itself in the result (the primary focus of this investigation), so we find it is preferable to keep the donor light curve the same for every injection. The signal-to-noise of the final light curve, then, is dictated entirely by the modeled transit, shallower transits having poorer signal-to-noise. We selected a single \textit{Kepler} light curve as the donor light curve footprint (Kepler-1625), and generated 10 ppm Gaussian noise around a flat line with flux count of 30,000. This approach obviates the need for detrending the light curve, as there are no trends present, so in the analysis that follows it is assumed that the light curves have been well-detrended. Generally speaking, we can achieve adequate detrending when light curves receive individual attention; light curves detrended in bulk tend to have more issues, owing to the one-size-fits-all approach to handling a range of astrophysical and instrumental features.  The benefit of adopting a real light curve footprint is that data gaps (both quarter-to-quarter and individual, stochastic dropouts) that are characteristic of all \textit{Kepler} light curves are present just as they would be in a real light curve. But by using the same footprint for each simulated light curve, we remove these gaps as a variable that might affect recoverability. Keeping the 30-minute \textit{Kepler} cadence, rather than adopting the higher cadence of (for example) TESS, ought to keep the results of this study more broadly applicable, as shorter cadence (higher frequency) observations may always be binned down to a longer cadence, but the reverse is of course not true. However, in utilizing 10 ppm precision, we are anticipating a future, larger aperture space telescope that can achieve precisions far better than recent and current space-based survey missions.

Using the donor \textit{Kepler} lightcurves and the stable system architecture parameters as inputs, we proceeded to generating the simulated light curves, using the photodynamical modeling code \textsc{Pandora} \citep{Pandora}. We made a few additional choices that were not simulated by \textsc{Rebound} but are required inputs for the transit modeling code:

\begin{itemize}
    \item The longitude of the ascending node $\Omega_{S}$ for the satellite system was chosen at random from a uniform distribution between 0 and 2$\pi$;

    \item The moon's inclination with respect to the planet's orbit was restricted to be $\pm 10$ degrees from the planet's orbital plane;
    
    \item The system's argument of periastron $\omega_{\mathrm{B}}$ was set to zero, but this plays no role as all the systems are on circular orbits.
    
    \item The \textsc{Pandora} input $\tau_{0,\mathrm{offset}}$ which fits an offset for linear ephemeris, was set to zero.
\end{itemize}

\textsc{Pandora} natively supports the generation of noise-free transit models containing either a planet by itself, or a planet hosting a single moon. In the present case, however, we must also generate systems with multiple moons (up to five). To create the transit signals for these systems using \textsc{Pandora}, we generated the transit signals of each planet-moon pair individually, modeling not just the transits themselves but also any associated transit timing variations (TTVs) produced by the presence of the moon. We used a cadence of 2.88 seconds (30000 data points per day).

The timing offset of a planet hosting multiple moons in any given epoch will simply be the sum total of timing offsets due to each individual moon. We therefore measured the TTVs of the planet produced by each individual moon and computed for each epoch a final, total offset from the system's barycentric linear ephemeris. \textsc{Pandora} natively outputs planet-only, moon-only, and planet+moon light curves, so it is straightforward to isolate planet and moon transit signals. To construct the final light curve, all transits (planet and moons) were placed individually in their appropriate locations with respect to the system's barycenter, which itself is traveling on a perfect Keplerian orbit, based on their calculated timing offsets. Mutual moon-moon occultations are not accounted for, though two or more moons transiting simultaneously will be accurately represented. 

We note that while the methodology above produces accurate TTVs, we made no attempt to model planet transit duration variations (TDVs) for these multi-moon systems, which is considerably more complex to compute. Because the single-moon model we ultimately fit to the data will include TDVs, we may see some slight tension in the mass and semi-major axis solutions (informed by both TTVs and TDVs in a real system), and in the inferred inclination of the satellite. The ratio of the velocity-induced TDV amplitude \citep[$\delta$TDV-V;][]{timingI}, to the TTV amplitude ($\delta$TTV) goes as $a_S^{-3/2}$, where $a_S$ is the semi-major axis of the moon, so the omission of TDV-Vs in the simulated data may result in overestimated solutions for the moon's semi-major axis, or a compensatory reduction in the moon's estimated mass. TDVs can also result from the presence of a moon with an orbit inclined with respect to the planet's orbital plane, which can have the effect of offsetting the planet north or south of the barycenter's transit chord, resulting in longer or shorter transit chords for the planet and correspondingly longer or shorter transit durations \citep[][]{timingII}. As a result, our model solutions may skew towards more co-planar solutions for the moon. These tendencies should be borne in mind in the subsequent analysis.

With the final high cadence models in hand, we binned them down to the \textit{Kepler} cadence and injected them into the donor light curve. For the final sample, we required that each light curve 1) contain at least three transits of the planet, 2) display moon transit features in at least 50\% of the the transit epochs, and that 3) the combined SNR for each moon (equal to a single transit SNR times the square root of the number of epochs where the moon appears) be at least 2. Of the 325 light curves that were generated, a total of 208 light curves met these requirements. This was our final sample for analysis. Demographics of these systems generated are shown in Figure \ref{fig:simulation_demographics}.


\section{Model fitting}
\label{sec:model_fitting}

\begin{table}
\begin{center}
\begin{tabular}{||c|c|c|c||}
\hline
Attribute & Symbol & Distribution & Range / Value \\ [0.5ex]
\hline \hline
    Planet radius & $R_P$ & normal & $\sigma = 0.1 R_P$\\
    \hline
    orbital period & $P_B$ & normal & $\sigma = 0.01 P_B$ \\
    \hline  
    semi-major axis & $a_B$ & normal & $\sigma = 0.1 a_B$ \\
    \hline
    impact parameter & $b_B$ & uniform & $[0,1]$ \\
    \hline 
    timing offset & $\tau_0,\mathrm{offset}$ & uniform & $[-1,1]$ \\
    \hline
    arg. peri. & $\omega_B$ & uniform & $[0^{\circ},360^{\circ}]$ \\
    \hline
    eccentricity & $e_B$ & uniform & $[0,0.99]$ \\
    \hline
    limb darkening & $q_1, q_2$ & uniform & $[0,1]$ \\
    \hline 
    Planet mass & $M_P$ & fixed & $M_P$ \\
    \hline
    Moon radius & $R_S$ & fixed & $10^{-8}$ \\
    \hline
    Moon mass & $M_S$ & fixed & $10^{-8}$ \\
    \hline
    Moon period & $P_S$ & fixed & 30 d \\
    \hline
    Moon phase & $\tilde{\phi}$ & fixed & 0. \\
    \hline
    long. asc. node & $\Omega_S$ & fixed & 0. \\
    \hline 
    Moon inclination & $i_S$ & fixed & 0. \\
    \hline
    
\end{tabular}
\end{center}
\caption{Priors and fixed inputs for the planet-only model fits.}
\label{tab:planet_priors}
\end{table}

\begin{table}
\begin{center}
\begin{tabular}{||c|c|c|c||}
\hline
Attribute & Symbol & Distribution & Range \\ [0.5ex]
\hline \hline
    Planet mass & $M_P$ & normal & $\sigma = 0.05 M_P$ \\
    \hline 
    Moon radius & $R_S$ & log-uniform & [$10^{-4}, 0.9$] $R_P$\\
    \hline
    Moon mass & $M_S$ & log-uniform & [$M_{S,\mathrm{min}}, 0.9 M_P]$\\
    \hline
    Moon period & $P_S$ & log-uniform & [$P_{\mathrm{Roche}}$, $P_{\mathrm{Hill}}$] \\
    \hline 
    Moon phase & $\tilde{\phi}$ & uniform & $[0, 1]$ \\
    \hline 
    long. asc. node & $\Omega_S$ & uniform & $[0^{\circ}, 360^{\circ}]$ \\
    \hline
    Moon inclination & $i_S$ & normal & $\mu = 90^{\circ}$; $\sigma = 5^{\circ}$ \\
    \hline
    \end{tabular}
    \caption{Additional priors for the planet+moon fits (fixed for the planet-only models).}
    \label{tab:moon_priors}
\end{center}
\end{table}

\begin{figure*}
    \centering
    \begin{subfigure}{\textwidth}
    \includegraphics[width=\linewidth]{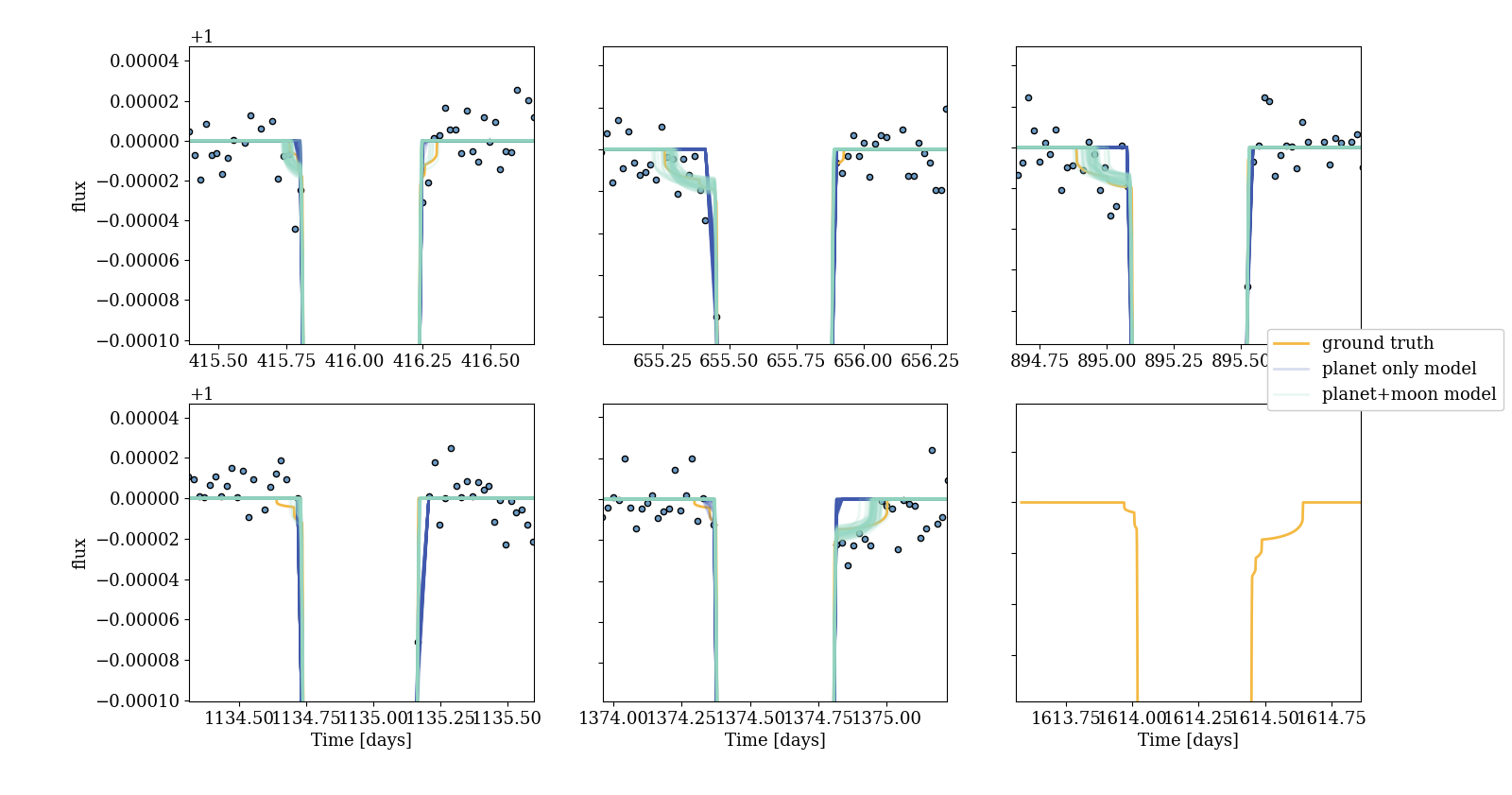}
    \end{subfigure}
    \begin{subfigure}{\textwidth}
    \includegraphics[width=\linewidth]{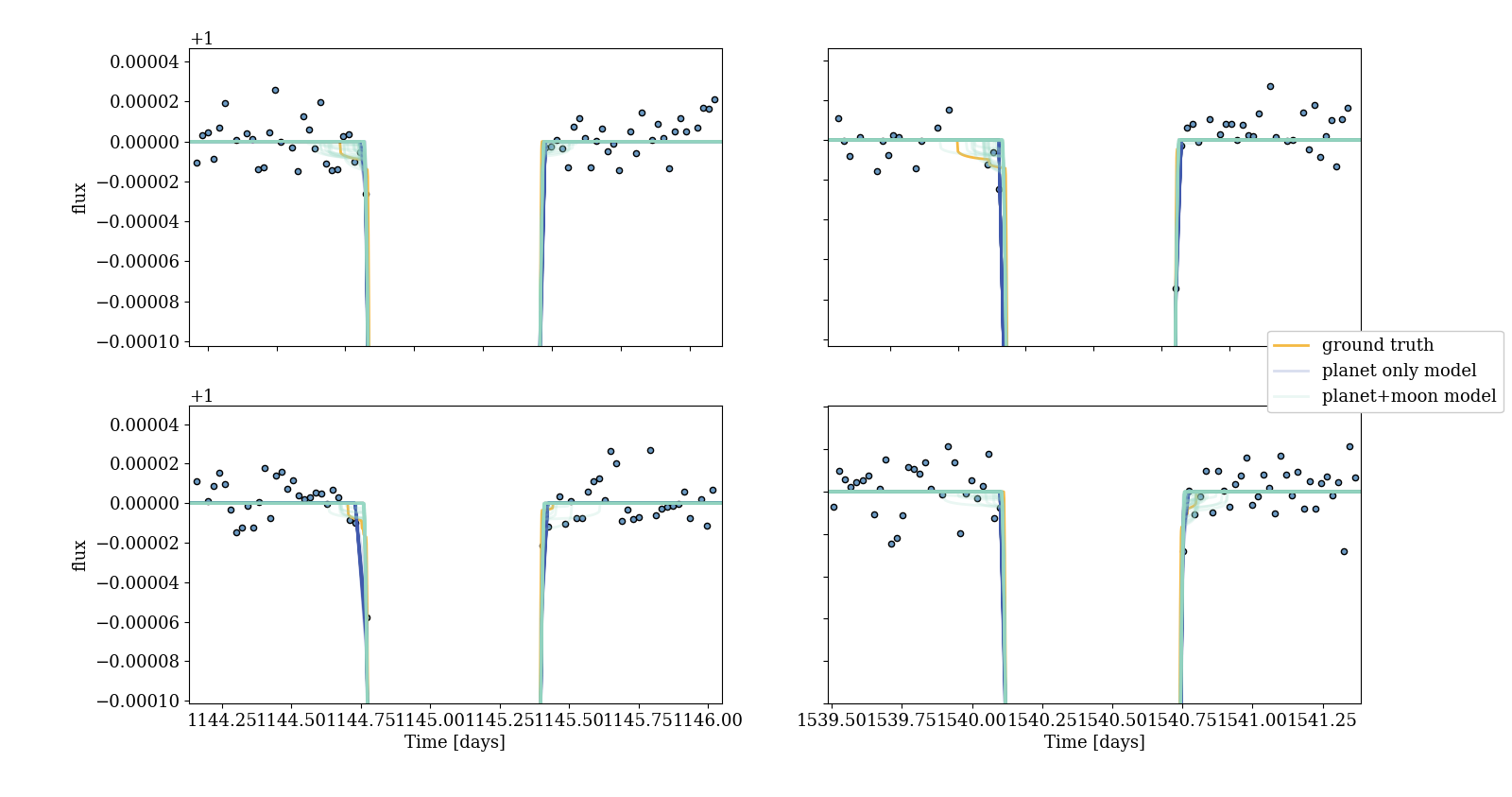}    
    \end{subfigure}
    \caption{\textit{Top six panels:} An example light curve fit with 50 draws of the planet+only (dark blue) and planet+moon (cyan) model posteriors, in the case where there is substantial evidence in favor of the moon. We have zoomed in on the out-of-transit moon dips for the sake of clarity, but the light curve also displays moon dips coinciding with the planet's transit. The ground truth light curve (including 4 transiting moons) is indicated in yellow. In this case, the single-moon model has accurately identified large moon transit dips in the first, second, third, and fifth epoch. The final panel shows a transit that fell in a data gap. \textit{Bottom four panels:} the same as above, for a system that did not return sufficient evidence in favor of a moon. This system contains three moons, and though the planet-moon model did detect the moon dips in some cases, it was not sufficient to return a verdict in favor of the moon.}
    \label{fig:example_lightcurve}
\end{figure*}

Once again, we utilize the \textsc{Pandora} code for generating the planet-only and planet+moon transit models that will be fit to the data. Because the objective of the present work is to examine the performance of a single-moon model on systems containing multiple moons, we require no modifications to the moon-model generation for this stage.

Model fitting is performed using the nested-sampling code \textsc{UltraNest} \citep{UltraNest}, which may be used for both parameter estimation and the computation of Bayesian evidences ($Z$). We fit 2 models to the simulated lightcurves: a planet-only model, and a planet+moon model. For the star- and planet-related parameters, we fit them using the priors shown in Table \ref{tab:planet_priors}. Here and throughout, $P$ and $B$ subscripts refer to the planet and the planet-moon barycenter, respectively. We will use the subscript $S$ to refer to the satellite.

Were this a real search, we would adopt previously established values from the NASA Exoplanet Archive as our priors, using a normal distribution centered on the catalogue value ($\mu$) and take the width ($\sigma$) from the published uncertainties. In the present case, we know the ground truth because we are generating artificial systems, but we should not use the ground truth as $\mu$ because that would neglect the uncertainties of a real measurement. As such, we simulate a real observation by perturbing the ground truth using typical uncertainties for the various parameters (sourced from the NASA Exoplanet Archive), and generate priors centered on these ``derived'' system attributes. 

The planet+moon model incorporates 7 additional parameters. In the case of a real search, we would have no previously catalogued system parameters to use as priors, so instead we adopt uninformative priors for all the moon's parameters, with the exception of the planet's mass (which has no bearing on the planet-only light curve and is therefore only included in the moon model). The priors for the additional parameters are shown in Table \ref{tab:moon_priors}. \noindent Note that \textsc{Pandora} natively uses the symbol $\tau$ for the normalized phase of the moon, but we opt to use $\tilde{\phi}$ so as not to confuse it with a transit time. 

Perhaps the most consequential of our choices in priors is to use a normal distribution for moon inclination around $0^{\circ}$ (co-planar to the planet's orbital axis), with $\sigma = 5^{\circ}$. This is a thumb on the scale to prefer roughly co-planar orbits solutions over inclined ones. We found from early model runs that a uniform prior allowing any inclination had the effect of nearly always tending towards highly inclined solutions, with little  to no need to display any moon transits in the data. Effectively, the model could account for the TTVs without needing any additional evidence in favor of moons. Of course, these architectures are possible, and if they are strongly favored by the data, the normal prior still allows it. The fit must simply be that much stronger to overcome the probability deficit associated with this inclined solution.

We adopted hyperparameters for \textsc{UltraNest} fitting directly from the \textsc{Pandora} documentation and tutorial.\footnote{\url{https://github.com/hippke/Pandora/blob/main/examples/injection_retrieval_simple_ultranest.ipynb}} We used the \texttt{ReactiveNestedSampler} with the \texttt{RegionSliceSampler}, setting \texttt{nsteps} to 4000, \texttt{adaptive\_nsteps} = \texttt{move-distance}, and using 1000 live points. We experimented early on with varying these hyperparameters, but we found no significant changes in model fitting performance or results so we chose to stick with the \textsc{Pandora} recommendations.

\section{Results}
\label{sec:results}

\subsection{Model fitting overview}
In all, 208 light curves were fit with planet-only and planet+moon models. These 208 represent those generated light curves that met the requirement of exhibiting at least three transits in the data and having a multi-transit integrated SNR of at least 2 for each moon. Of these 208 systems, 59 host single moons, 60 host two moons, 44 host three moons, 25 host four moons, and 20 host five moons. The diminishing number of moons at higher values of $N$ is due to the fact that these moons must generally be smaller to survive in multi-moon systems \citep[][]{Teachey:2021}, and thus there are fewer of them that meet our moon SNR requirement. A total of 416 model fits were run: a planet-only and planet+moon model for each system. Of these 208 systems, only 141 systems had converged moon solutions after multiple 5-day runs. As we require both planet-only and planet+moon models to run to completion for model selection, this further reduced our sample to 141. Figure \ref{fig:example_lightcurve} shows two systems with models drawn from the posteriors overlaid, the first of which returned substantial evidence in favor of a moon, while the second one did not.

We further screened the results with the following cuts: first we removed any systems for which either the planet or moon's $\log Z < -10^4$, which indicate clearly demonstrably bad fits -- this removed 22 planet fits and 10 moon fits. We then fit a line to scatter plots of $\log Z$ as a function of the number of the data points, 
and iteratively kicked out outliers one at a time and refit the line until the number of $\log Z$ values above and below the line were equal to within 5\%. 
This choice was motivated by the observation that there is a clear linear relationship between $\log Z$ and the number of flux measurements, with the exception of a small minority of results that are far off this trend. And indeed, this is what we should expect, as the calculation of the Bayesian evidence (marginal likelihood) incorporates the likelihood function, which is itself just $-0.5 \chi^2$ and scales with the number of data points.

\begin{figure}
    \centering
    \includegraphics[width=\columnwidth]{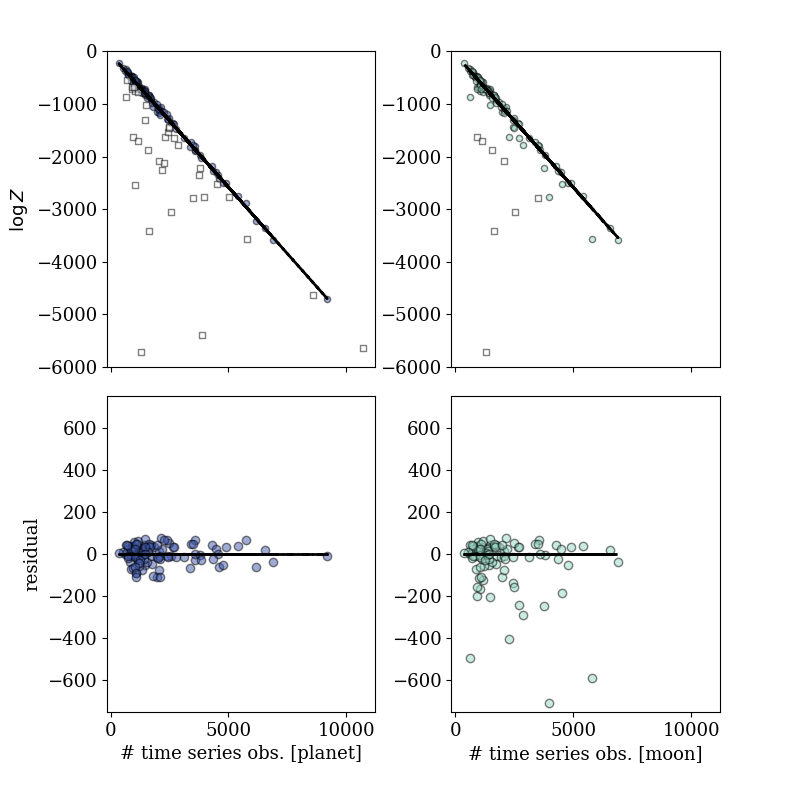}
    \caption{$\log Z$ values for the planet-only (left) and planet+moon (right) fits used in our final sample, as a function of  the number of flux measurements in the light curve. Discarded models are indicated with empty squares.}
    \label{fig:logz_vs_ndata}
\end{figure}

Final fits and their residuals can be seen in Figure \ref{fig:logz_vs_ndata}. As we require both planet-only and planet+moon fits to be available, the final number will consist of those systems for which both models ran to completion. Following these cuts, we were left with a total of 104 usable systems. This left us with 27 one-moon systems, 26 two-moon systems, 26 three-moon systems, 14 four-moon systems, and 11 five-moon systems.

Let us now examine some of the trends in the results, or lack thereof.

\subsection{No trend based on the number of moons present}
\begin{figure}
    \centering
    \includegraphics[width=\columnwidth]{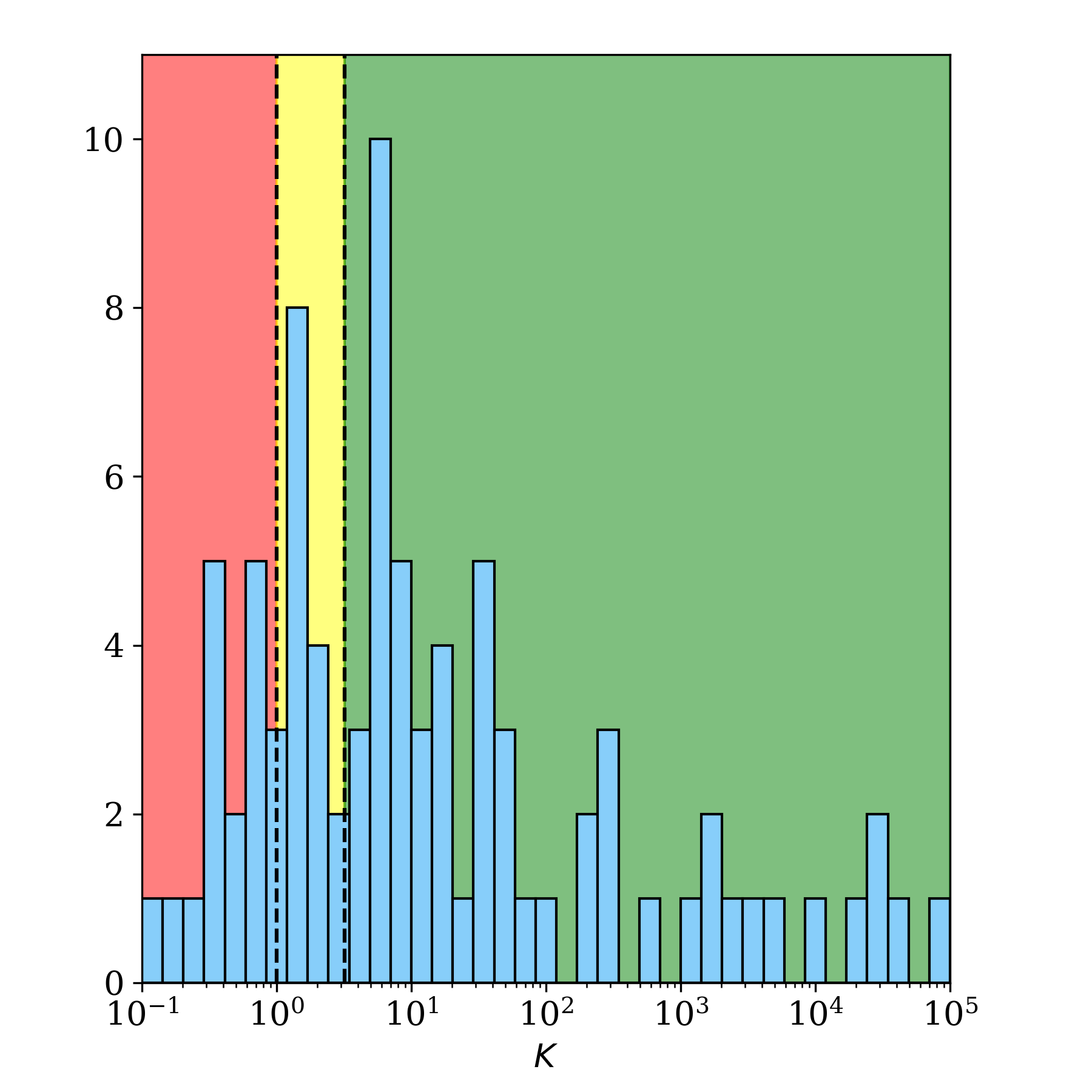}
    \caption{Distribution of Bayes factors $K = \mathrm{exp} [\Delta \log Z]$ comparing planet-only and planet+moon models. The red region on the left indicates Bayes factors that are solidly against the moon hypothesis ($K < 1$). The yellow region in the middle indicates Bayes factors that are marginally in favor of the moon hypothesis, but would generally not be considered sufficiently robust to warrant further investigation ($1 \leq K \leq 3.2$). The green region at right shows those systems that do show significant evidence in favor of the moon hypothesis ($K > 3.2$).}
    \label{fig:bayes_factor_distribution}
\end{figure}

\begin{figure}
    \centering
    \includegraphics[width=\columnwidth]{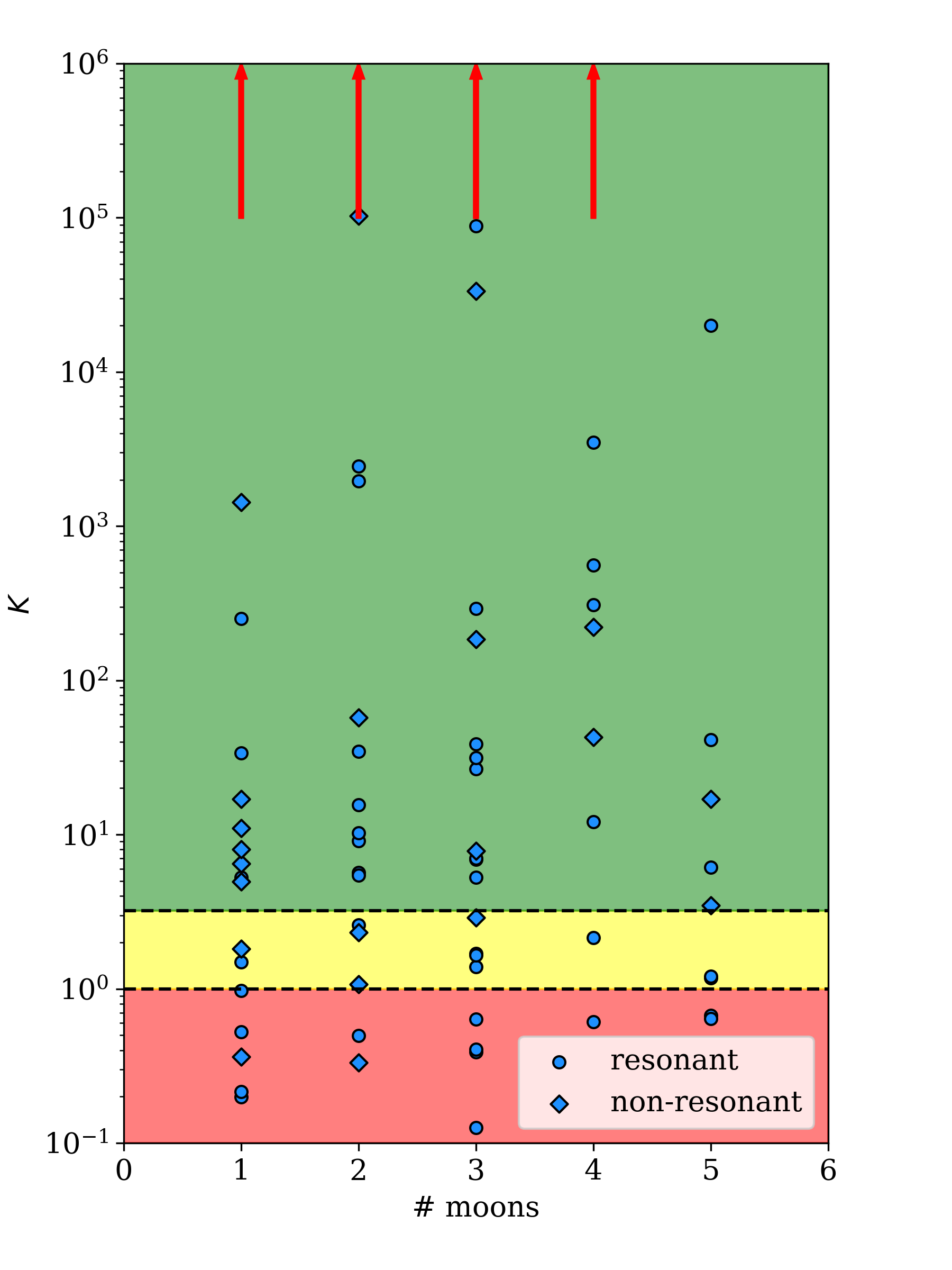}
    \caption{The evidence in favor of moons as a function of the number of moons in the system. No trend is apparent. The green shaded region (top) indicates values of $K$ that are associated with substantial evidence for a moon. The yellow region (middle) indicates positive but weak evidence for a moon. The red region (bottom) indicates evidence against the presence of a moon.}
    \label{fig:bayes_factor_vs_nmoons}
\end{figure}

\begin{table}
\begin{center}
\begin{tabular}{||c|c|c|c|c||}
\hline
\# moons & \# systems & \# used & \# $K > 3.2$ & \# $K > 10^4$ \\ [0.5ex]
\hline \hline
    1 & 59 & 27 (45.7\%) & 18 (66.6\%) & 6 (22.2\%) \\
    \hline 
    2 & 59 & 26 (44.1\%) & 19 (73.1\%)  & 6 (23.1\%) \\
    \hline
    3 & 44 & 26 (59.1\%) & 17 (65.4\%) & 6 (23.1\%) \\
    \hline
    4 & 25 & 14 (56.0\%) & 12 (85.7\%) & 4 (28.6\%) \\
    \hline 
    5 & 20 & 11 (55\%) & 6 (54.5\%) & 1 (9.1\%) \\
    \hline
    \hline
    total & 208 & 104 (50\%) & 72 (69.2\%) & 23 (22.1\%) \\ 
    \hline
    \end{tabular}
    \caption{Results of the the model fitting. The \# of systems represents the number of systems run, while the \# used is the number of systems that were deemed usable after various quality cuts. The \# with $K > 3.2$ indicates the number of systems for which significant evidence for a moon exists, while the \# with $K > 10^4$ are those with exceedingly high evidence in favor of a moon, possibly resulting from a poor planet fit. These percentages are from the number of used systems.}
    \label{tab:nmoon_stats}
\end{center}
\end{table}

Figure \ref{fig:bayes_factor_distribution} shows the distribution of Bayes factors $K$ from the final sample of 104 systems. Systems in the red (left) indicate evidence against moons ($K < 1$), while those in the yellow (middle) are those with ``marginal'' evidence for a moon ($1 \leq K \leq 3.2$), and those in green (right) showing substantial evidence ($K > 3.2$). The distribution peaks in the green, and includes a long tail into very high Bayes factors. These metrics for interpreting the Bayes factor are adopted from the widely-cited \citealt{Bayes_Factors} framework. Based on this metric alone, the strength of the favored model scales with the value of $K$, higher numbers corresponding to greater confidence that that model is correct. However, it should be borne in mind that extremely high values of $K$ may also be indicative that the fit of the disfavored model is exceptionally poor, rather than the fit of the favored model being exceptionally strong.

Figure \ref{fig:bayes_factor_vs_nmoons} goes on to show these Bayes factors as a function of the number of moons present in the system. Visual inspection suggests a slight trend in the plot, indicating that the more moons that are added to the system, the weaker the evidence for the presence of a moon. This would accord well with our expectations, particularly if more moons generally means smaller moons.

However, Table \ref{tab:nmoon_stats} tells a somewhat different story. At least half of all systems returned significant evidence for the presence of a moon, regardless of architecture, while roughly a quarter of each architecture (with the exception of $N=5$) showed exceptionally high evidence of the presence of a moon. The outlier here is the four-moon systems, which showed an $\sim85$\% recovery rate. 

There are competing factors that may be at play here. On the one hand, systems with fewer moons generally allow each moon to have larger radii, as they can be more massive, both from the perspective of getting a greater percentage of the total mass budget, and also because systems with fewer moons are more stable at larger mass ratios (see \citealt{Teachey:2021}). On the other hand, systems with more moons can show overlapping transit features, which when combined may mimic a single, deeper transit feature. At the same time, these multi-moon systems can show transit features in multiple locations during a single epoch, which may have the effect of giving the model more options to find \textit{something}, and perhaps also more flexibility in the model fitting. As we will see later, the models are finding evidence of moons, but not necessarily correctly identifying the moons, which suggests that this flexibility is perhaps a confounding issue also.

\subsection{Few predictive features for recovery}

\begin{figure*}
    \centering
    \begin{subfigure}{0.45\textwidth}
    \includegraphics[width=\linewidth]{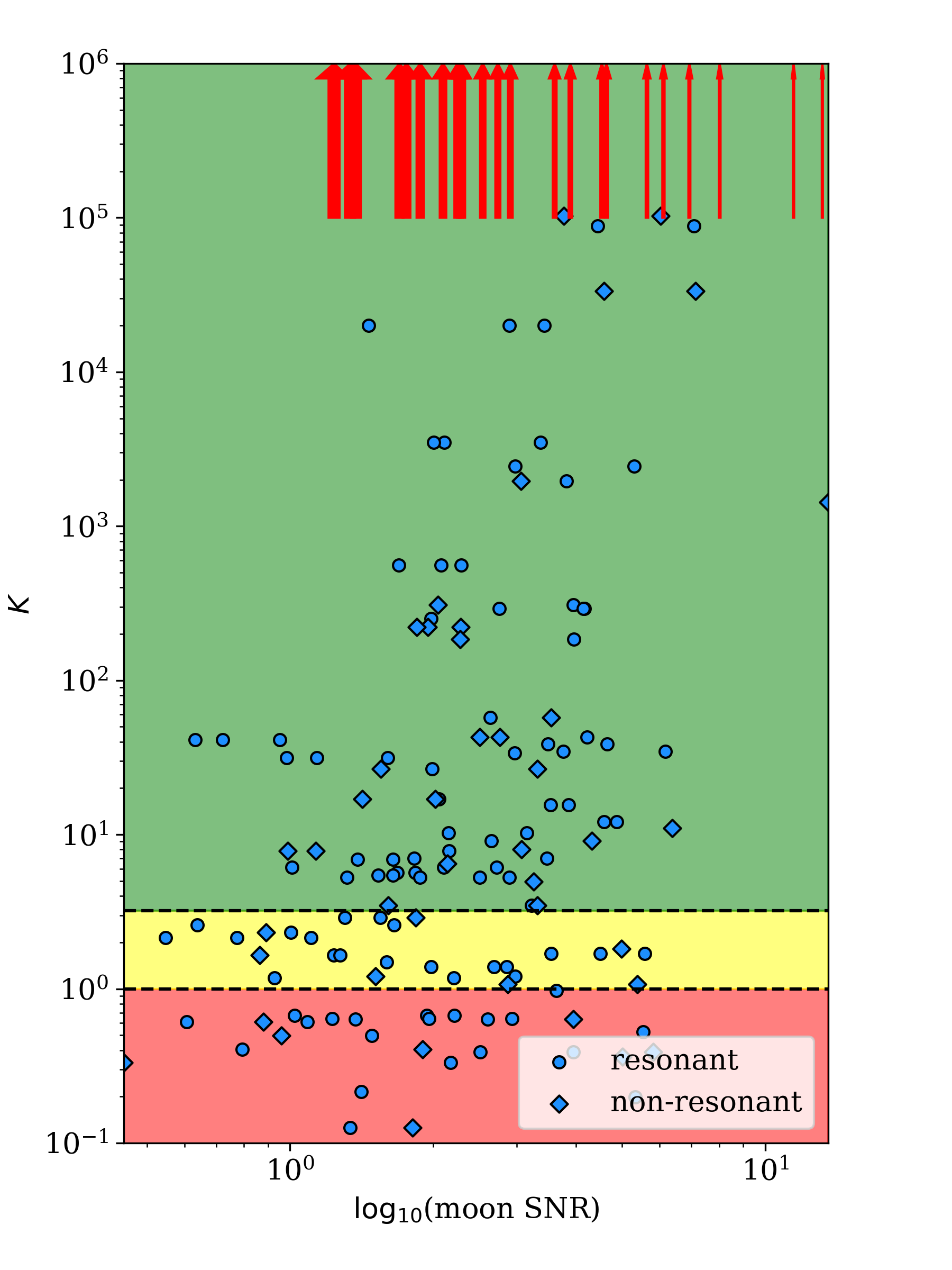}
    \end{subfigure}
    \begin{subfigure}{0.45\textwidth}
     \includegraphics[width=\linewidth]{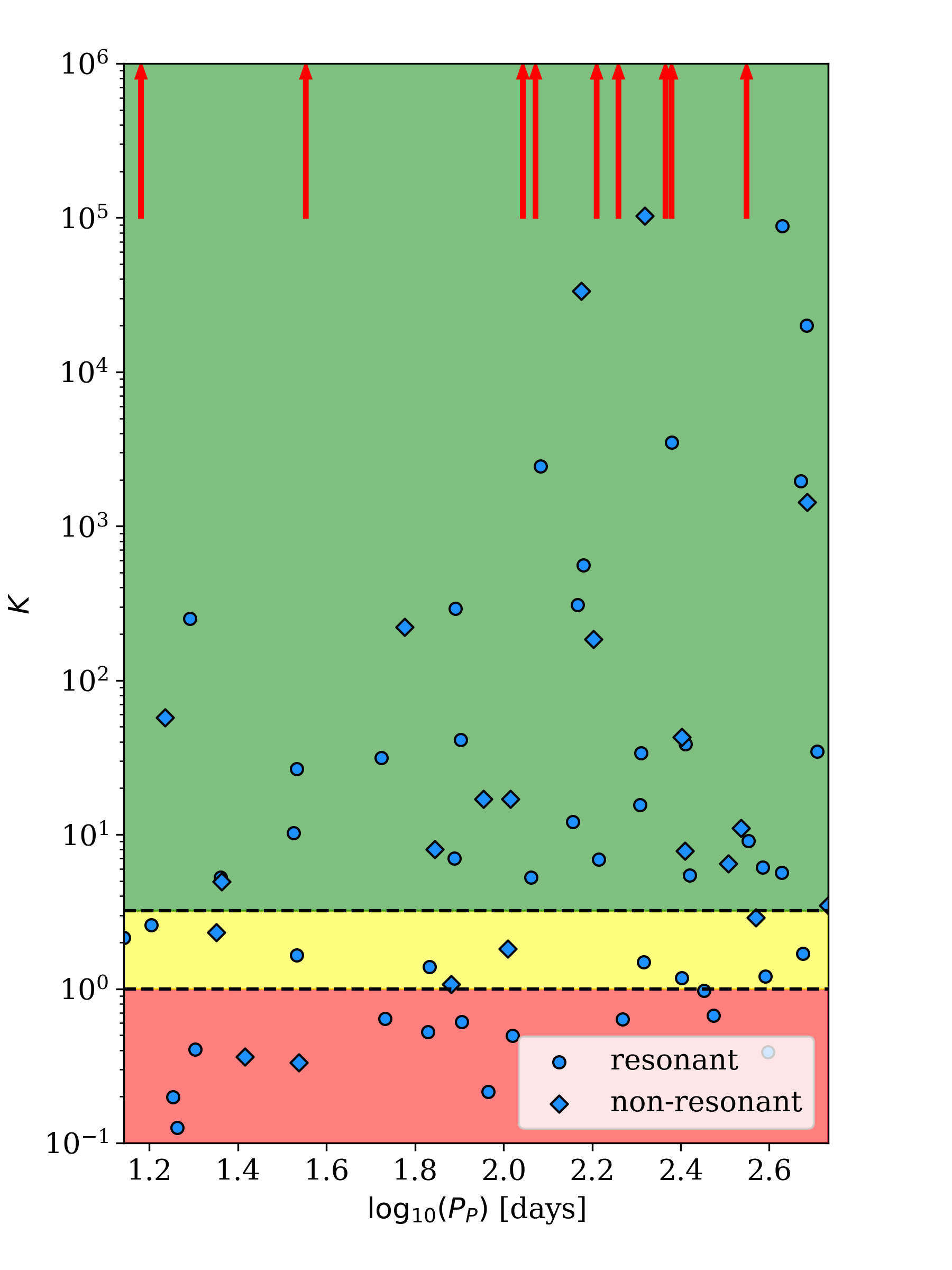}
    \end{subfigure}
    \begin{subfigure}{0.45\textwidth}
    \includegraphics[width=\linewidth]{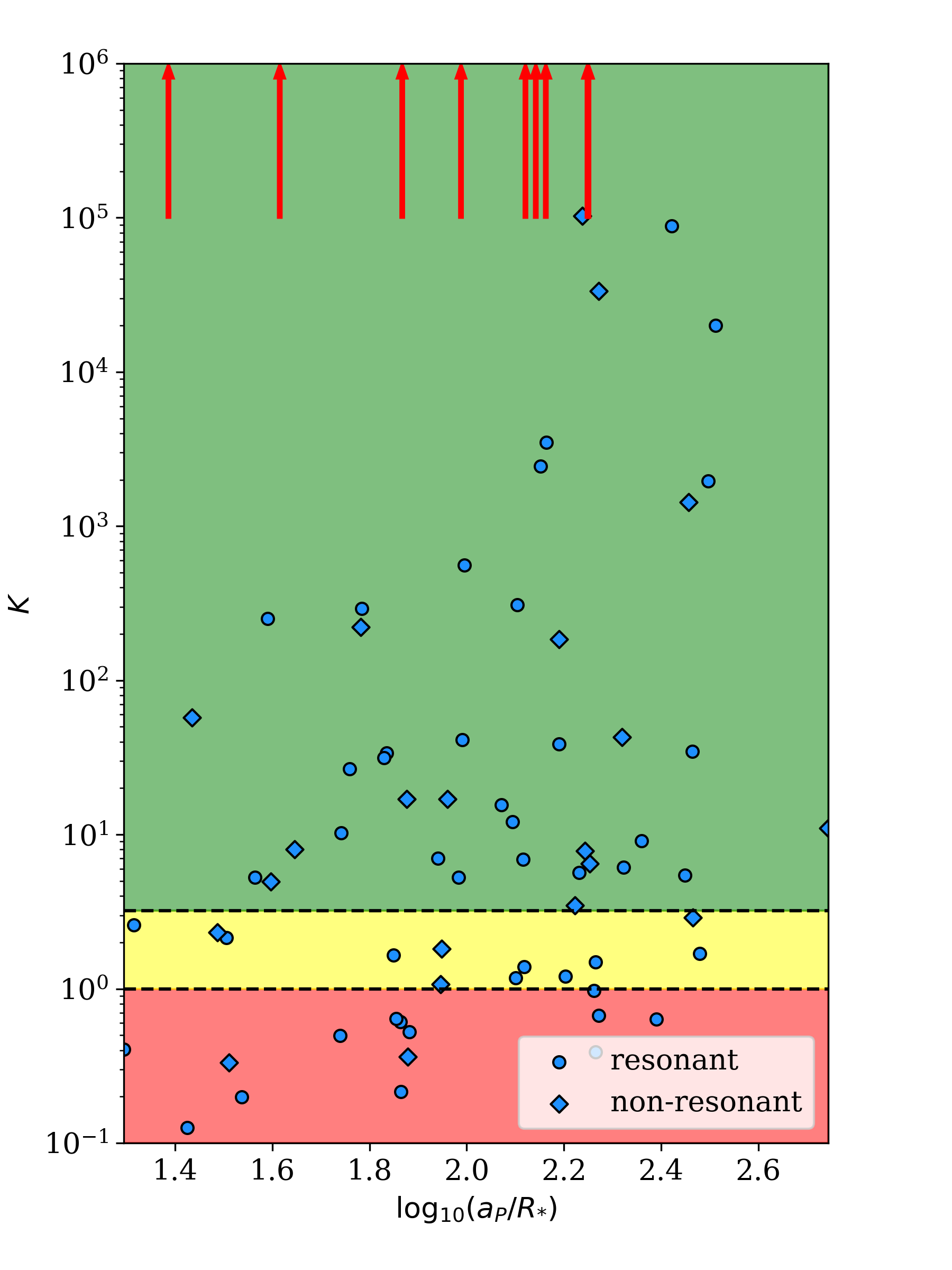}
    \end{subfigure}
    \begin{subfigure}{0.45\textwidth}
    \includegraphics[width=\linewidth]{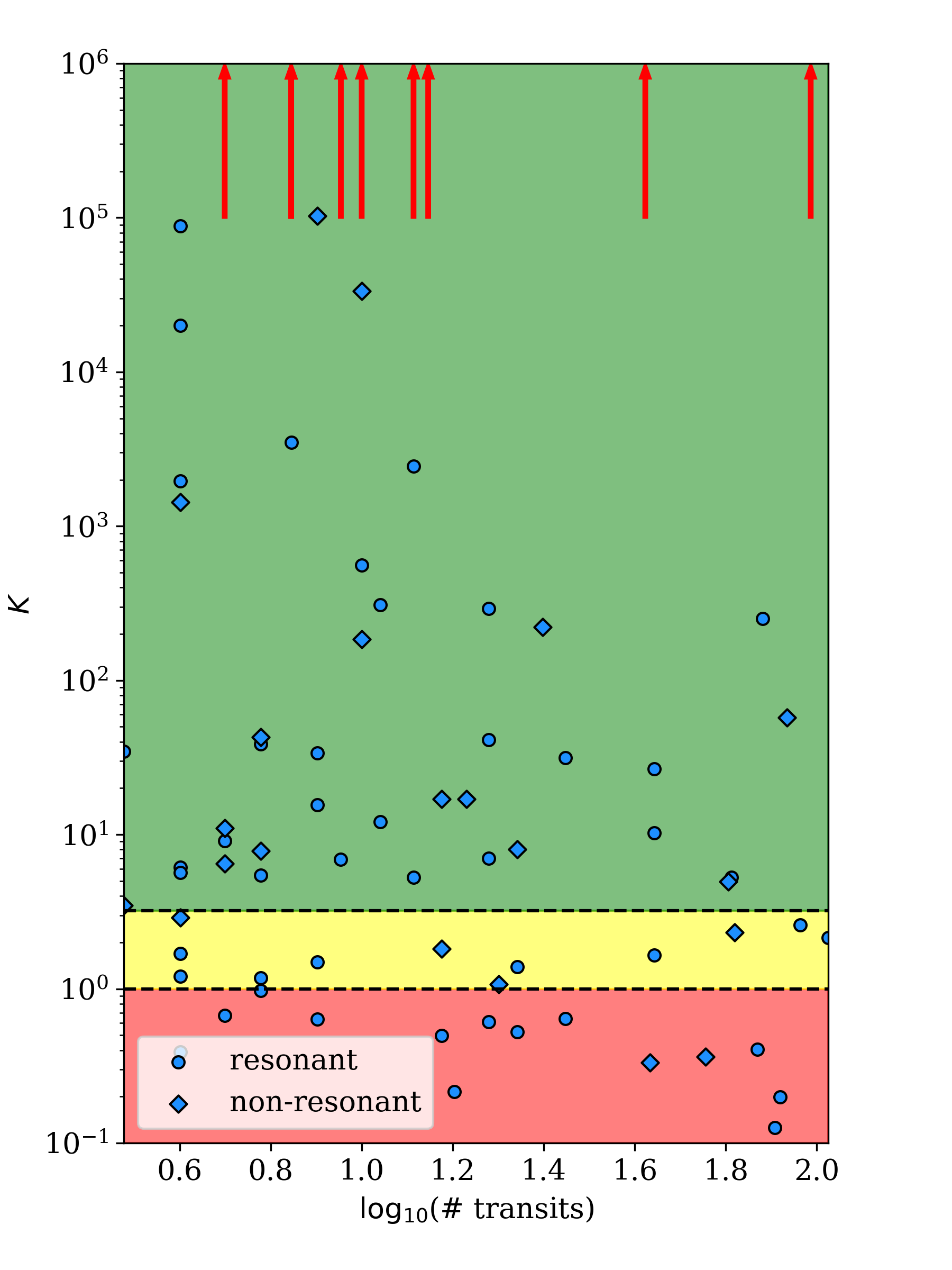}
    \end{subfigure}

    \caption{\textit{Top left:} The Bayes factors $K$ as a function of moon SNRs. \textit{Top right:} $K$ as function of the planet's orbital period. \textit{Bottom left:} $K$ as a function of the planet's semi-major axis divided by the stellar radius $a/R_{*}$. \textit{Bottom right:} $K$ as a function of the number of planet transits in the data. All of these relationships show considerable scatter and are not exceptionally predictive on their own.}
    \label{fig:predictive_features}
\end{figure*}

In examining the strength of evidence for or against the moon as a function of the various ground truth system parameters, we find that nearly all of these features by themselves play little-to-no role in whether the model fitting returns evidence in favor of a moon. Of course, we expect the system features to play \textit{some} role in the chances of recoverability or the strength of evidence in favor of a moon, but in most cases it seems likely that it will be some combination of these factors that determine recoverability. 

The clearest relationship between the Bayes factor and a system attribute -- though hardly definitive -- is based the SNR of the moon transits (Figure \ref{fig:predictive_features}). The SNR is computed from the depth and duration of the transits, so this is quite naturally associated with moon radii and their respective masses, as well as planet orbital periods / semi-major axes (longer period planets displaying correspondingly longer transit durations). But none of these features on their own show trends that are as clear, owing to the fact that 1) some small moons around small stars may have deeper transits than larger moons around larger stars, and 2) moons with longer transit durations have fewer transits present in the data, while those with shorter transits have more transits present, and these two effects are countervailing. However, we do see some indication that longer planet periods, associated with larger semi-major axes and fewer transits, show somewhat higher evidences (Figure \ref{fig:predictive_features}). It is worth noting, however, that there is considerable scatter in all of these relationships, so we do not consider any of these features on their own especially predictive.

\subsection{Resonant vs. non-resonant architectures}
Of the 104 systems used in the final analysis, 61 were resonant chain systems while 43 were not. Of the 61 resonant chain systems, 37 (60.7\%) showed substantial evidence for a moon ($K > 3.2)$. For comparison, 35 of the 41 non-resonant systems (85.4\%) showed substantial evidence for a moon.

\subsection{A shift towards higher impact parameters}
\begin{figure*}
    \centering
    \begin{subfigure}{0.32\textwidth}
    \includegraphics[width=\linewidth]{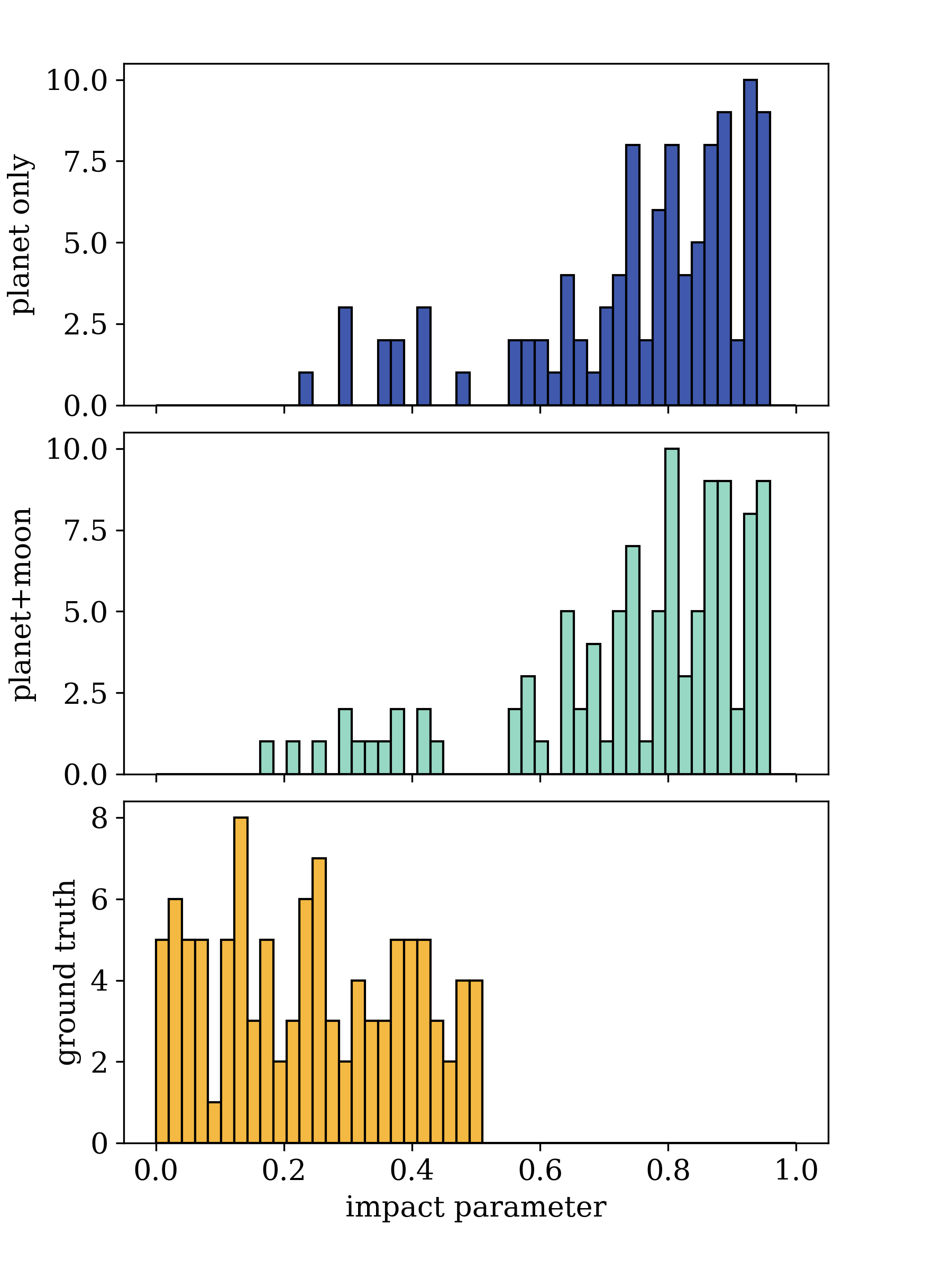}
    \end{subfigure}
    \begin{subfigure}{0.32\textwidth}
     \includegraphics[width=\linewidth]
     {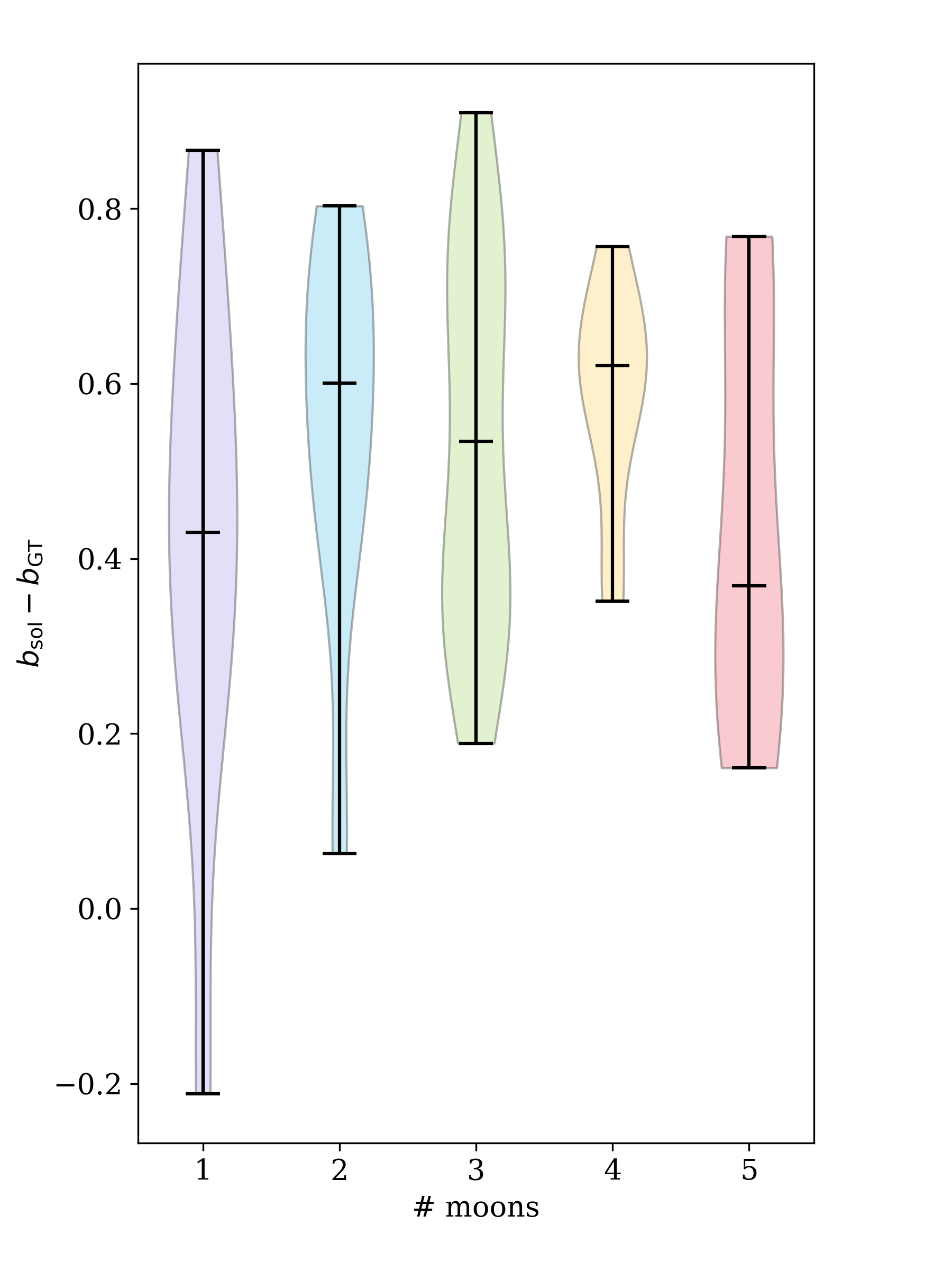}
    \end{subfigure}
    \begin{subfigure}{0.32\textwidth}
     \includegraphics[width=\linewidth]{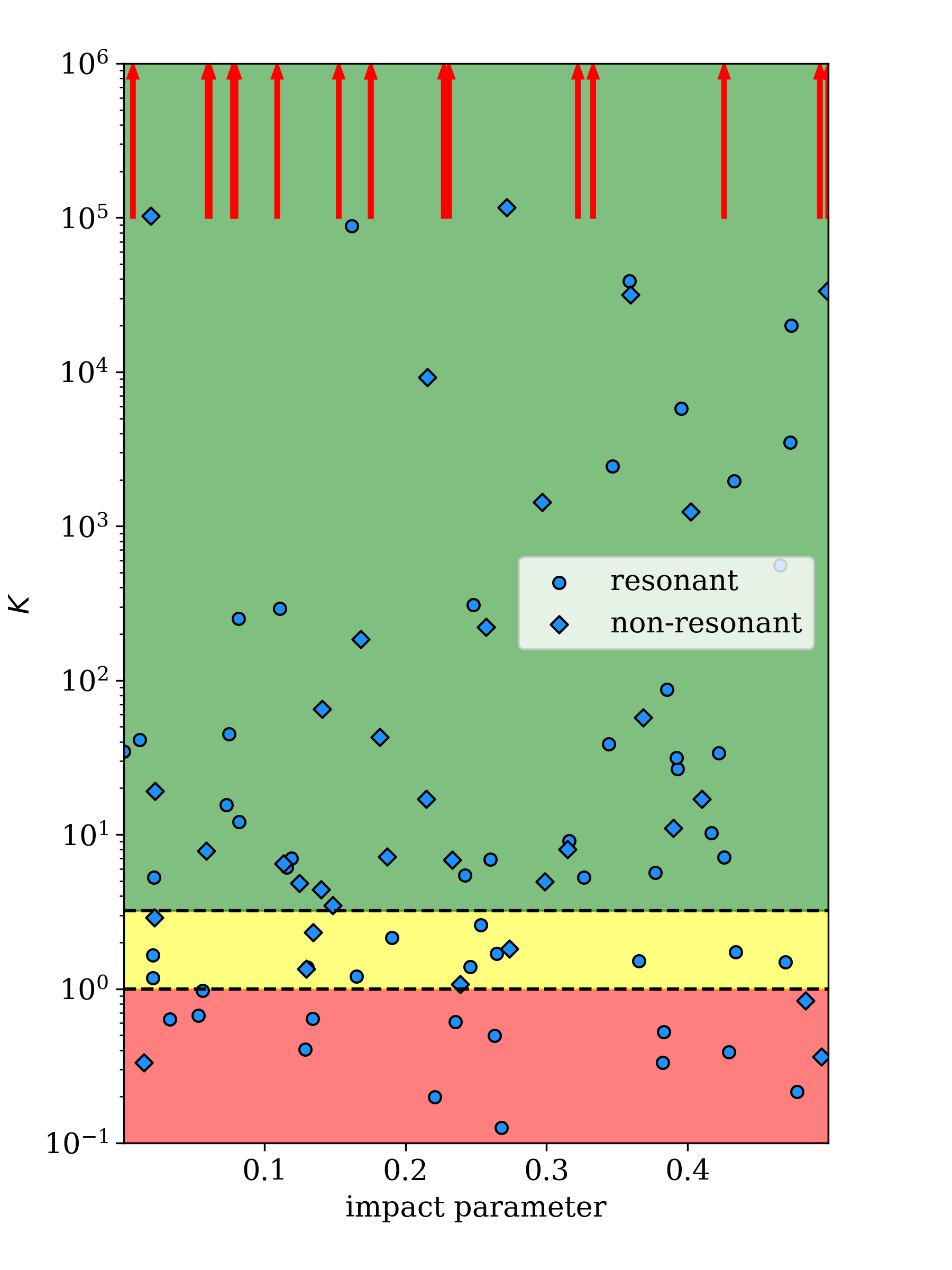}    
    \end{subfigure}
    \caption{\textit{Left:} Histogram showing planet-only model solutions for the planet's impact parameter (top), the planet+moon model solutions (middle), and the ground truth values (bottom). The results indicate a clear trend skewing towards higher impact parameters. \textit{Middle:} The distribution of impact parameter shifts $\Delta b$ (solution minus ground truth) as a function of the number of moons present in each system. \textit{Right:} Bayes factors as a function of ground truth impact parameters.}
    \label{fig:impact_parameter_plots}
\end{figure*}

A notable feature of both planet-only and planet+moon model solutions is that impact parameter solutions are consistently skewed towards higher values (see  left panel of Figure \ref{fig:impact_parameter_plots}). Recall that our artificial systems were limited to having an impact parameter $b \leq 0.5$ in order to prevent grazing transits that run the risk of hiding the moons. Figure \ref{fig:impact_parameter_plots} makes clear, however, that the vast majority of model solutions push the planet to impact parameters greater than 0.5, with only a handful at lower values.

At first glance, this result may be somewhat surprising. Higher impact parameters have the effect of reducing transit durations, and generally speaking we would expect systems displaying moon transits to mimic somewhat longer transit durations. In order to accommodate these higher impact parameters, the model solution will have to compensate somehow, which will see shortly.

However, it is perhaps not so surprising when we consider the other effect caused by higher impact parameters, that is, changing of the transit's morphology, as the planet passes over more limb-darkened area of the stellar disk. This generally has the effect of spreading and rounding the transit shape. Of course, we know that these artificial systems all have moons present, and those moon transits on the wings of the planet's transit could very well mimick this flattening effect. This certainly could explain the planet-only model's preference for higher impact parameters.

On the other hand, we might expect the planet-moon models to be able to account for this by placing a moon transit in the photometry, but there is a key difference here: Whereas a single moon will show up either before or after the planet's transits, multiple moons can display features on both sides of the planet's transit. A single moon model will be unable to account for both of these flux deficits, and so even with a moon present, a higher impact parameter is a convenient way to account for the morphological change. Curiously, we see a tendency to overestimate impact parameters even in single moon systems, though to a lesser degree. The middle panel of Figure \ref{fig:impact_parameter_plots} suggests a slight trend towards higher impact parameter skews as we add more moons, and only in the single moon systems do we see any \textit{underestimates} of impact parameter.

Fortunately, this tendency provides an observational test. Transiting systems for which we suspect one or more moons to be present can have their inclinations constrained through radial velocity monitoring: with these data in hand, the eccentricity of the planet and argument of periastron may be obtained, which in turn may be used to make a more precise prediction of the planet's transit duration, which is of course also a function of impact parameter. For systems displaying a mismatch between a photometric impact parameter solution and a radial velocity impact parameter solution, the presence of one or more moons may be the culprit and would be worth further investigation.

The right panel of Figure \ref{fig:impact_parameter_plots} plots the Bayes factor as a function of ground truth impact parameters. There appears to be a slight trend towards higher moon evidences for higher impact parameters, but this effect appears to be weak, and certainly, there are also high values of $K$ at lower impact parameters as well. Indeed, it is not clear why a higher ground truth impact parameter might contribute to higher evidences. We can only guess that it again has something to do with transit morphology.

\subsection{Inferring non-zero eccentricities}

\begin{figure*}
    \begin{subfigure}{0.45\textwidth}
    \centering
    \includegraphics[width=\columnwidth]{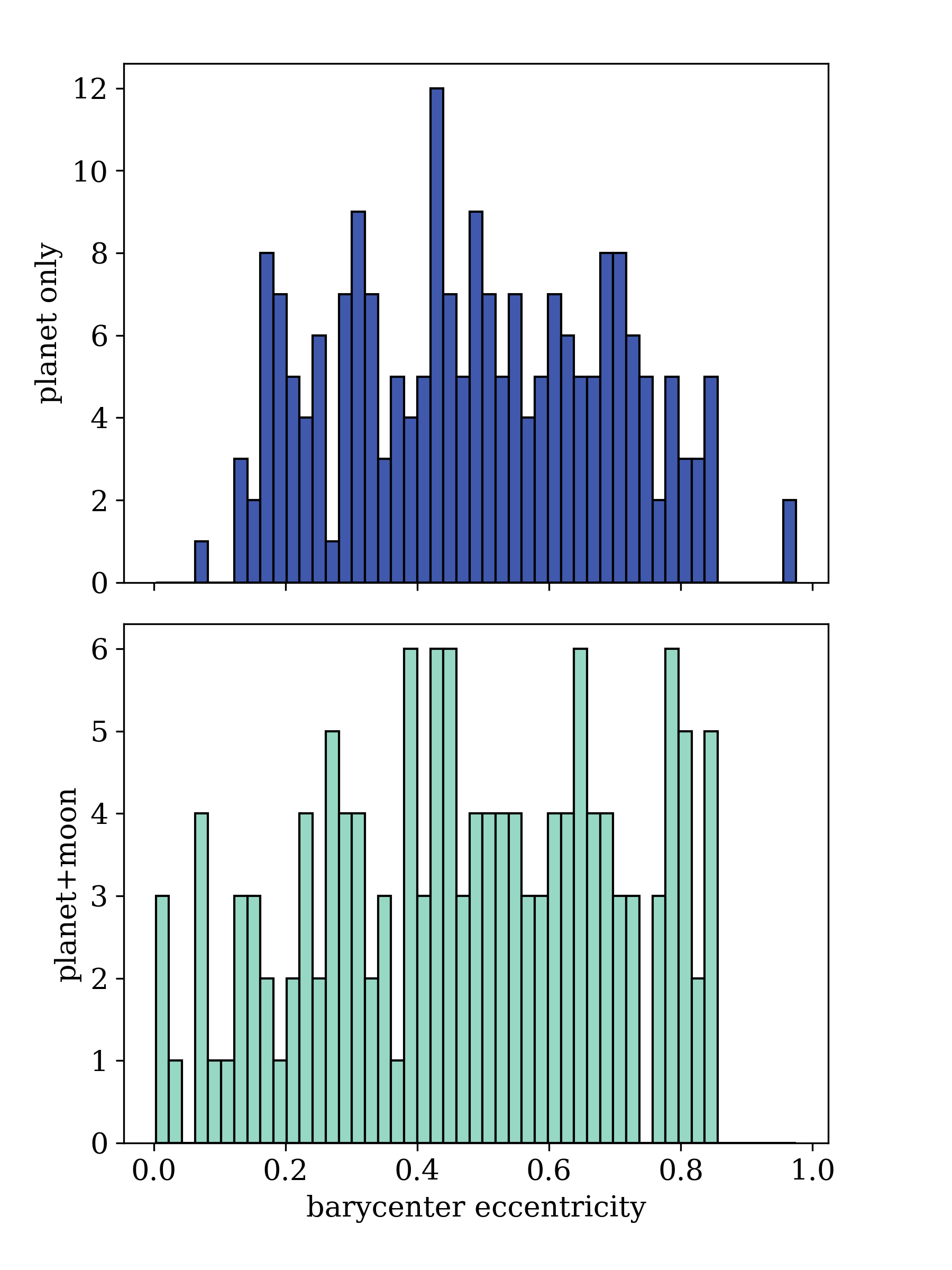}
    \end{subfigure}
    \begin{subfigure}{0.45\textwidth}
     \includegraphics[width=\columnwidth]{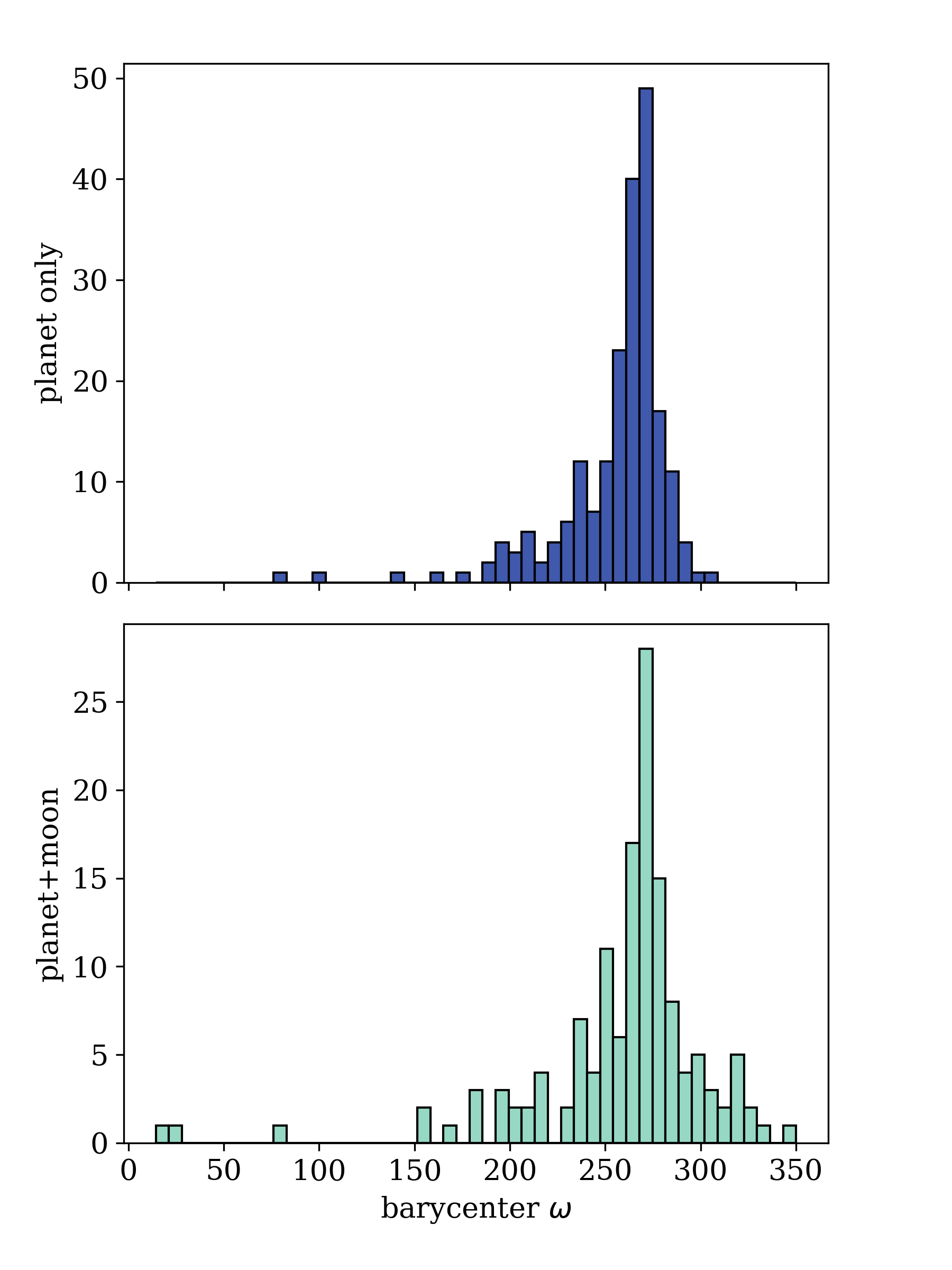}       
    \end{subfigure}
    \caption{\textit{Left:} Distribution of eccentricity solutions for the planet-only and planet+moon values (ground truth values were all zero). These results despite the fact that every system was set to have eccentricity equal to zero. \textit{Right:} Distribution of arguments of periastron $\omega$, showing a pile-up around 270 degrees. This despite the fact that every model utilized a uniform prior on $\omega$ acrosss the full 360 degree range.}
    \label{fig:ecc_bary_histogram}
\end{figure*}

Another significant result of our modeling is the consistent emergence of non-zero eccentricity solutions for the system (Figure \ref{fig:ecc_bary_histogram}), this despite the fact that every single system we generated was fixed at $e=0$. What is the explanation for this consistent skew? Again, we turn to consider the influence that moons will have on the morphology of the planet transit: the moon transits on the wings of the planet transits may mimick the effect of having a longer transit duration. It is straightforward to calculate the expected transit duration of a planet on a circular orbit, but when a planet is on an eccentric orbit, the transit duration will now be shorter or longer if we are seeing the planet closer to periastron or apoastron, respectively. At the same time, planets with eccentric orbits will display assymetric transits, owing to the fact that they will be slightly faster or slower at ingress than they will be at egress \citep[][]{photoeccentric}. As we have seen, model solution consistently tend towards higher impact parameters, which also shorten the duration. As a result, the model may compensate by pushing the planet to significant eccentricities, with a transit seen to be near the further extreme of the planet's orbit. We see evidence for this in a consistent pile-up of arguments of periastron around 270 degrees (Figure \ref{fig:ecc_bary_histogram}), despite every model having a uniform prior on this value spanning the full 360 degree range.

Fortunately, this too provides an observational test. Again, through radial velocity observations, the eccentricity of the planet may be well constrained, as can the argument of periastron. As before, if we see a significant discrepancy between the photometric model solution and the radial velocity model solution with regard to eccentricity, the presence of moons is a potential explanation that deserves investigation.

\subsection{Possibility of unphysical limb darkening solutions}

\begin{figure*}
    \begin{subfigure}{.33\textwidth}
    \centering
    \includegraphics[width=1\linewidth]{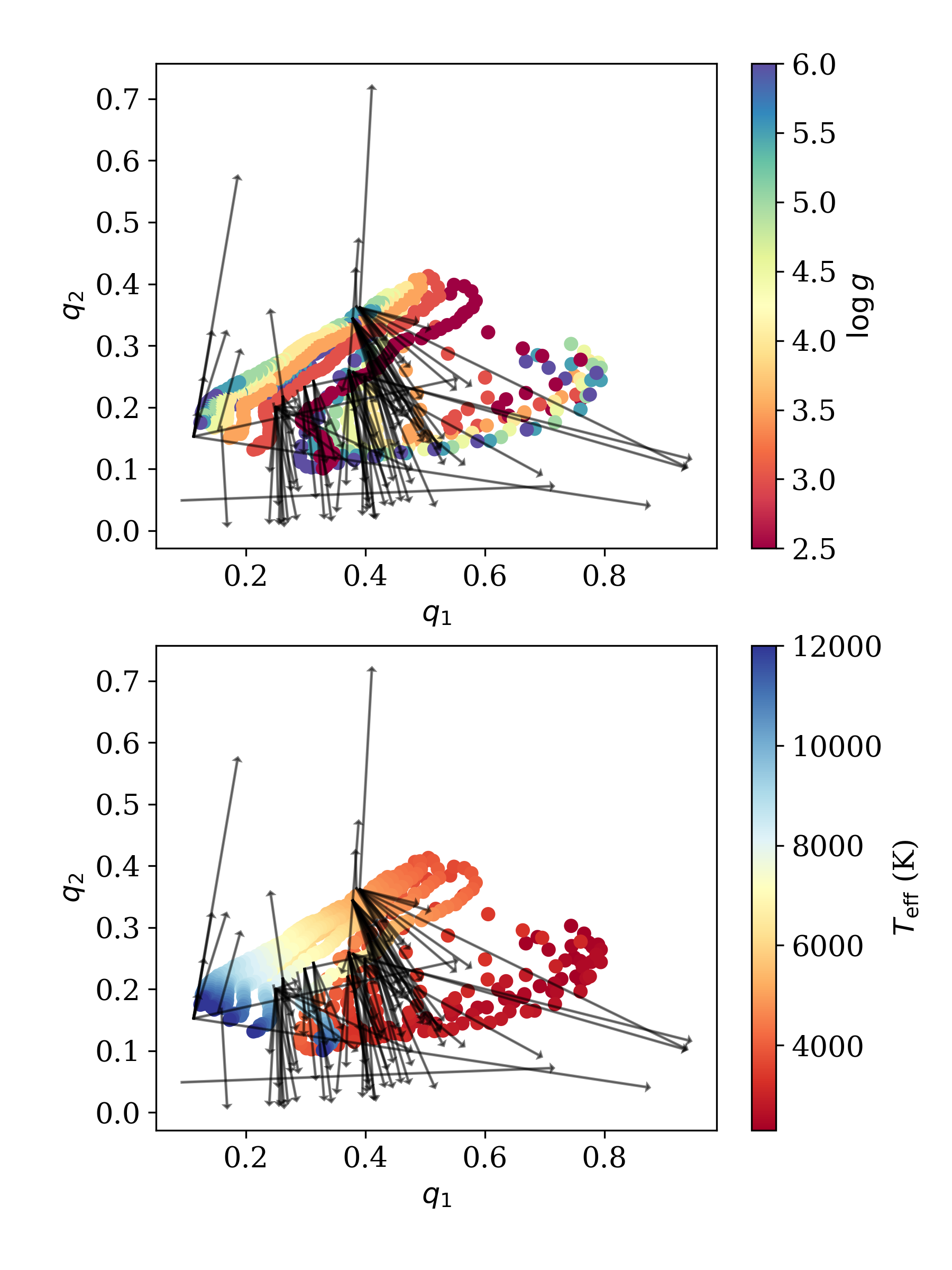}
    \end{subfigure}
    \begin{subfigure}{.33\textwidth}
    \centering
    \includegraphics[width=1\linewidth]{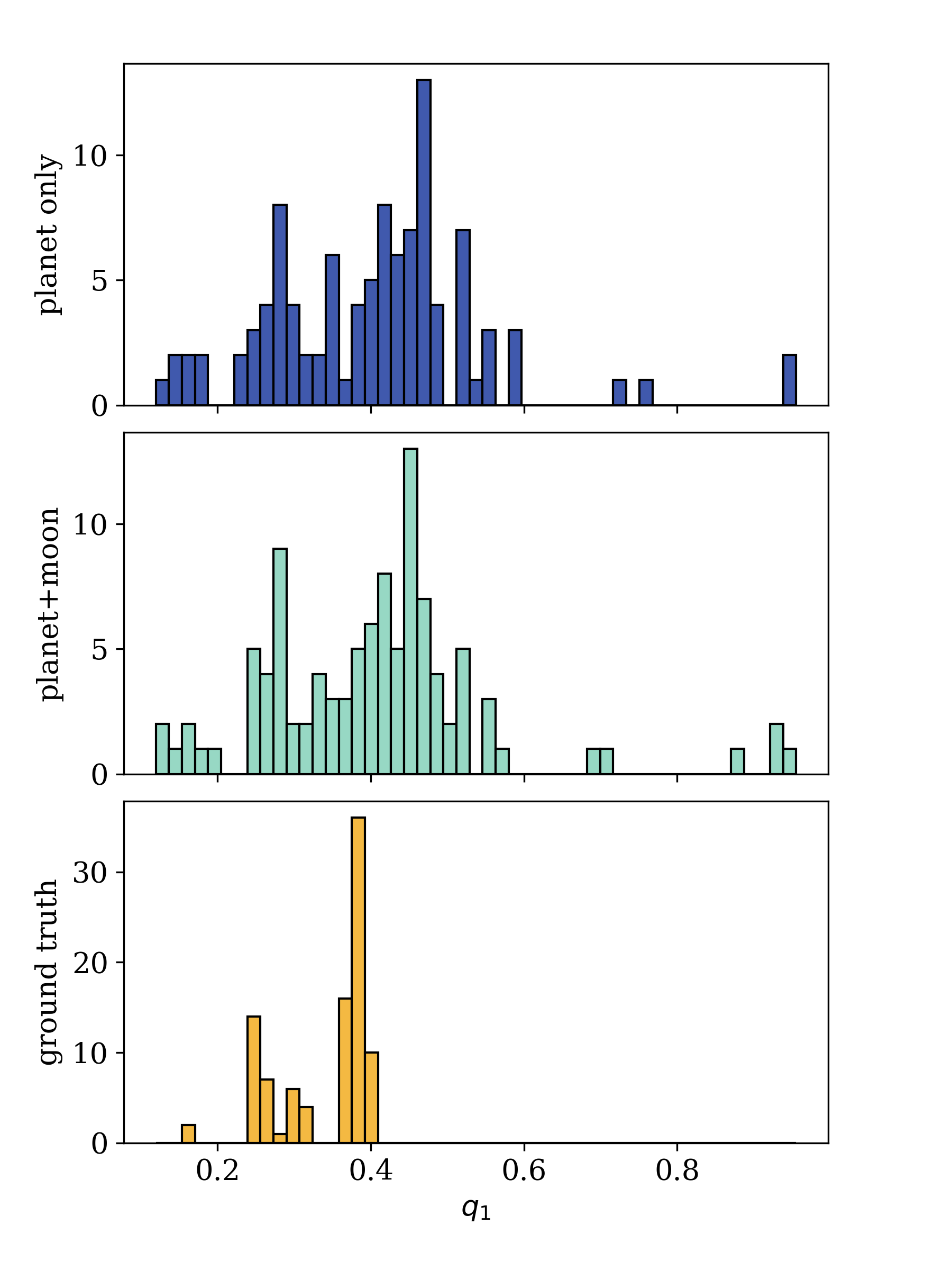}
    \end{subfigure}
    \begin{subfigure}{.33\textwidth}
    \centering
    \includegraphics[width=1\linewidth]{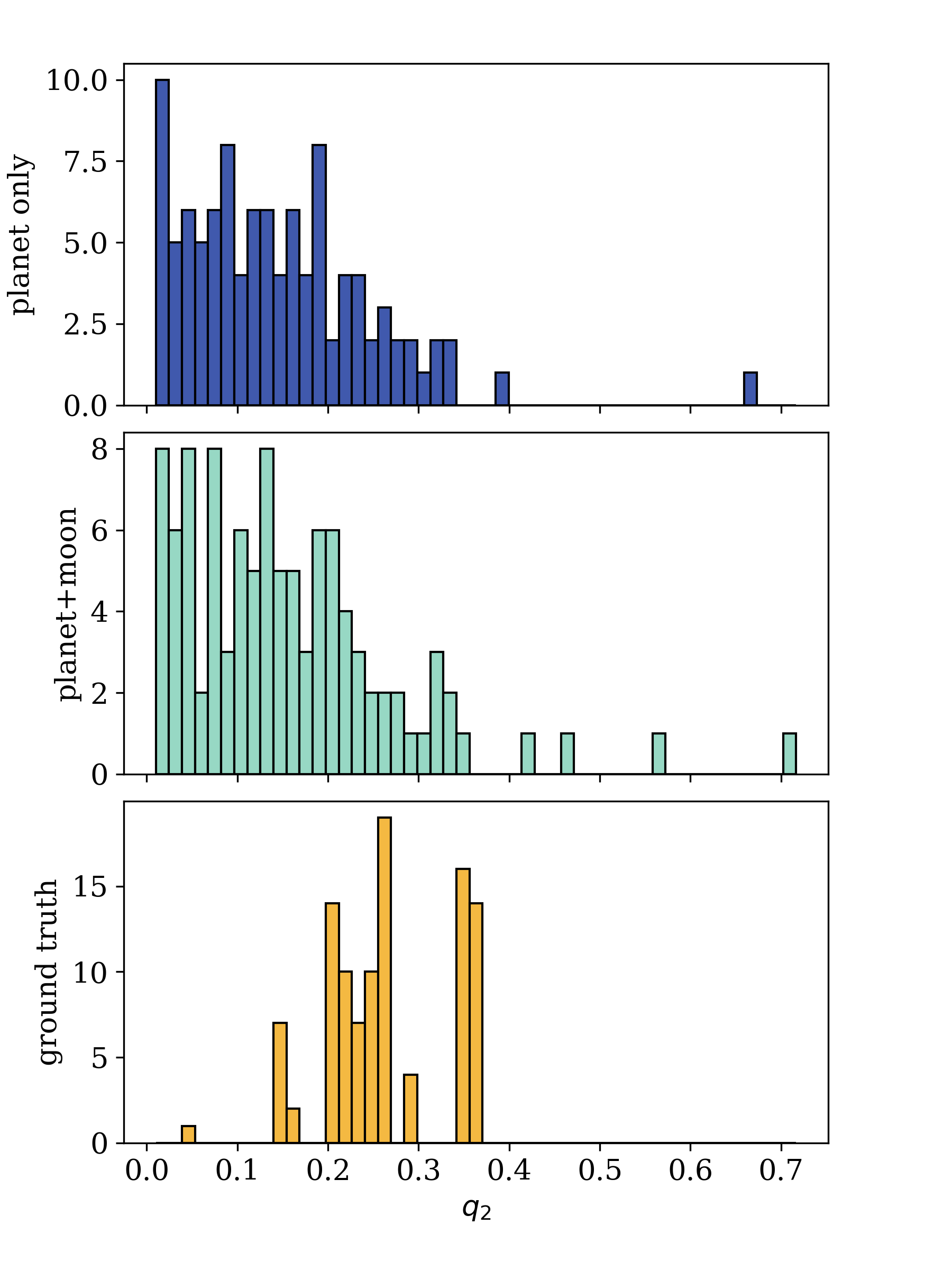}
    \end{subfigure}
    \caption{\textit{Left:} Limb darkening coefficients taken from \citep{Claret:2011} and re-parametrized \citep{LDCs}, color-coded by surface gravity $\log g$ and effective temperature. Arrows represents the change from the ground truth limb darkening values to those that were fit by the models. \textit{Middle and right:} Distribution of limb darkening coefficients $q_1$ and $q_2$ for the planet-only model, planet+moon model, and ground truth.}
\label{fig:limb_darkening}
\end{figure*}

\begin{figure}
    \centering
    \includegraphics[width=\columnwidth]{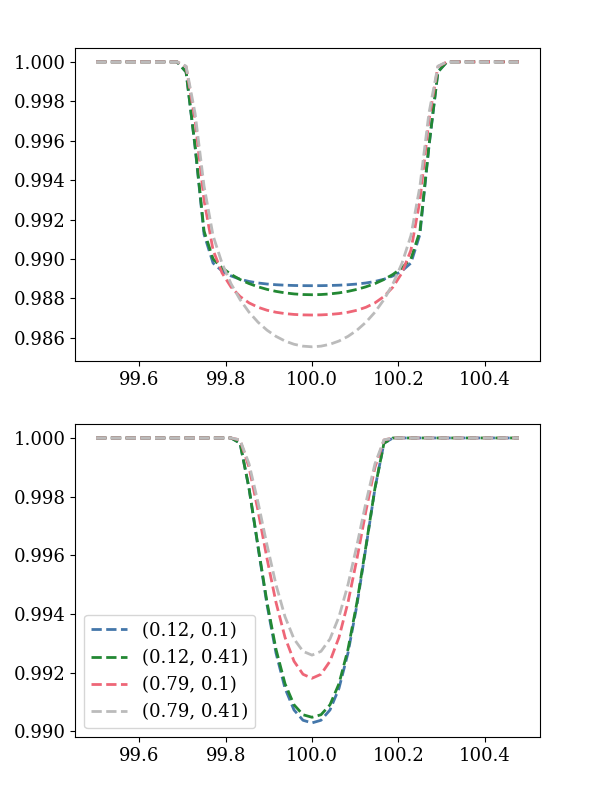}
    \caption{The effect of limb darkening across the range of $(q_1,q_2)$ values (adapted from \citealt{Claret:2011} using the \citealt{LDCs} reparametrization) for impact parameters $b=0$ (top) and $b=0.9$ (bottom). The planet's parameters are consistent across all four models. Limb darkening will effect both the shape and depth of the transit.}
    \label{fig:effect_of_limb_darkening}
\end{figure}

Our artificial light curves were all generated using quadratic limb darkening coefficients taken from \citealt{Claret:2011}, using $\log g$ and $T_{\mathrm{eff}}$ values from the stars drawn from the NASA Exoplanet Archive, and re-parametrized as $q_1$ and $q_2$ (see \citealt{LDCs}) using the \textsc{Pandora}'s built-in conversion tool. As we have seen, the presence of moons can alter the shape of the planet's ingress and egress, apparent transit duration, and its transit depth. As these same features are also impacted by limb darkening, estimates of limb darkening coefficients may also be skewed by the presence of one or more moons.

The left side of Figure \ref{fig:limb_darkening} shows the distribution of limb darkening coefficients catalogued by \citealt{Claret:2011} for a range of surface gravities and stellar effective temperatures, overplotted with arrows indicating the change from the ground truth values to the best fitting model solution. The results indicate a general trend shifting towards higher values of $q_1$ and lower values of $q_2$. Figure \ref{fig:limb_darkening} also shows model solution distributions for both planet-only and planet+moon models. 

Notably, the left panel of Figure \ref{fig:limb_darkening} shows that many of these model solutions are outside the range of values catalogued by \citealt{Claret:2011}, suggesting the possibility that some of these solutions are mathematically allowed but could be unphysical or at least highly unusual in nature. The modeling code does not check for a physically plausible solution, it merely ensures it is within the prior boundaries (0-1), and homes in on a solution that best fits the transit morphology. Here again, we see the opportunity for using anomalous solutions as a proxy indicator that something may be altering the shape of the planet's transit, namely, one or more moons. In a real search, tighter constraints on the priors might be preferable, in order to achieve more accurate limb darkening solutions; but at least in the case of this experiment, it is interesting to see what the model ``prefers'' to do.

Worth noting also is that the left panel of Figure \ref{fig:limb_darkening} is plotting the planet+moon model solutions. As such, it's evident that even in the presence of a moon in the model, limb darkening solutions may run into unphysical territory. It could be that the presence of the additional moons are confounding, pushing into unphysical parameter space because a single moon is inadequate to account for the transit morphology by itself. However, there is also a less interesting possible explanation, which is that the limb darkening is simply poorly constrained by the data in some cases, and because the priors allow the model to roam freely into unphysical territory, it does so. Even so, we would naively expect such a scenario to show more-or-less random drift, but as it happens, the drift definitely occurs in a preferred direction (towards higher values of $q_1$ and lower values of $q_2$).

Looking to Figure \ref{fig:effect_of_limb_darkening}, we see that limb darkening affects both the roundness of the transit as well its depth, the second of these dramatically so, and interplay with the impact parameter solution. A tendency towards higher values of $q_1$ then suggests a deepening of the transit depths when the impact parameter solution is low, and a shallowing of transit depths at greater impact parameters. A skew in $q_1$ then might be associated with the priors that were placed on planet radii. As we mentioned, these ``derived'' radius solutions came from perturbing the ground truth values based on typical uncertainties. A more realistic approach might have been to fit these planets first with planet models, then take those radius solutions as the basis of our radius prior for the moon modeling. In so doing, the shift towards higher $q_1$ values might be attenuated to some degree.

\subsection{Unphysical moon densities}
\begin{figure}
    \centering
    \includegraphics[width=\columnwidth]{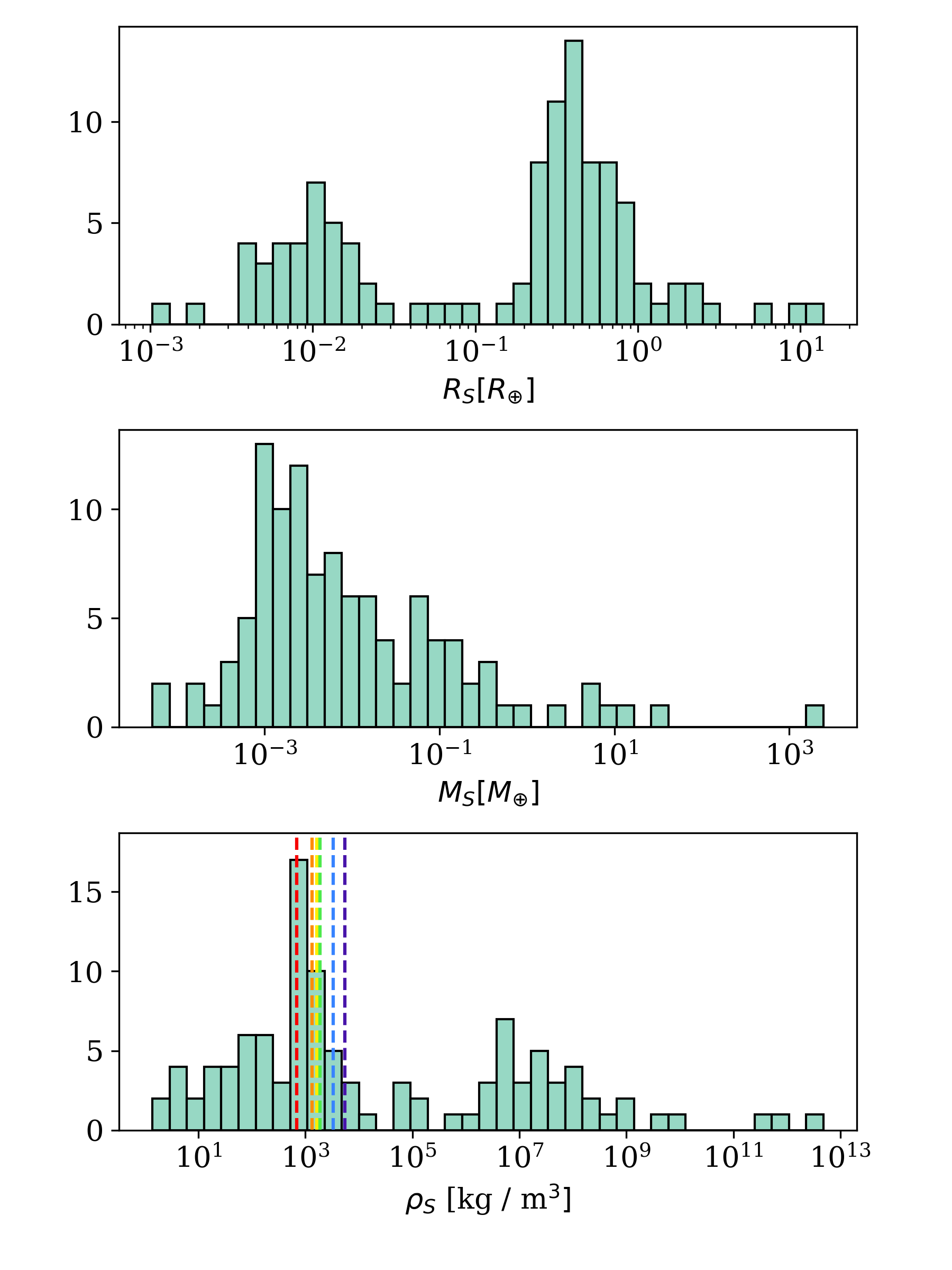}
    \caption{Model solutions for systems where the moon is favored with $K \geq 3.2$. \textit{Top:} distribution of moon solution radii, in Earth radii. \textit{Middle:} mass solutions for the moons, in Earth masses. \textit{Bottom:} Moon densities computed from the mass and radius solutions of the model runs. The vertical dashed lines indicate the mean densities of Saturn (red), Jupiter (orange), Neptune (yellow), Titan (green), The Moon (blue), and Earth (purple). A sizeable number of moon solutions have unphysically high density solutions, while others are unphysically low.}
    \label{fig:moon_densities}
\end{figure}

A number of our model runs for which the presence of a moon is the favored solution show unphysically high or low densities (Figure \ref{fig:moon_densities}). This is the result of having uninformative priors on both moon radius and moon mass that are uncoupled from one another. If we were to model these systems with a prior on density instead of letting mass and radius be independent, we would effectively force the fitter to look for physical solutions and, in the event that physical solutions cannot be found, the evidence might come out more strongly against the presence of moons.

However, letting these two parameters be fit independently yields some insight into what may be happening during the fitting: radius estimates derive from whatever moon transits may be found by the model fitting, while mass estimates will stem from whatever transit timing variations may be present. If densities are low, this is of course either because masses are too small or radii are too large. But it is exceptionally sensitive to changes in the latter, since volume goes as the cube of the radius. In Figure \ref{fig:moon_densities}, we can see that there are two model solution clusters in radius, one around $10^{-2}$ and one around $0.3 \, R_{\oplus}$. Comparing this with Figure \ref{fig:simulation_demographics}, which shows the simulation demographics, we see that indeed some of our radius solutions are skewing well below their ground truth values. The mass solutions similarly skew below the ground truth distribution (which peaks strongly around $10^{-2} \, M_{\oplus}$). But again, the radius underestimates will dominate the density calculation, resulting in some exceptionally high densities in some cases. Indeed, the radius plot in Figure \ref{fig:moon_densities} looks almost like a mirror image of the density plot.

Unphysical density solutions run the risk of scuttling any further analysis of a real system, as the system may be dismissed as showing some sort of anomaly that makes it an unattractive target for further inquiry. But these results suggest unphysical density estimates combined with evidence in favor of moons could potentially be clues that multiple moons are present. Further analysis may be warranted even if model solutions appear spurious.

\subsection{Examining the four possible outcomes}

As we pointed out in the introduction, there are four possible outcomes to the model fitting:

\begin{enumerate}[(1)]
    \item the single-moon model correctly identifies one of the moons in the system;
    \item the single-moon model finds evidence for a moon, but the parameters derived for that moon are illusory (a ``phantom'' moon);
    \item the single-moon model returns multi-modal posteriors, with the various modes attributable to the different moons in the system;
    \item the single-moon model finds that the moon hypothesis is disfavored over a planet-only explanation of the data. 
\end{enumerate}

Let us now examine these results in turn.

\begin{figure*}
    \centering
    \includegraphics[width=\textwidth]{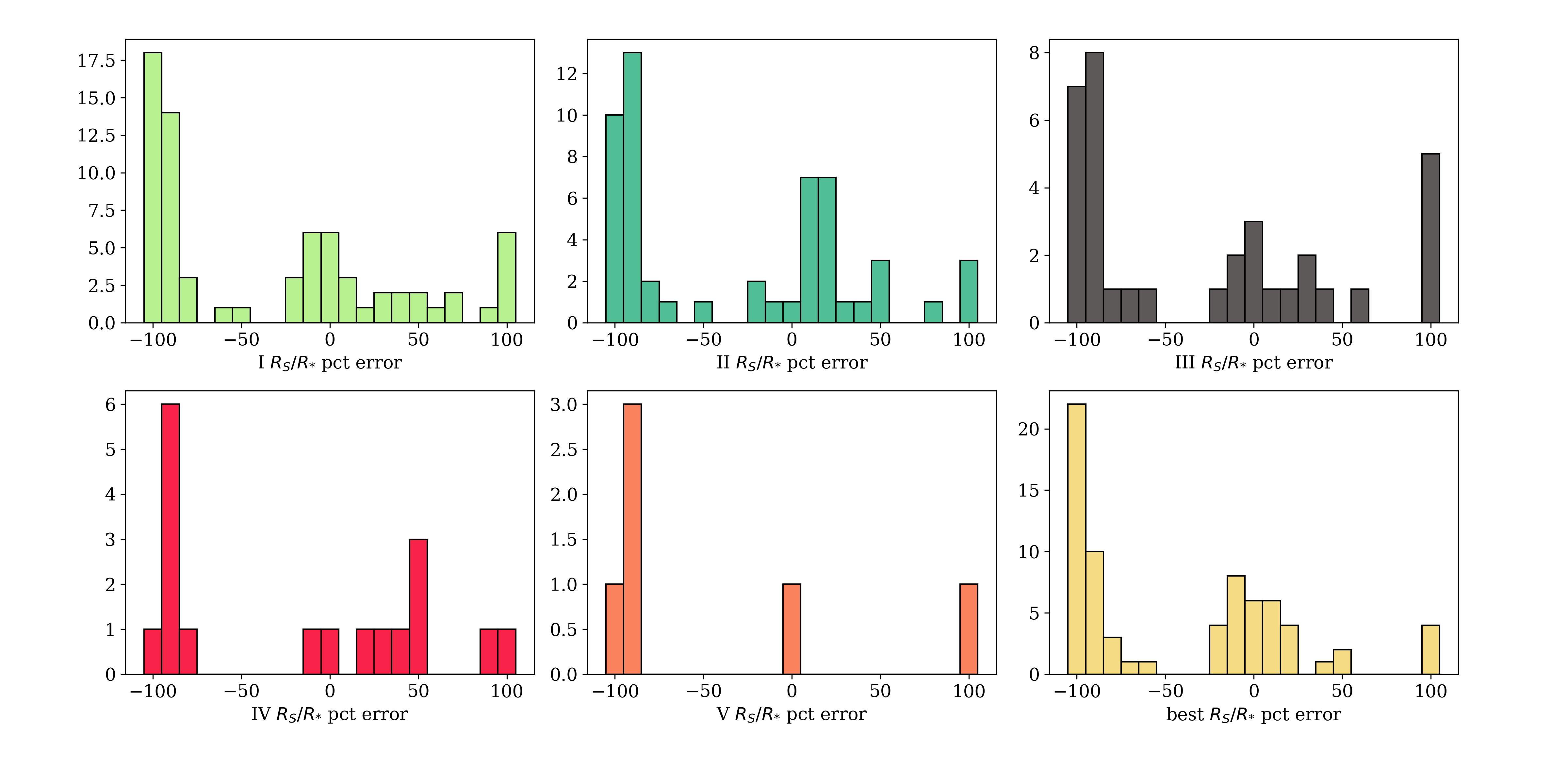}
    \caption{Histogram of percent errors between the planet+moon model solution for the moon's radius (here represented as $R_S / R_{*}$) and the ground truths for each satellite (I, II, III, IV, and V). Positive values represent overestimates while negative values represent underestimates. The satellite  are numbered from closest to farthest from the host planet. The percent error for the closest-matching satellite in each system is bottom right.}
    \label{fig:RsRstar_pcterr_histogram}
\end{figure*}

\begin{figure*}
    \centering
    \includegraphics[width=\textwidth]{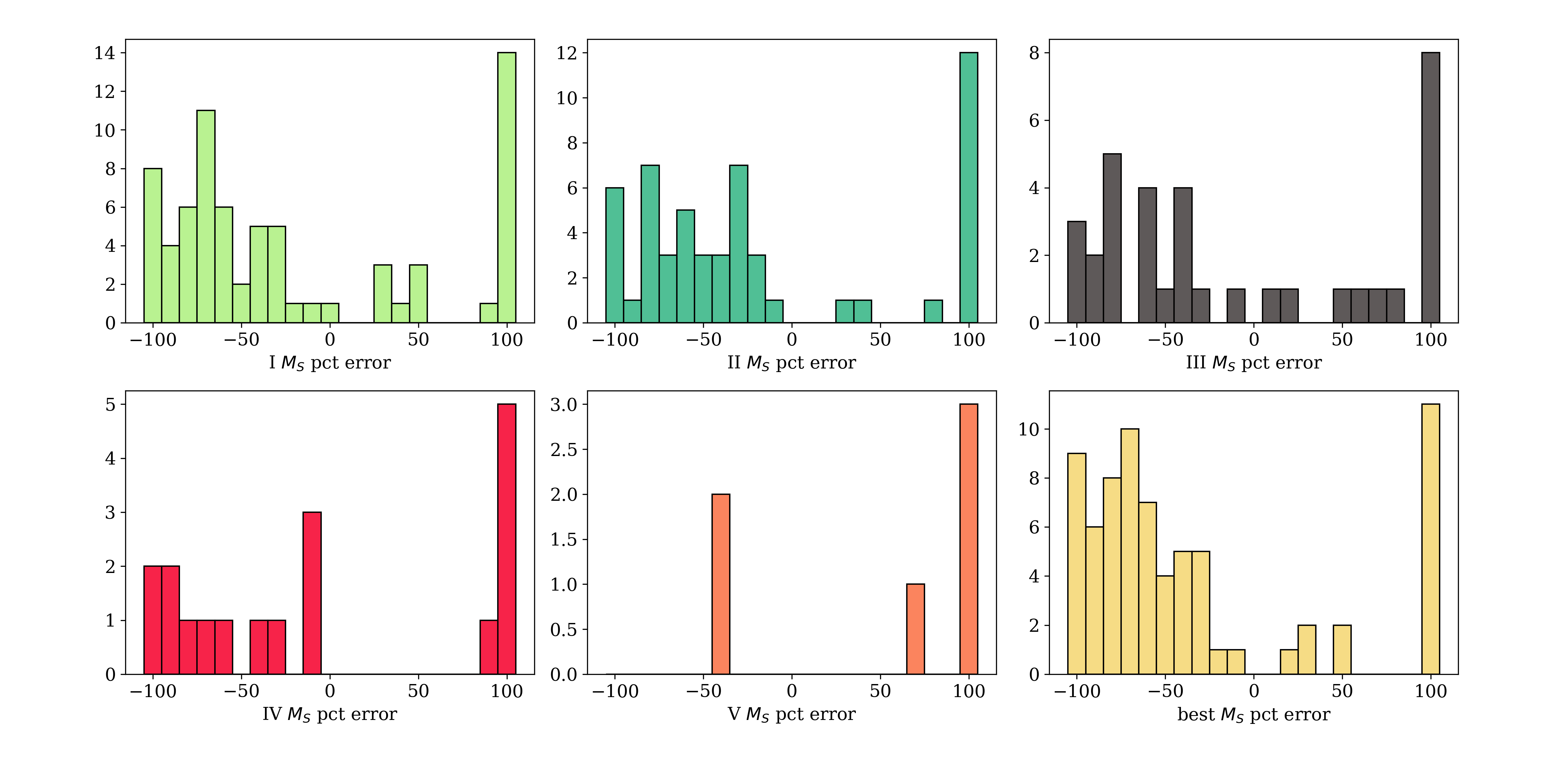}
    \caption{Histogram of percent errors between the planet+moon model solution for the moon's mass and the ground truths for each satellite (I - V). Positive values represent overestimates while negative values represent underestimates. The satellite are numbered from closest to farthest from the host planet. The percent error for the closest-matching satellite in each system is bottom right.}
    \label{fig:mmoon_pcterr_histogram}
\end{figure*}

\begin{figure*}
    \centering
    \includegraphics[width=\textwidth]{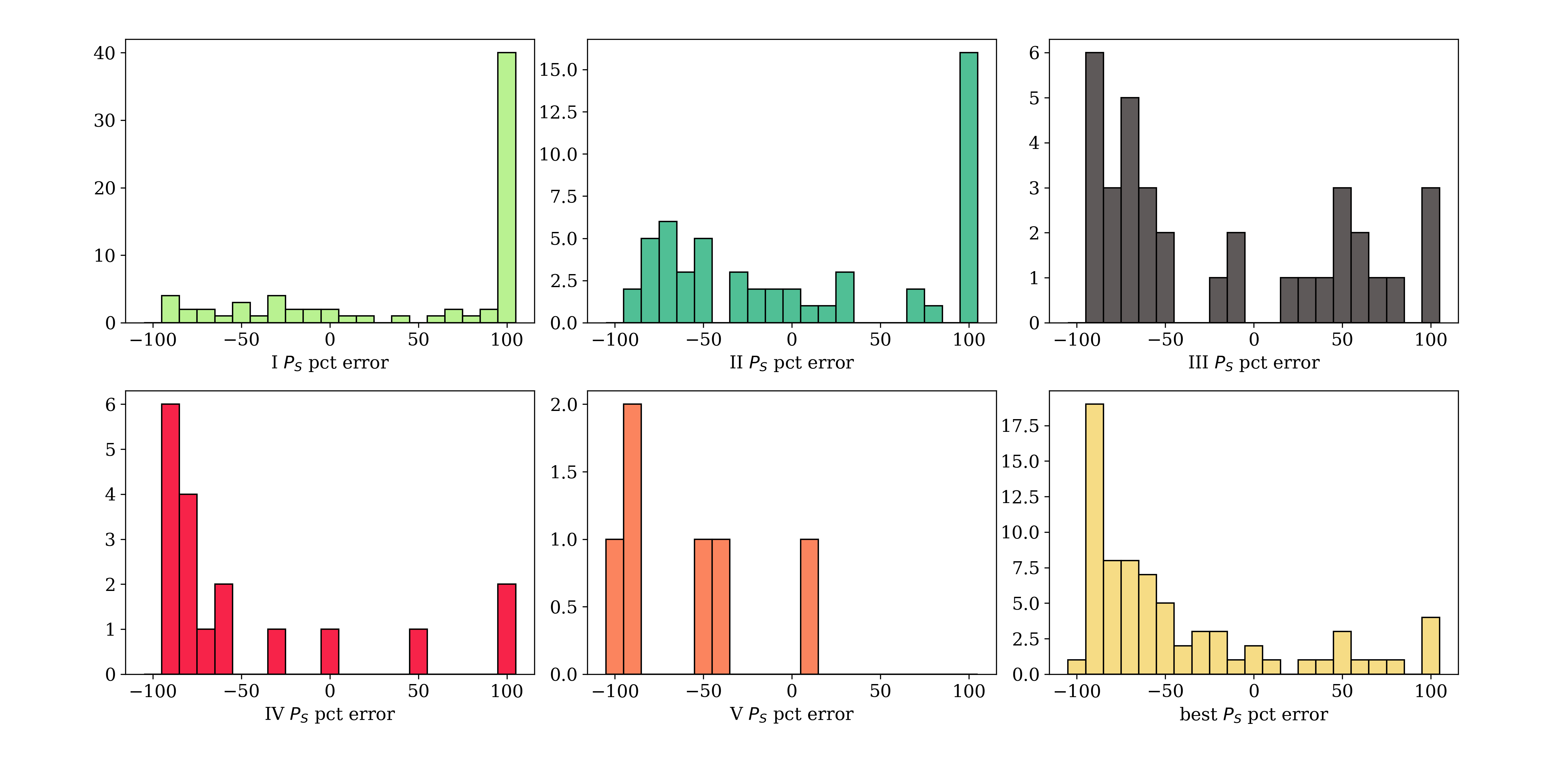}
    \caption{Histogram of percent errors between the planet+moon model solution for the moon's orbital period and the ground truths for each satellite (I - V). Positive values represent overestimates while negative values represent underestimates. The satellite are numbered from closest to farthest from the host planet. The percent error for the closest-matching satellite in each system is bottom right.}
    \label{fig:permoon_pcterr_histogram}
\end{figure*}

\begin{figure*}
    \begin{subfigure}{.33\textwidth}
    \centering
    \includegraphics[width=1\linewidth]{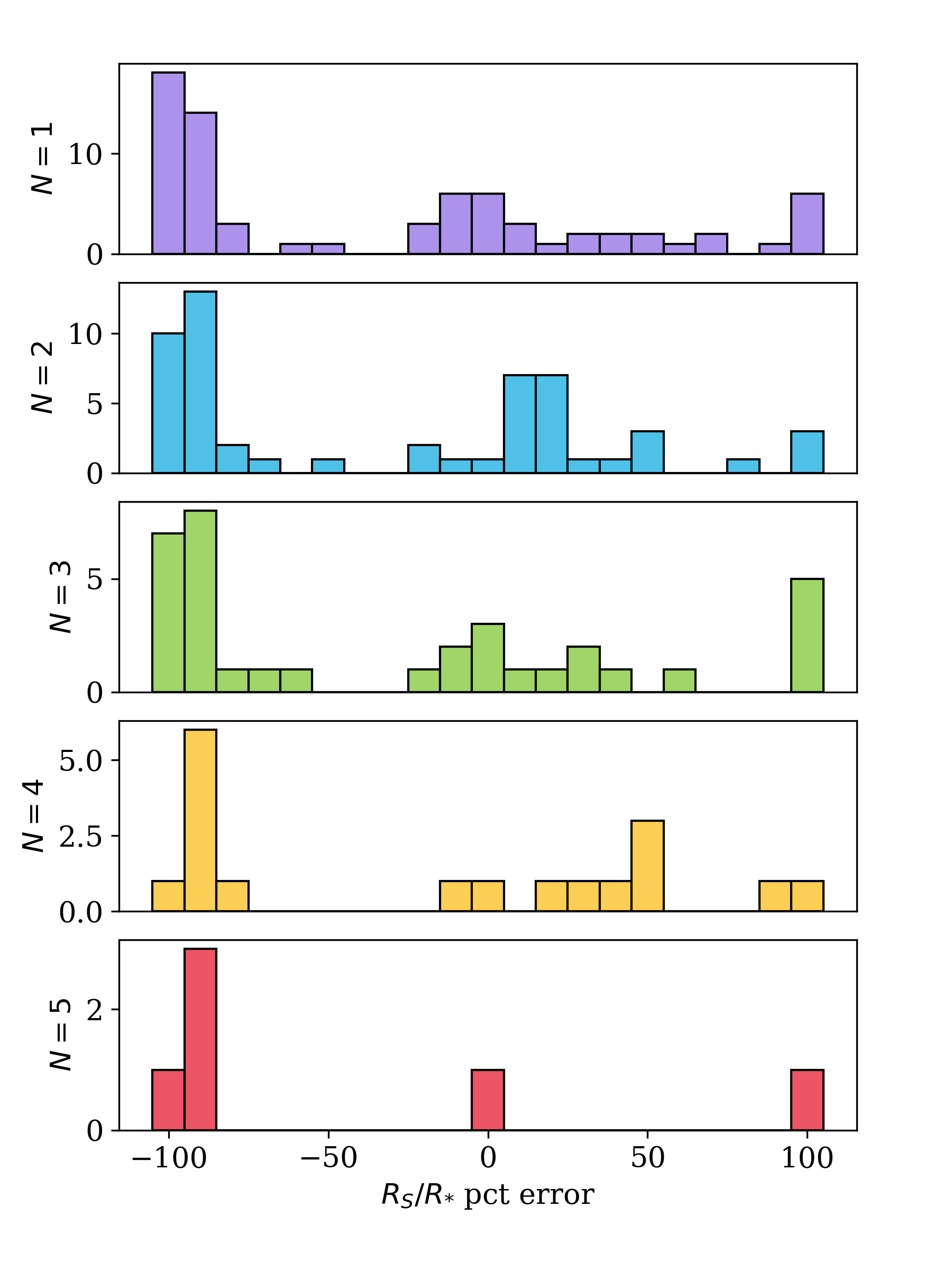}
    \end{subfigure}
    \begin{subfigure}{.33\textwidth}
    \centering
    \includegraphics[width=1\linewidth]{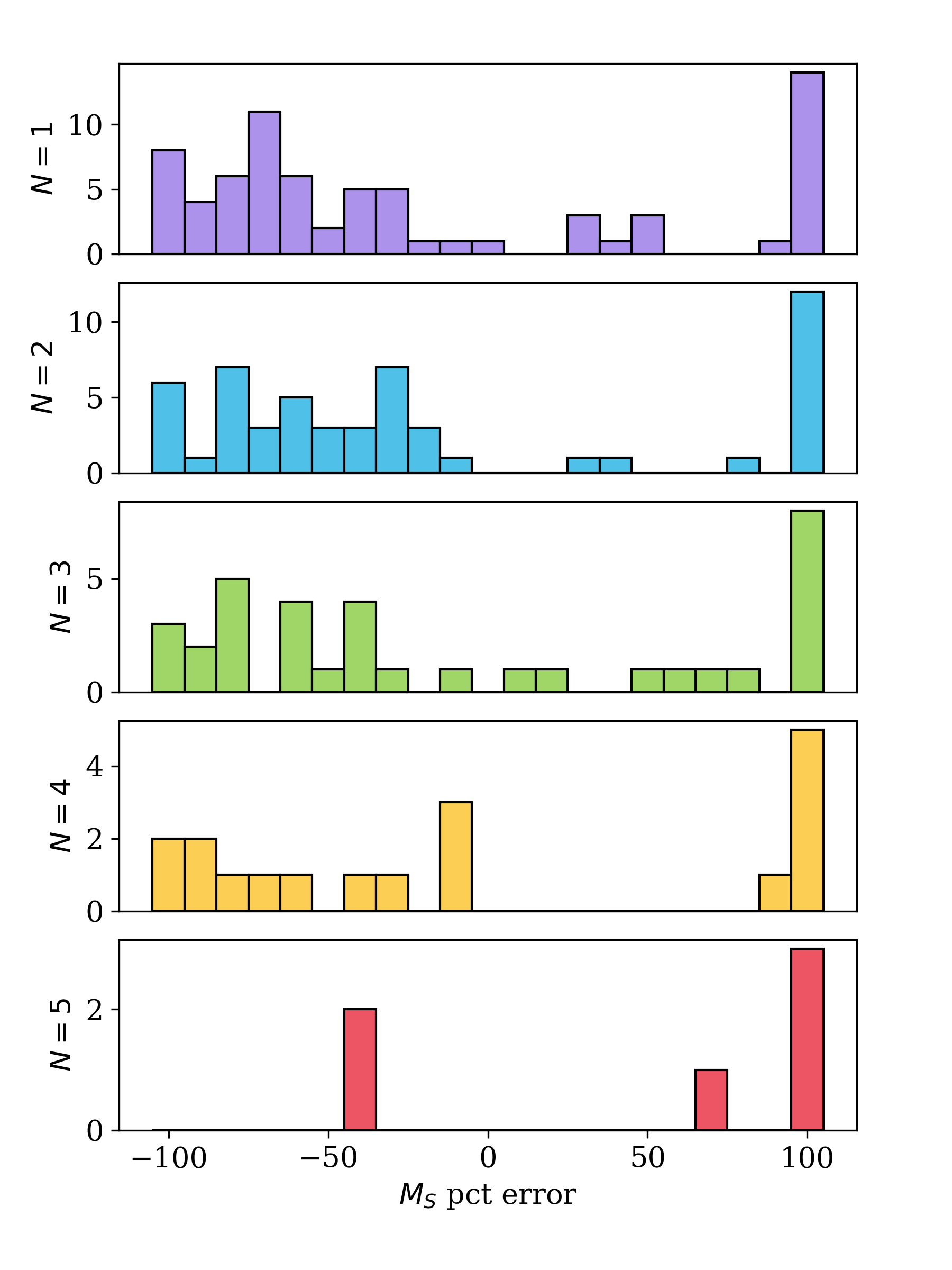}
    \end{subfigure}
        \begin{subfigure}{.33\textwidth}    
        \centering
    \includegraphics[width=1\linewidth]{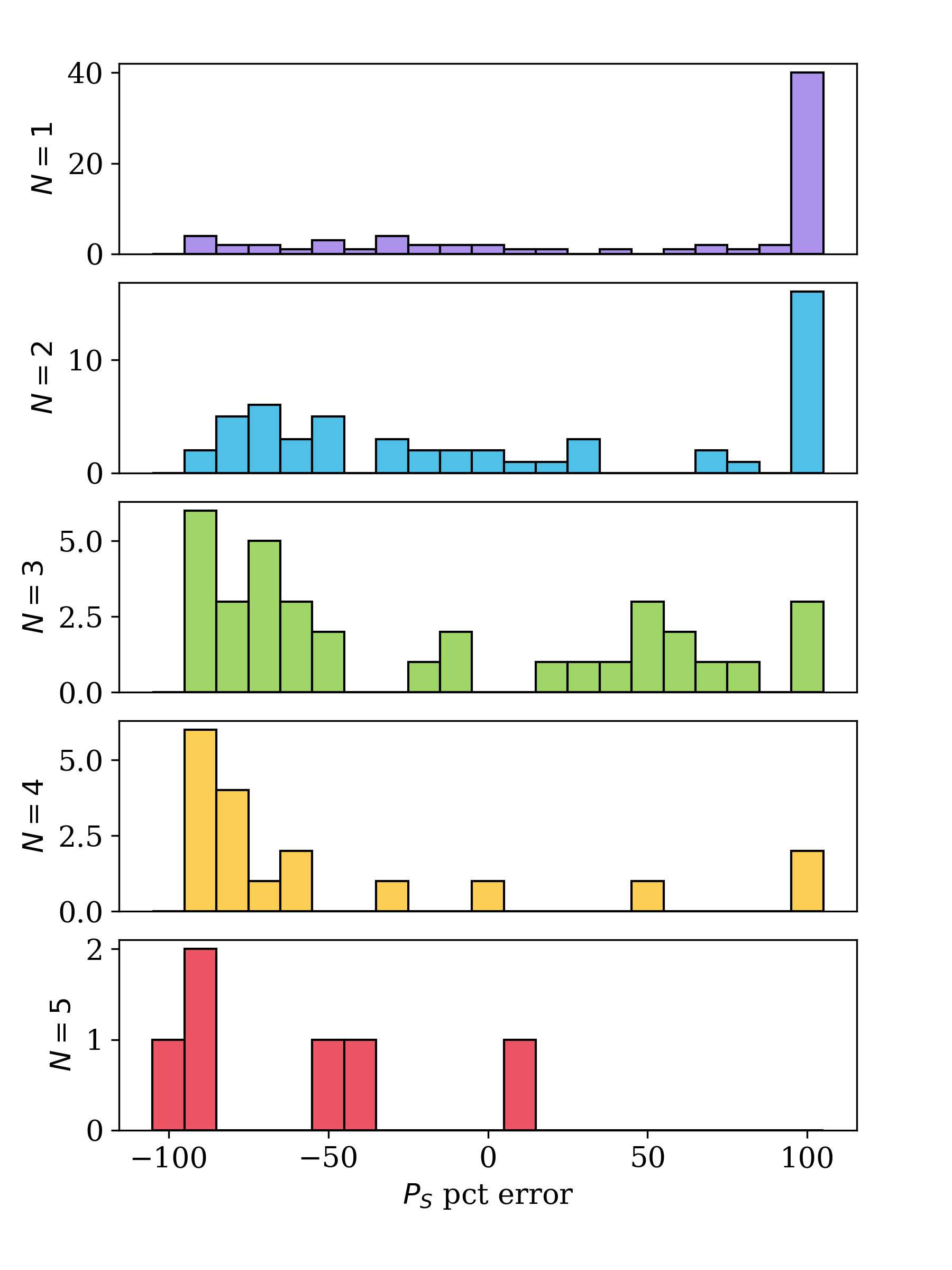}
    \end{subfigure}
    \caption{\textit{Left:} percent error of $R_S / R_{*}$ for the closest matching satellite in each system architecture (from $N=1$ to $N=5$ moons). Positive values represent overestimates while negative values represent underestimates. For example, if $N=3$, there are three satellites with ground truth $R_S / R_{*}$ values, each with their own percent errors based on the model solution; we plot the smallest of these percent errors. \textit{Middle:} The same as before, but for closest-matching moon mass $M_S$. \textit{Right:} As before, for the closest-matching moon orbital period $P_S$.}
    \label{fig:nsats_per_system_histograms}
\end{figure*}

\begin{figure*}
    \includegraphics[width=\textwidth]{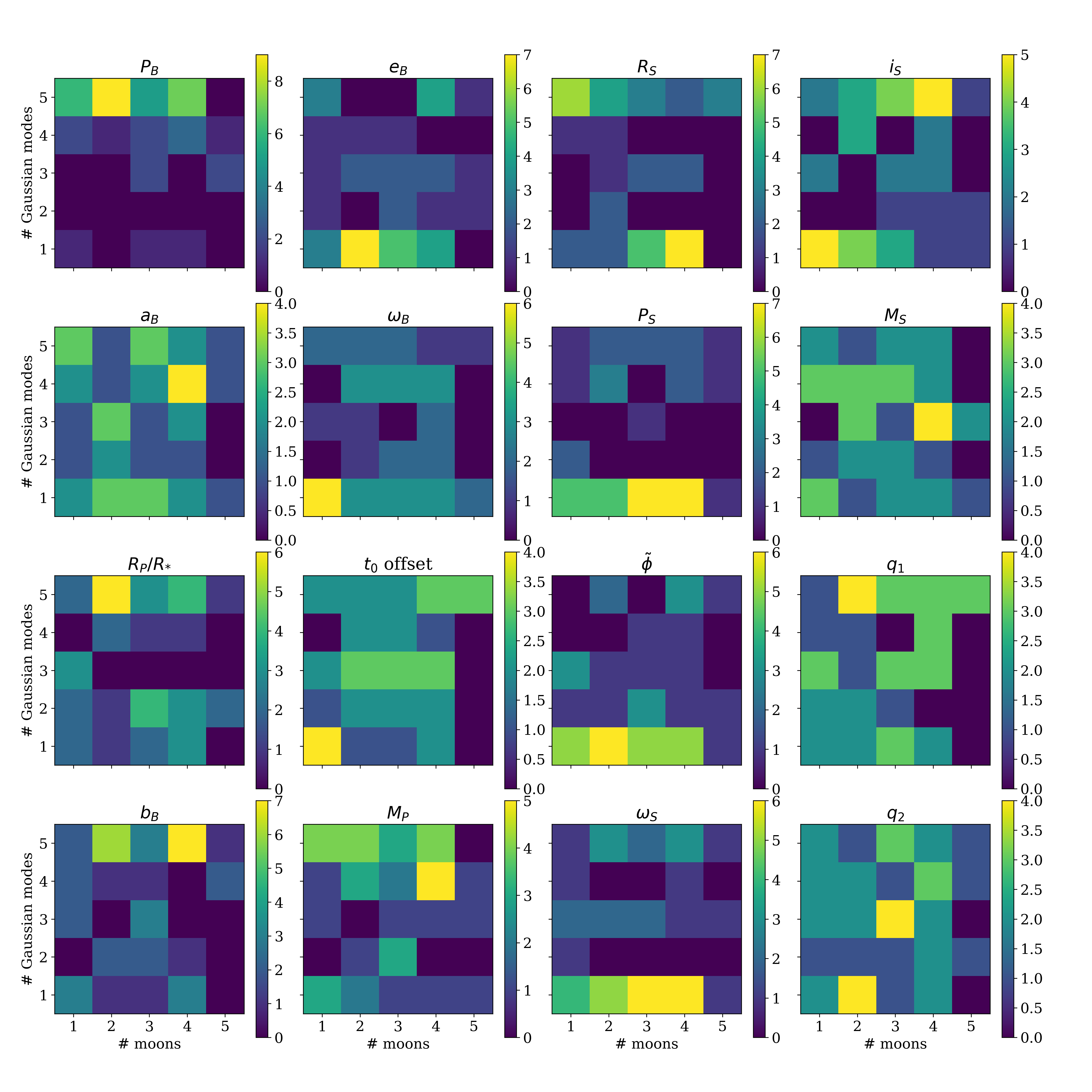}
    \caption{Number of posterior peaks as a function of the number of moons in each system, for the systems that showed positive evidence for the presence of a moon. If the posteriors were showing evidence of multimodality corresponding to the moons, we should see some diagonality in these plots.  the colorbar for each plot indicates the number of systems within each bin.}
    \label{fig:npeaks_vs_nmoons}
\end{figure*}

\subsubsection{Possibility (1) - one moon sought, one moon found}
In the case of a system hosting multiple moons, we want the single moon model to find at least one authentic moon. This has been the working assumption until now; the most easily detectable moon should be found using a single-moon model, and the other, more difficult moons, may be found later.

Figures \ref{fig:RsRstar_pcterr_histogram}, \ref{fig:mmoon_pcterr_histogram}, and \ref{fig:permoon_pcterr_histogram} show the percent errors on the model solutions of systems for the 72 systems where the moon model is preferred ($K > 3.2$) for arguably the three most important parameters of the moon fit, namely, the moon's radius, its mass, and its orbital period. Percent error is defined as

\begin{equation}
    \mathrm{\% \, error } = \frac{\mathrm{best \, solution} - \mathrm{ground \, truth} }{\mathrm{ground \, truth}} \times 100
\end{equation}

\noindent so that, in the aforementioned figures, a value greater than zero means the solution is overestimating the ground truth value, and vice versa. We also note that the bins at each extreme ($\pm 105 - 95\%)$ are inclusive of every value in excess of those extremes; peaks at either end therefore do not indicate a pile-up of solutions that are off by a factor of 2, but rather the presence of solutions more than 105\% off.

Of course, the model is looking for a single moon, so in the case of a system hosting multiple moons, the moon solution could apply to any one of them. We must therefore look at how well the model solution accords with each one of them, and also whether it has found \textit{any} of them. The first five panels in Figures \ref{fig:RsRstar_pcterr_histogram} - \ref{fig:permoon_pcterr_histogram} show the percent errors for each moon (I - V), while the bottom right panel shows the distribution for the closest matching moon in each system. Figure \ref{fig:RsRstar_pcterr_histogram} shows that there are 6 systems (8.3\%) for which the moon model fits come within $\pm 5\%$ of one of the moons in the system. Adding up the $\pm 5\%$ bins for each satellite, however, we see a total of 18 moons (25\%) with a model solution that comes this close to the ground truth, so in some cases it is unclear which moon is actually being modeled, or the moon solution could be modeling one or more of the moons simultaneously. For systems containing multiple moons, overlapping moon transits could have the effect of mimicking a single, larger moon (producing a deeper transit). But if such were the case, we should be overestimating the size of the real moons (showing up on the right side of the plot). 

In the case of the mass solutions (Figure \ref{fig:mmoon_pcterr_histogram}), none of the moon solutions come within $\pm 5\%$ accuracy. But we also see something of a pile-up on the left side of the plots, indicating an underestimation of moon masses in these cases, and peaks at the right end of the plots, indicating mass overestimates in excess of 100\%. Underestimates may not be very surprising, as the multiple moons may be working in concert to mitigate the TTV amplitudes, thereby making the moon mass appear smaller. Recall that in the modeling code, the mass and radius are fit independently without any regard for a physically plausible density, so the mass solutions are controlled entirely by the transit timings and the period solution of the moon.

And for the satellite periods (Figure \ref{fig:permoon_pcterr_histogram}), we again see very few solutions within 5\% of any of the moons in the system. The interpretation of these results is a bit less straightforward, however, due to the undersampling of moon orbits. Posteriors of a moon's period are often multi-modal, showing harmonics of the orbital frequency; a percent error of 100\% in such a case would not be so egregious, then, as it could simply be one of the aliases of the true period. In any case, we rarely see a good solution. Note that in Figure \ref{fig:permoon_pcterr_histogram}, we see a sustantial pile-up at the high end for satellites I and II, indicating an overestimate of the period, a mix of under- and over-estimates for the outer satellite III, and underestimates for the outer sallites IV and V. But this can be understood when we remember that the satellites are labeled in order of distance from the host planet. It appears, then, that the generally prefers some intermediate period solution, one that overshoots the inner satellite periods and undershoots those of the outer moons.

Figure \ref{fig:nsats_per_system_histograms} provides another look at these important parameters, this time split by individual architectures ($1 \leq N \leq 5$ moons) instead of individual moons, and plotting the closest matching moon in each architecture (akin to the lower right panel in Figures \ref{fig:RsRstar_pcterr_histogram} - \ref{fig:permoon_pcterr_histogram}). We would hope to see excellent results particularly for the single-moon ($N=1$) systems, since that is what we are modeling after all. But it is not always the case. The single-moon systems for which our results are poor are generally among the more difficult to retrieve, owing to low SNRs and / or presenting fewer transits to model. Even so, all of these systems have been classified as showing evidence for a moon, so in a real search they would certainly get a closer look, though ultimately the model solutions would not hold up to scrutiny.

The same general behavior is evident as above: we do reasonably well with radii solutions, but mass solutions are consistently off the mark, and period solutions are either high or low depending on the architecture.

\subsubsection{Possibility (2) -- one moon sought, imaginary moon found}
As we have just seen in the preceding section, the possibility that we have ``found'' a moon, but one whose properties do not reflect any of the actual moons in the system, seems to be the case for a large fraction of the systems we've analyzed. On the one hand, we can be glad that the presence of moons in these systems is being detected more often than not. But the fact that our model solutions for them are frequently inaccurate and would not hold up to additional scrutiny should give us considerable pause. If a more thorough, subsequent analysis of a given system finds system properties that are entirely inconsistent with the model solutions of the initial claim, it could be argued that the initial claim did not actually find the moon at all.

\subsubsection{Possibility (3) -- one moon sought, multiple moons found}

Evidently, our maximum likelihood solutions often do not match ground truth values for any of the moons in the systems. But might we see evidence in the posteriors that suggest the presence of multiple moons? This could potentially manifest in the form of multi-modal posteriors.

To investigate this question, we will not attempt to see whether the modal peaks are consistent with the satellites attributes themselves, as we have already seen that even in the best case scenario, the ground-truth values are often recovered with only moderate success. Instead, we will simply examine whether there is any evidence that, say, a system containing 3 moons shows three modes in the posteriors.

For each posterior, we fit five different models, containing a sum of 1 - 5 Gaussians, checking which model comes closest to matching the distribution, computing a Bayesian Information Criterion (BIC) for each fit and taking the minimum BIC as the best fitting model. Visual inspection indicated that these fits performed reasonably well, but far from perfectly. In any case, we may use it as a proxy for multi-modality.

Figure \ref{fig:npeaks_vs_nmoons} shows the number of modes found versus the number of moons for each model parameter. Were we to see a relationship between the number of moons present and the number of modes, we ought to see it manifest as a concentration of solutions along the diagonals. There may be some slight hints of such behavior, but it is barely there at all. 
From this result, we conclude that fitting a single-moon model to a multiple-moon system will generally not show evidence of multiple moons, even in cases where the posteriors are multi-modal.

\subsubsection{Possibility (4) -- one moon sought, no moons found} 
Our results indicate that 30.8\% of the final system set did not return sufficient evidence for a moon. Looking back at Table \ref{tab:nmoon_stats}, we can see that this holds across all architectures, even $N=1$ systems. So while this is not exactly a promising return, we also cannot readily attribute this to the presence of multiple moons.

\subsection{Slow Convergence of Moon Models}
\begin{figure*}
    \begin{subfigure}{.23\textwidth}
    \centering
    \includegraphics[width=1\linewidth]{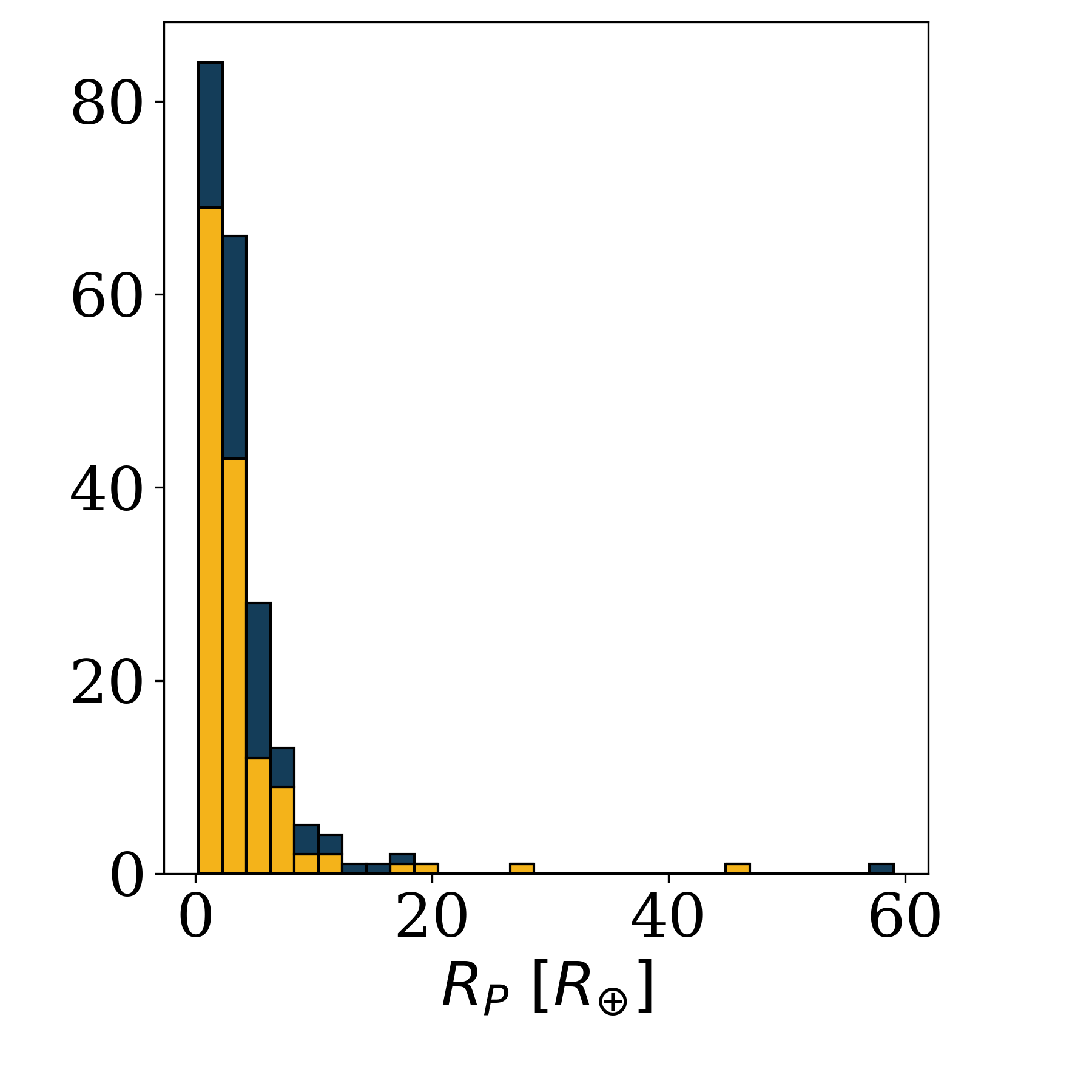}
    \end{subfigure}
    \begin{subfigure}{.23\textwidth}
    \centering
    \includegraphics[width=1\linewidth]{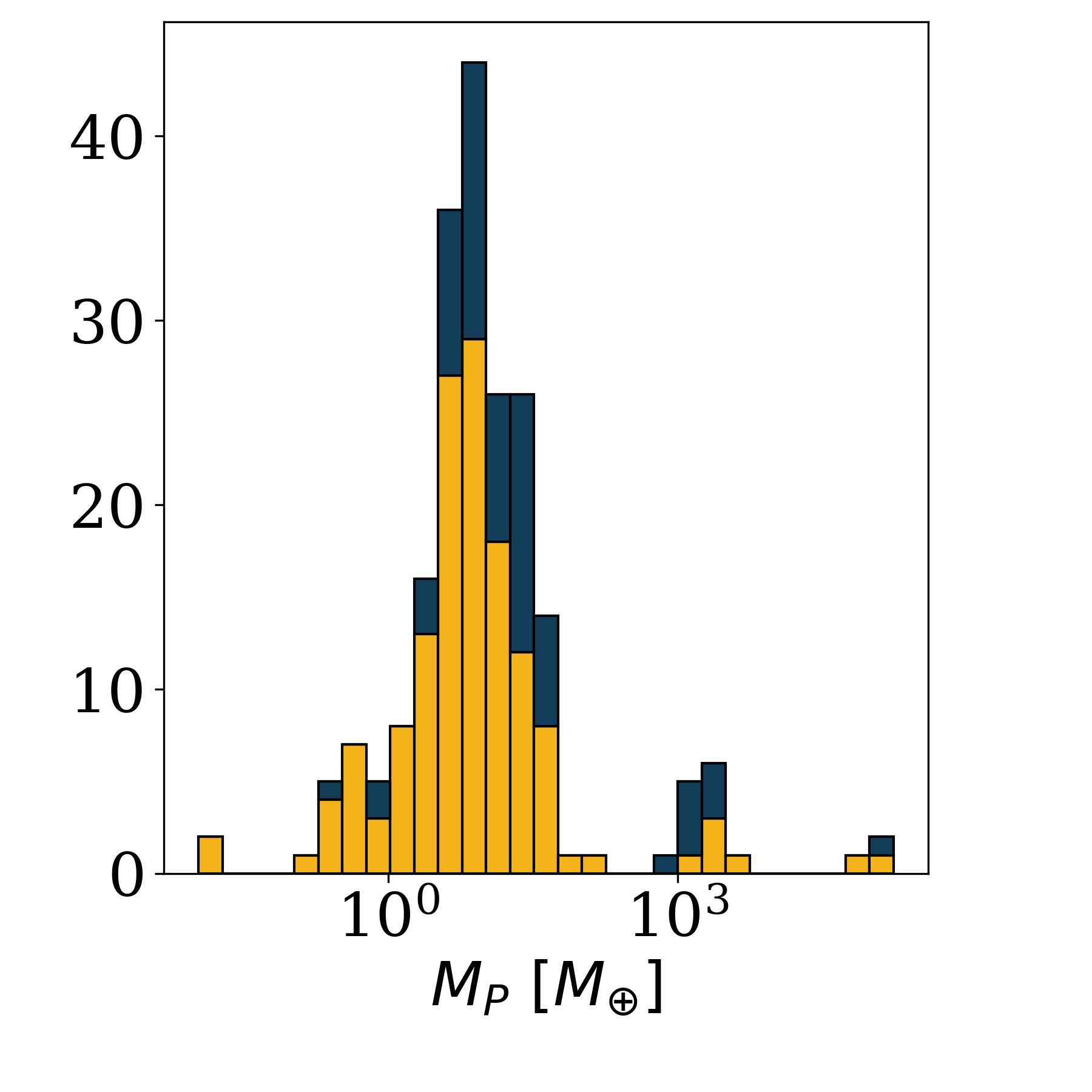}
    \end{subfigure}
        \begin{subfigure}{.23\textwidth}    
        \centering
    \includegraphics[width=1\linewidth]{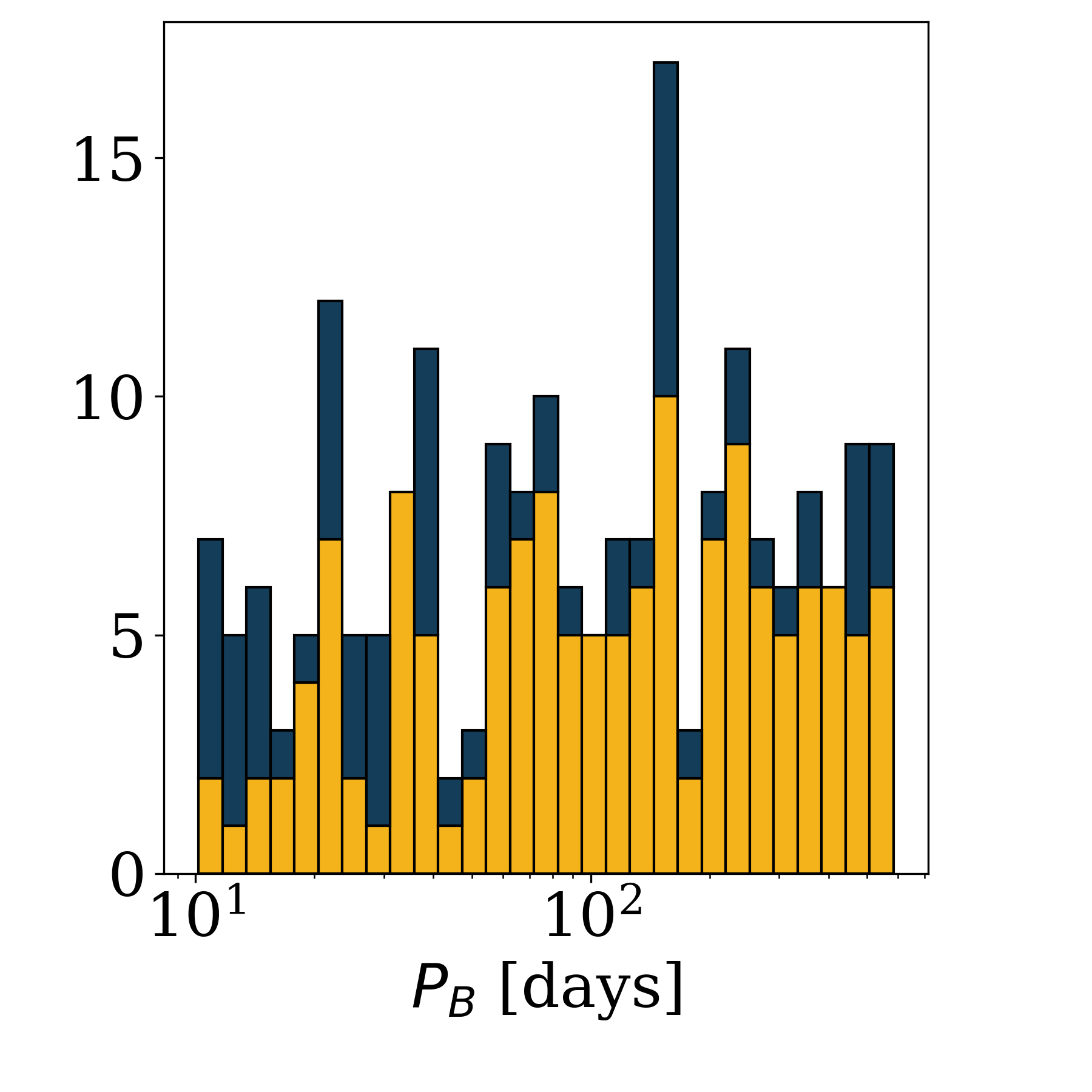}
    \end{subfigure}
    \begin{subfigure}{.23\textwidth}
    \centering
    \includegraphics[width=1\linewidth]{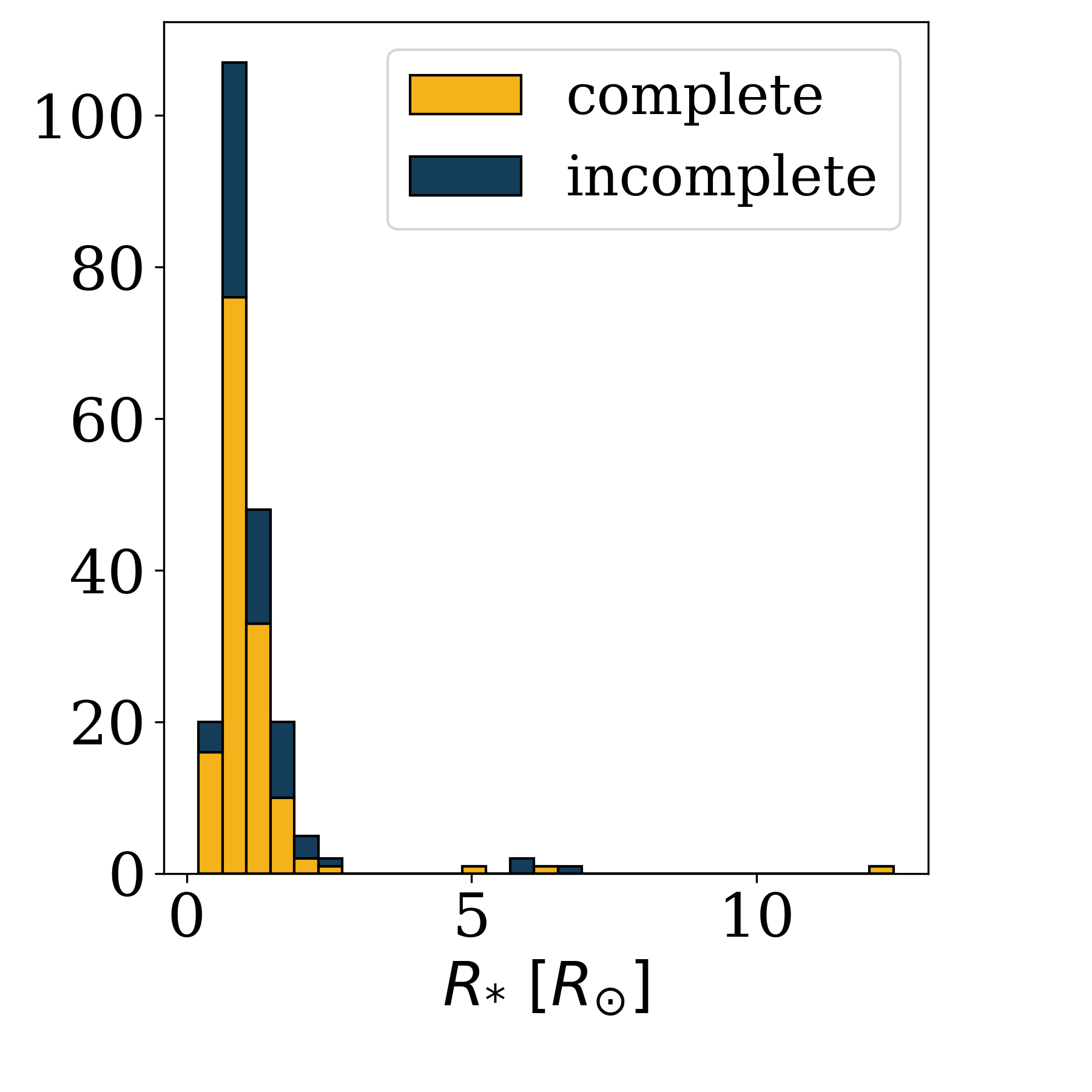}
    \end{subfigure}
    \begin{subfigure}{.23\textwidth}
    \centering
    \includegraphics[width=1\linewidth]{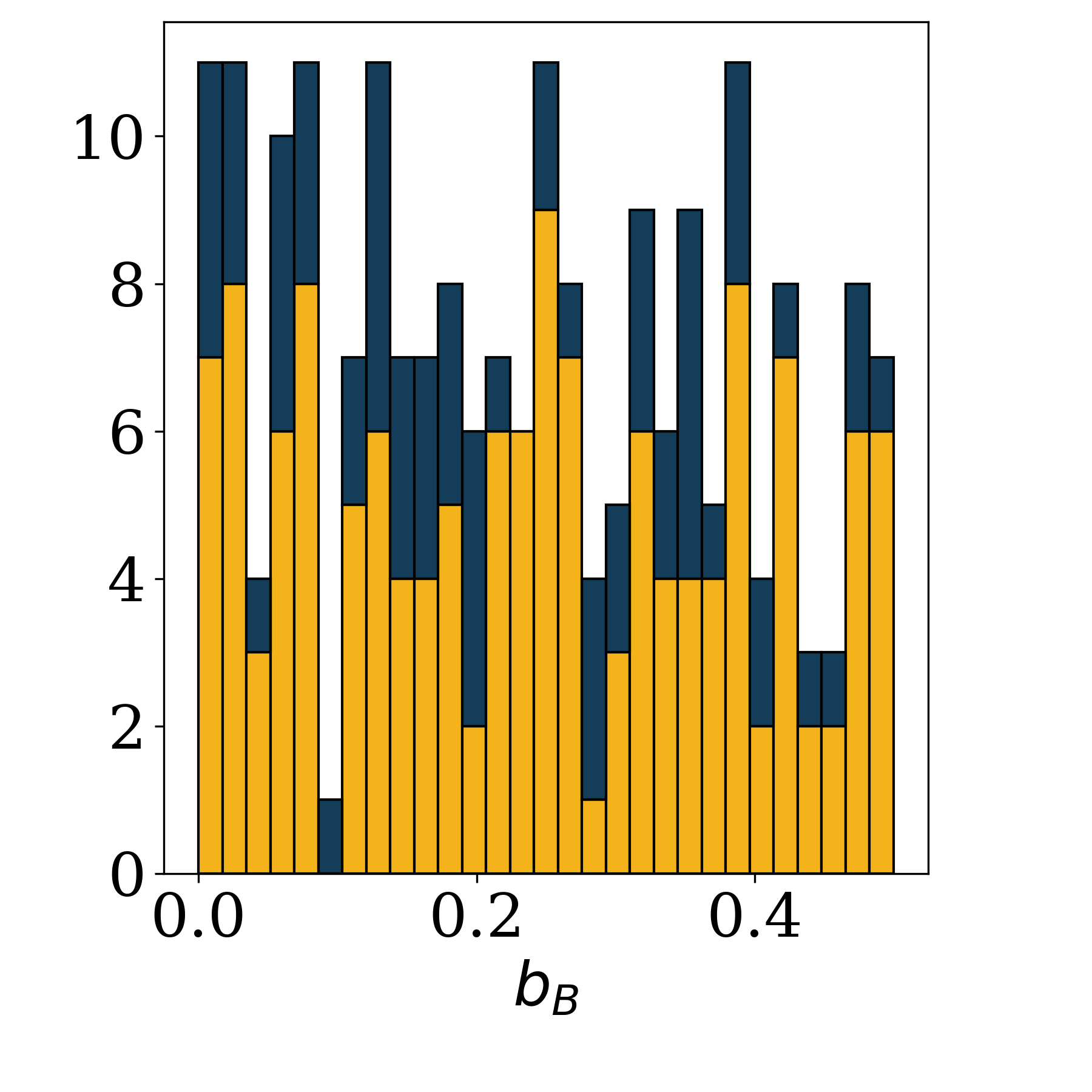}
    \end{subfigure}
    \begin{subfigure}{.23\textwidth}
    \centering
    \includegraphics[width=1\linewidth]{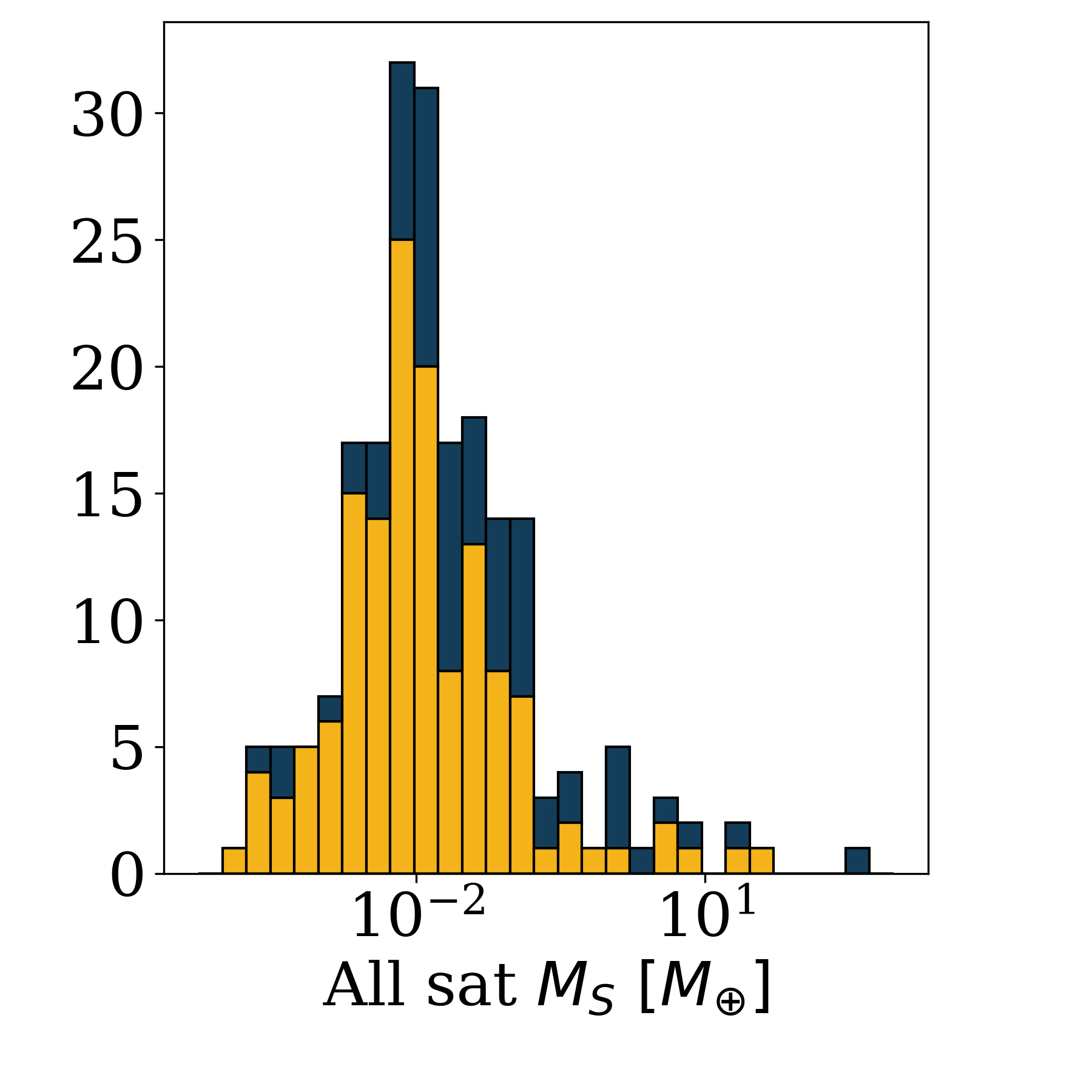}
    \end{subfigure}
    \begin{subfigure}{.23\textwidth}
    \centering
    \includegraphics[width=1\linewidth]{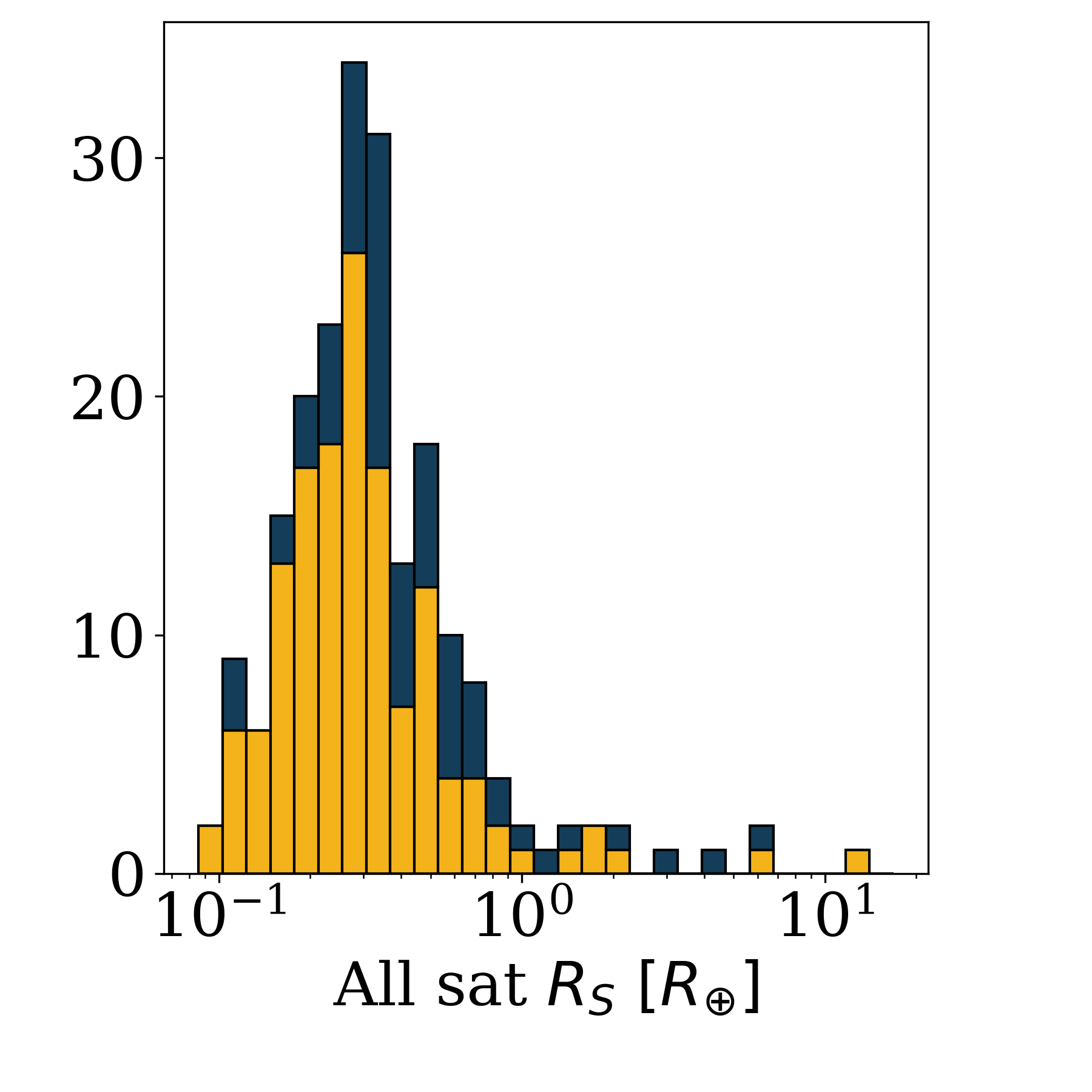}
    \end{subfigure}
    \begin{subfigure}{.23\textwidth}
    \centering
    \includegraphics[width=1\linewidth]{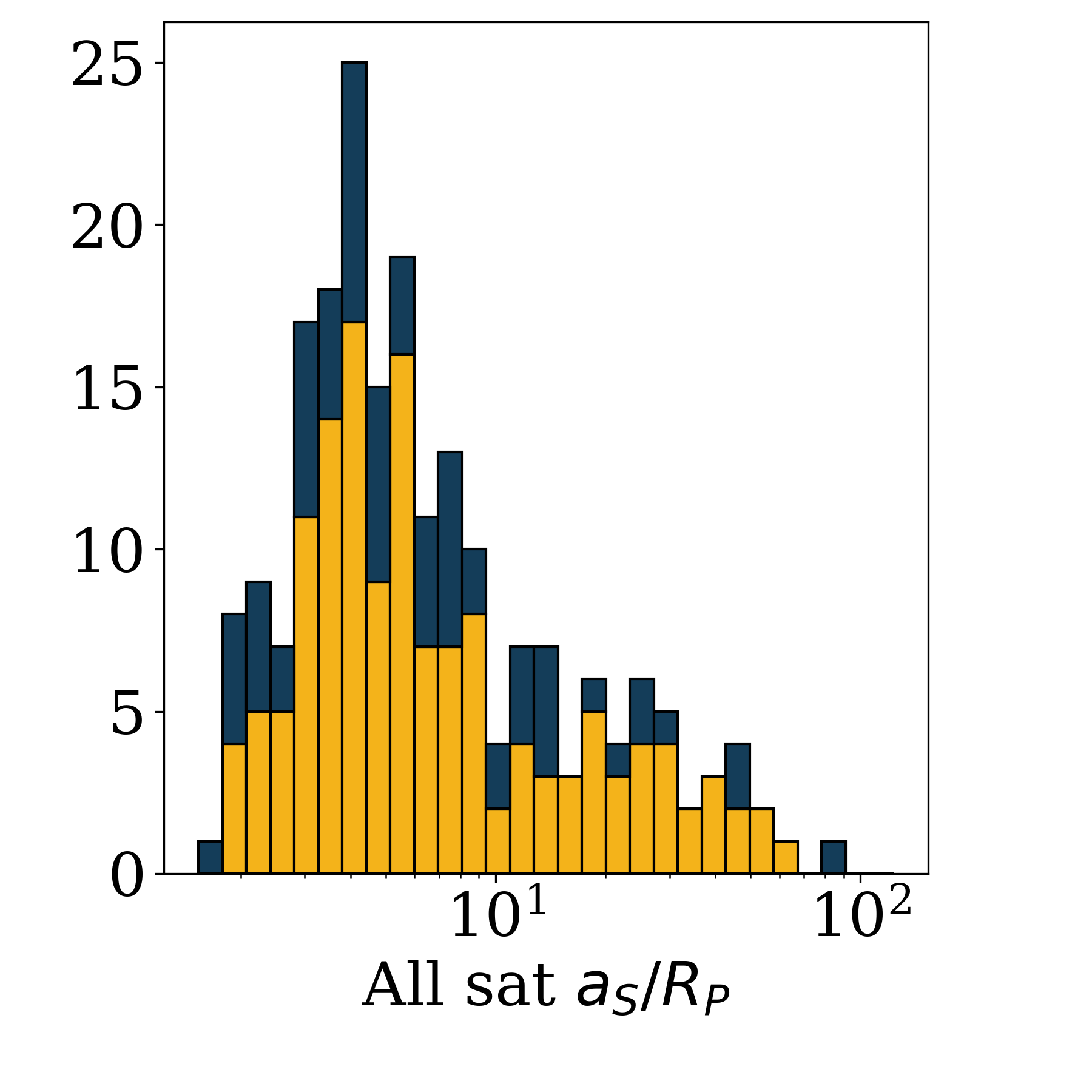}
    \end{subfigure}
    \begin{subfigure}{.23\textwidth}
    \centering
    \includegraphics[width=1\linewidth]{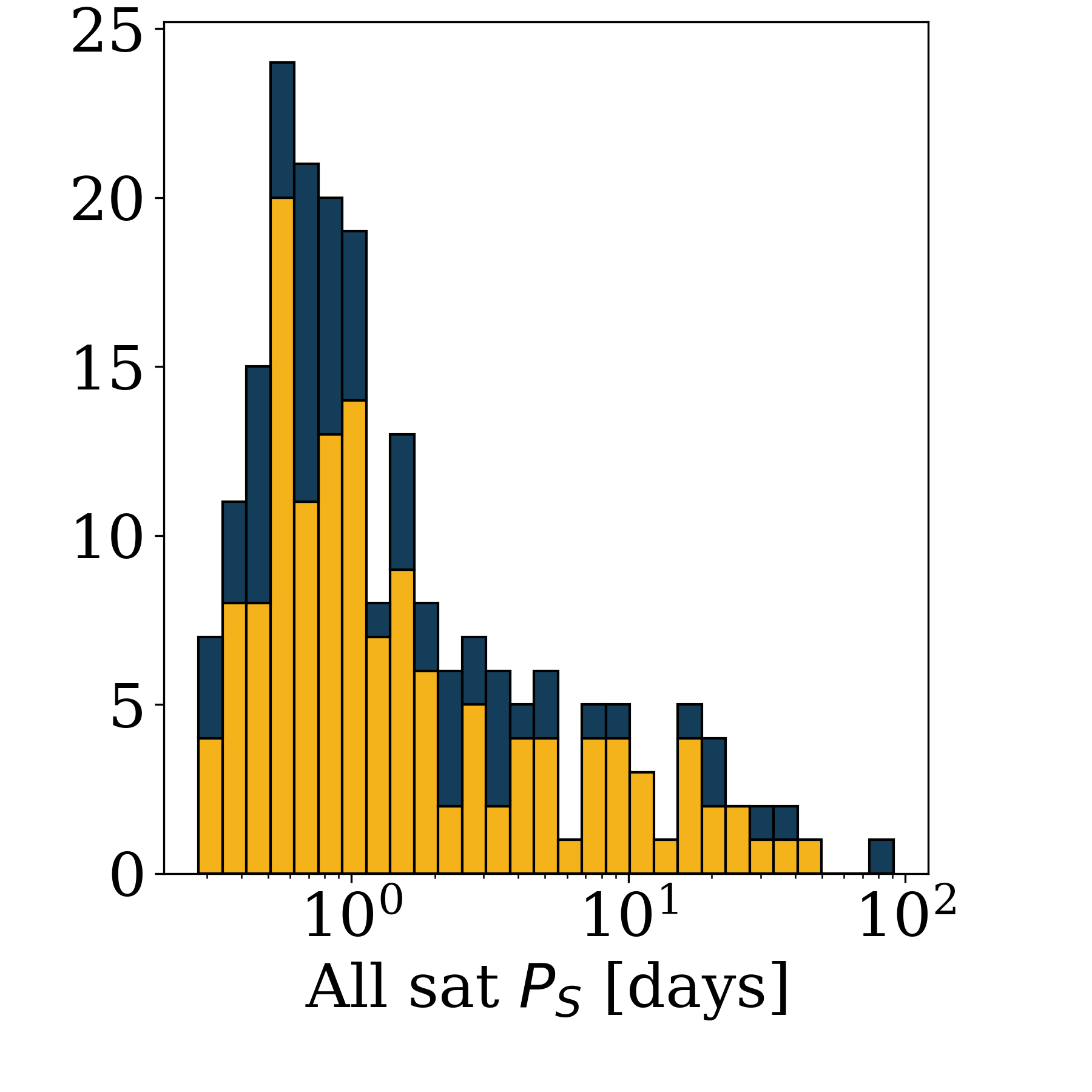}
    \end{subfigure} 
    \begin{subfigure}{.23\textwidth}
    \centering
    \includegraphics[width=1\linewidth]{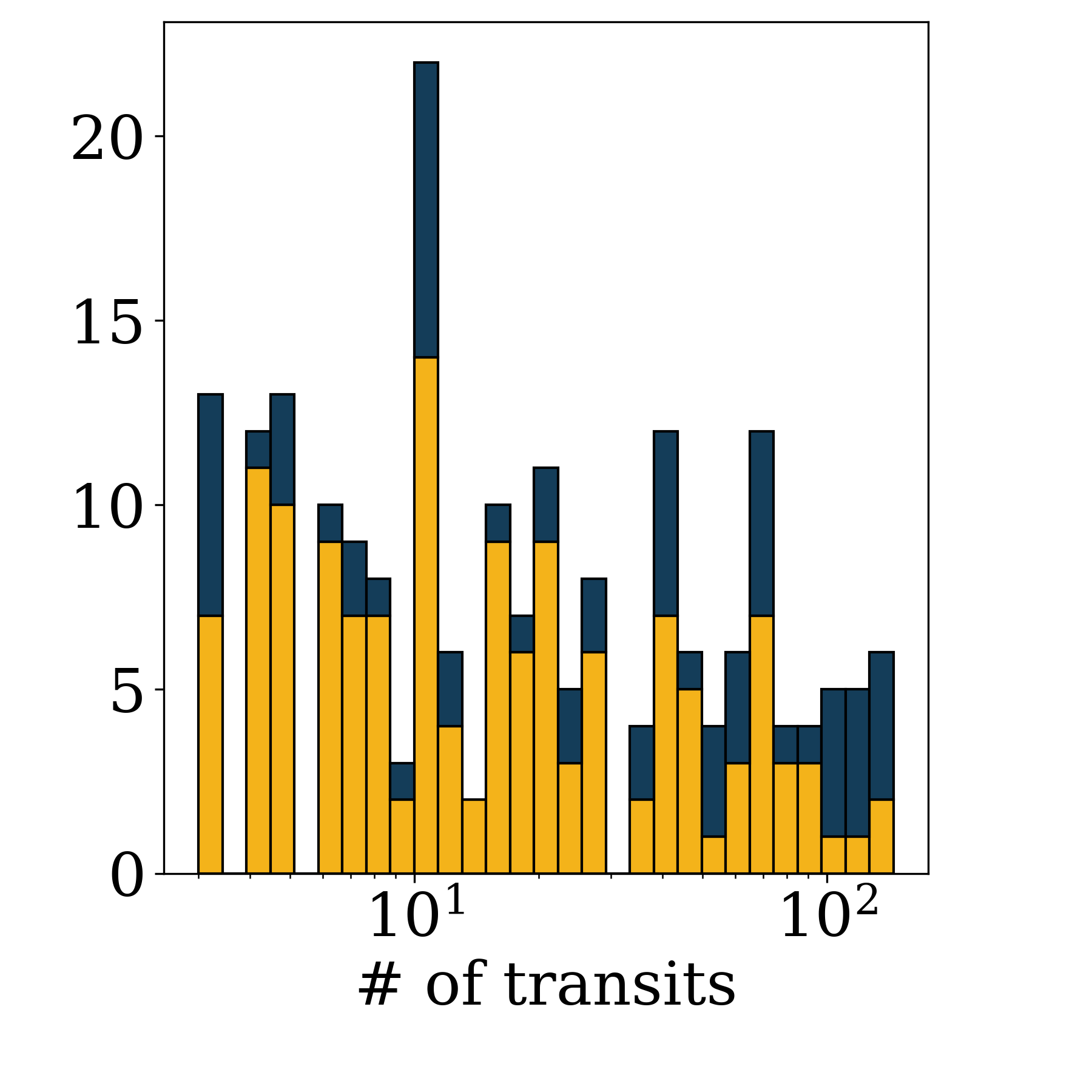}
    \end{subfigure} 
    \begin{subfigure}{.23\textwidth}
    \centering
    \includegraphics[width=1\linewidth]{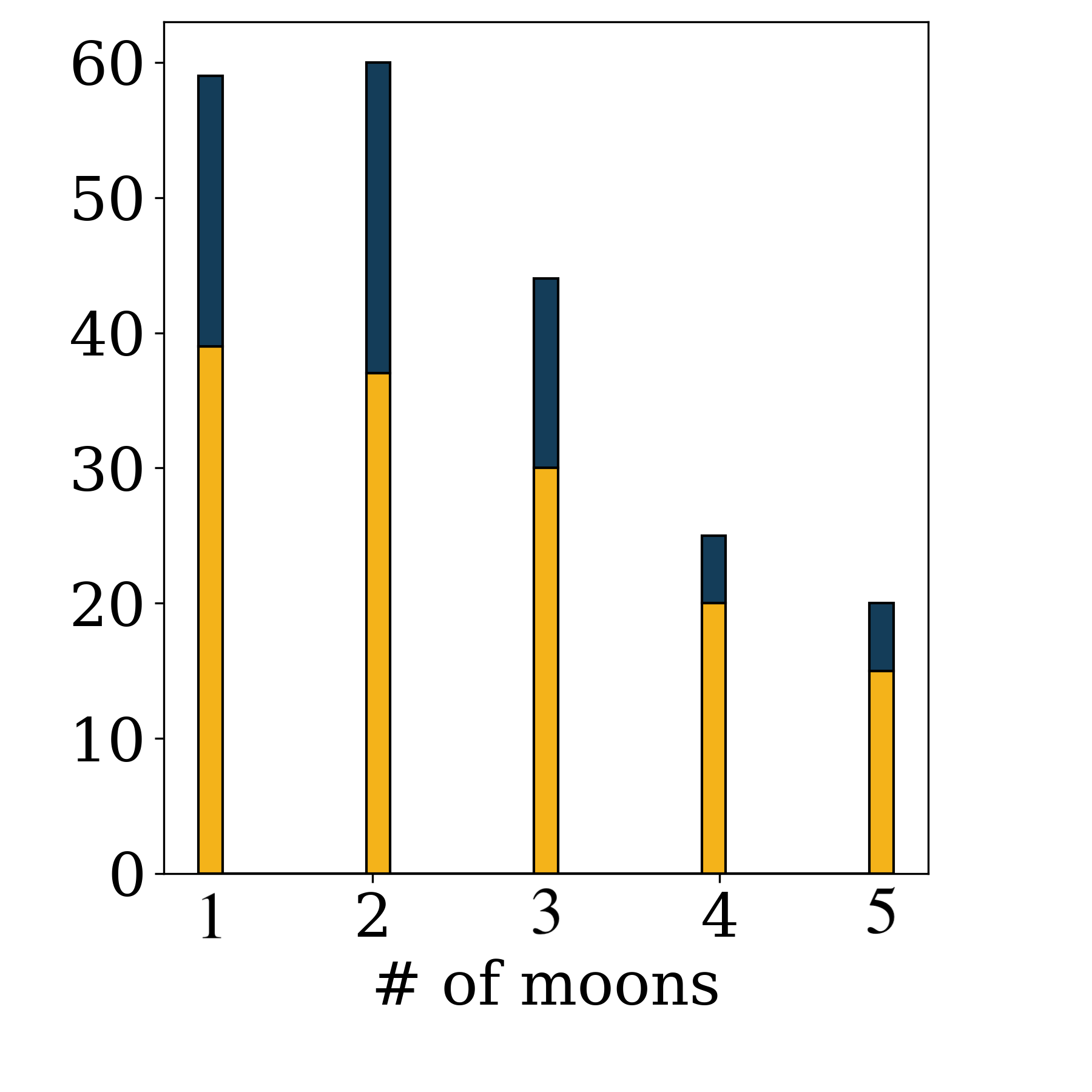}
    \end{subfigure} 
    \begin{subfigure}{.23\textwidth}
    \centering
    \includegraphics[width=1\linewidth]{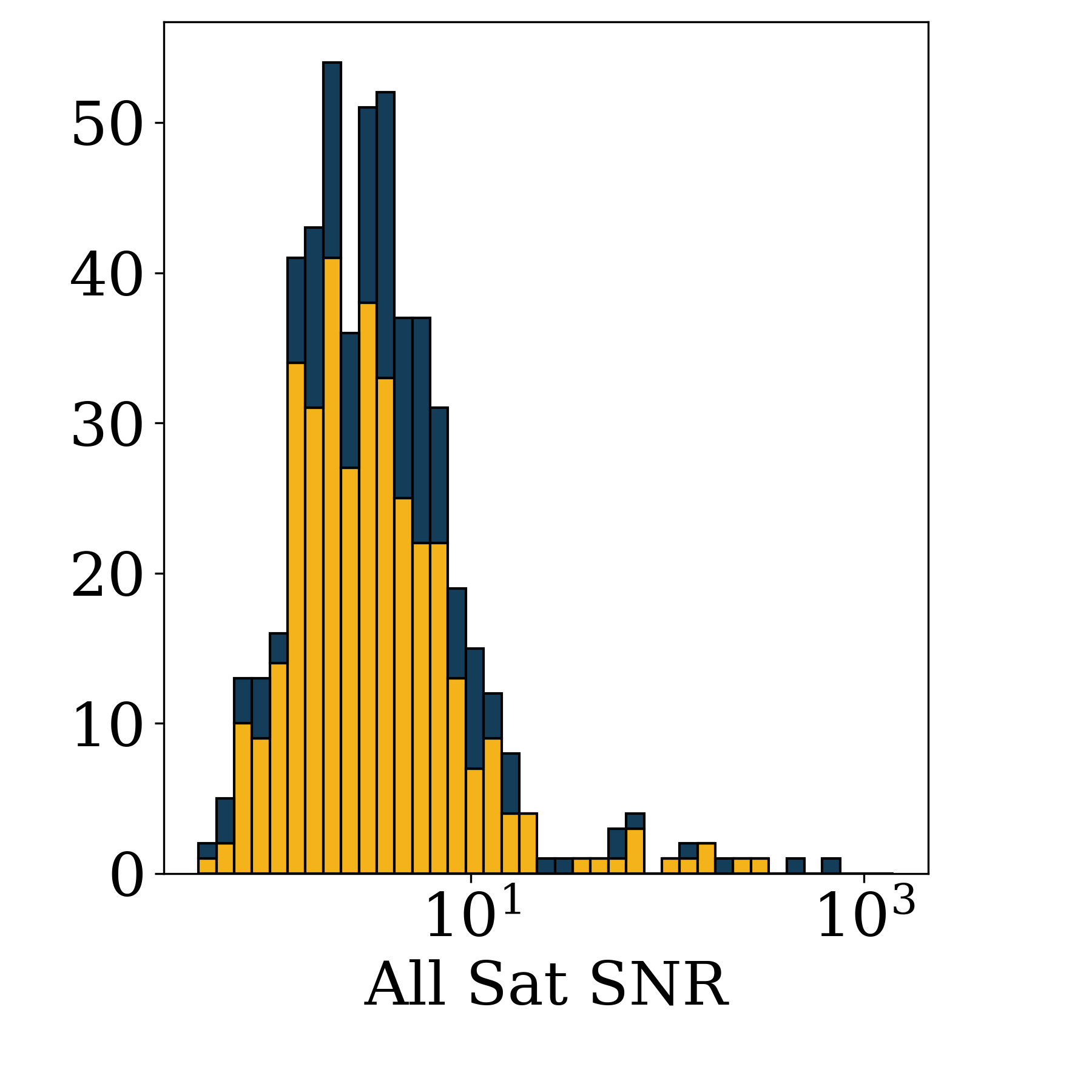}
    \end{subfigure} 
    \caption{Distributions of various ground truth features of the artificial systems generated in this work, showing those systems for which both planet-only and planet+moon models converged in yellow, and those for which the moon model failed to converge in blue. There is are no obvious features to distinguish these two cases. \textit{Top row:} planetary radii $R_P$, in Earth radii; planetary masses $M_P$, in Earth masses; orbital periods of the planetary system barycenter $P_B$, in days; and radius of the star $R_{*}$, in Solar radii. \textit{Middle row:} impact parameter of the system barycenter $b_B$; masses of all the satellites $M_S$, in Earth masses; radii of all the satellites $R_S$, in Earth radii; and the semimajor axes of the moons in units of planetary radii ($a_S / R_P$). \textit{Bottom row:} orbital periods of the moons $P_S$, in days; the number of transits for each planet; the number of moons in each system; and the signal-to-noise ratios SNR for each moon.}
\label{fig:complete_vs_incomplete}
\end{figure*}

Each model run used a single processor on the \texttt{XL} cluster at the Academia Sinica Institute of Astronomy \& Astrophysics. We chose not to parallelize the modeling, though \textsc{UltraNest} supports it, as it would have requiring sharing the time with multiple other users within the institute, actually slowing the rate of completion. It was therefore faster to run dozens of single-core jobs simultaneously, for which there was very little competition for time on the cluster.

In the course of our modeling, we found that the vast majority of our moon model fits and a good fraction of the planet model fits failed to converge after 96 hour (5 day) runs on a single core, which was the maximum walltime allowed for a single job. These fits would be resumed upon timeout, but a sizeable fraction of the moon models in particular failed to converge even after multiple 5-day runs. By contrast, the vast majority of the planet runs were completed after fewer than 10 days (two job submissions). This may be an indicator of the difficulty of fitting these multi-moon systems with single-moon systems. That is, these moon models may be attempting to fit all the moon dips, and struggling to converge, whereas the planet runs essentially do not bother with these complications.

Curiously, many of the light curves for which the moon models failed to converge had exceptionally large moon transit signals evident in the data. This is peculiar, because we would generally expect the large moon transits to be easier to identify than the much smaller transits. However, it could be these large dips become increasingly difficult for the moon model to reconcile, particularly if the transit timing variations appear to be inconsistent with a single moon. But this is also not an entirely satisfactory explanation, because the moon's mass and radius are entirely unconstrained by physically plausible limits; the model should be able to produce either extremely low or extremely high density moon solutions -- ones that we would reject in a real search -- and these solutions could conceivably take care of any discrepancy between transit depth and timing variations.

\subsection{Analysis of unfinished runs}
Figure \ref{fig:complete_vs_incomplete} shows the distribution of system parameters once again, but now a distinction is made between those systems for which both planet-only and planet+moon models converged, and those for which the planet+moon models failed to run even after multiple 5-day runs. The hope is that we may see some feature in these plots that would be a strong predictor of a failure to converge. Unfortunately, we see no such clear features that might suggest a planet+moon model will struggle.

\section{Conclusions}
\label{sec:conclusions}
In this work we have investigated the impact of applying single-moon models to systems containing multiple moons. We modeled realistic multi-moon systems with $N-$body simulations using \textsc{Rebound}, generated artificial light curves from these simulations with \textsc{Pandora}, and applied a model fitting code (\textsc{UltraNest}) to perform parameter estimation and model selection by computing Bayesian evidences and Bayes factors.

We found that the evidence for moons is hardly impacted by the number of moons in the system, but there may be other important knock-on effects that can be identified and potentially leveraged in search of a moon. In particular, we found that these systems show a consistent skew towards higher impact parameters and higher eccentricities. Densities for these systems, meanwhile, are often badly characterized, and potentially unphysical. Limb darkening coefficients may likewise pushed into non realistic territory.

We analyzed four possibilities with respect to moon solutions, namely: 1) whether one real moon is found when searching for one moon, 2) whether one spurious moon is found during the search for one moon, 3) whether multiple moons may be identified or at least suggested by the presence of multi-modal posteriors, or 4) whether moons will go unrecovered due to the confounding presence of multiple moons. We found that we were able to recover a bit more than half of all moons, but that their properties were not often accurately measured. We found little evidence to support the notion that multiple-moon systems will be apparent from multi-modal posteriors.

These results taken together will be important to keep in mind for future exomoon searches, particularly if a single-moon model continues to be utilized. We find that using a single-moon model can still be used, even in multiple moon systems, and may remain an attractive option in view of its limited computational demands. At the same time, the detections made may not always hold up to scrutiny, and there may be side-effects that ultimately lead to an erroneous rejection of the system's candidacy. It will be worth exploring in future work whether the application of multiple-moon models may improve our detection capabilities. Fortunately, the skews in impact parameter, eccentricity, densities, and limb darkening coefficients provide some useful tools that may help in identifying anomolous systems in the future -- systems that may be hosting exomoons. 

\section*{Acknowledgements}
We thank the anonymous reviewer who provided valuable comments which improved the clarity of this work. This research has made use of the NASA Exoplanet Archive, which is operated by the California Institute of Technology, under contract with the National Aeronautics and Space Administration under the Exoplanet Exploration Program. This research has made use of NASA's Astrophysics Data System. This paper includes data collected by the \kepler\ mission and obtained from the MAST data archive at the Space Telescope Science Institute (STScI). Funding for the \kepler\ mission is provided by the NASA Science Mission Directorate. STScI is operated by the Association of Universities for Research in Astronomy, Inc., under NASA contract NAS 5–26555. We also gratefully acknowledge the developers of the following software packages which made this work possible: \texttt{Astropy} \citep{astropy1, astropy2}, \texttt{NumPy} \citep{numpy1, numpy2}, \texttt{Pandas} \citep{pandas}, \texttt{Matplotlib} \citep{matplotlib}, and \texttt{SciPy} \citep{scipy}.

\section*{Data Availability}
The code used to produce the simulations and perform the analysis are hosted at \url{https://github.com/alexteachey/multimoon_modeling_final_repo}.





\begin{thebibliography}{99}
\bibitem[\protect\citeauthoryear{Akeson et al.}{2013}]{NEA} Akeson R.~L., Chen X., Ciardi D., Crane M., Good J., Harbut M., Jackson E., et al., 2013, PASP, 125, 989. doi:10.1086/672273

\bibitem[\protect\citeauthoryear{Astropy Collaboration et al.}{2013}]{astropy1} Astropy Collaboration, Robitaille T.~P., Tollerud E.~J., Greenfield P., Droettboom M., Bray E., Aldcroft T., et al., 2013, A\&A, 558, A33. doi:10.1051/0004-6361/201322068

\bibitem[\protect\citeauthoryear{Astropy Collaboration et al.}{2022}]{astropy2} Astropy Collaboration, Price-Whelan A.~M., Lim P.~L., Earl N., Starkman N., Bradley L., Shupe D.~L., et al., 2022, ApJ, 935, 167. doi:10.3847/1538-4357/ac7c74

\bibitem[\protect\citeauthoryear{B{\'e}ky et al.}{2014}]{Beky:2014} B{\'e}ky B., Holman M.~J., Kipping D.~M., Noyes R.~W., 2014, ApJ, 788, 1. doi:10.1088/0004-637X/788/1/1

\bibitem[\protect\citeauthoryear{Buchner}{2021}]{UltraNest} Buchner J., 2021, JOSS, 6, 3001. doi:10.21105/joss.03001

\bibitem[\protect\citeauthoryear{Carter \& Agol}{2013}]{QATS} Carter J.~A., Agol E., 2013, ApJ, 765, 132. doi:10.1088/0004-637X/765/2/132

\bibitem[\protect\citeauthoryear{Chen \& Kipping}{2017}]{forecaster} Chen J., Kipping D., 2017, ApJ, 834, 17. doi:10.3847/1538-4357/834/1/17


\bibitem[\protect\citeauthoryear{Cilibrasi et al.}{2021}]{Cilibrasi:2021} Cilibrasi M., Szul{\'a}gyi J., Grimm S.~L., Mayer L., 2021, MNRAS, 504, 5455. doi:10.1093/mnras/stab1179

\bibitem[\protect\citeauthoryear{Claret \& Bloemen}{2011}]{Claret:2011} Claret A., Bloemen S., 2011, A\&A, 529, A75. doi:10.1051/0004-6361/201116451

\bibitem[\protect\citeauthoryear{Dobos et al.}{2021}]{Dobos:2021} Dobos V., Charnoz S., P{\'a}l A., Roque-Bernard A., Szab{\'o} G.~M., 2021, PASP, 133, 094401. doi:10.1088/1538-3873/abfe04


\bibitem[\protect\citeauthoryear{Domingos, Winter, \& Yokoyama}{2006}]{Domingos:2006} Domingos R.~C., Winter O.~C., Yokoyama T., 2006, MNRAS, 373, 1227. doi:10.1111/j.1365-2966.2006.11104.x


\bibitem[\protect\citeauthoryear{Gordon \& Agol}{2022}]{Gefera} Gordon T.~A., Agol E., 2022, AJ, 164, 111. doi:10.3847/1538-3881/ac82b1

\bibitem[\protect\citeauthoryear{Harris et al.}{2020}]{numpy2} Harris C.~R., Jarrod Millman K., van der Walt S.~J., Gommers R., Virtanen P., Cournapeau D., Wieser E., et al., 2020, arXiv, arXiv:2006.10256

\bibitem[\protect\citeauthoryear{Heller et al.}{2016}]{Heller:2016} Heller R., Hippke M., Placek B., Angerhausen D., Agol E., 2016, A\&A, 591, A67. doi:10.1051/0004-6361/201628573


\bibitem[\protect\citeauthoryear{Heller, Rodenbeck, \& Bruno}{2019}]{Heller:2019} Heller R., Rodenbeck K., Bruno G., 2019, A\&A, 624, A95. doi:10.1051/0004-6361/201834913

\bibitem[\protect\citeauthoryear{Hippke \& Heller}{2022}]{Pandora} Hippke M., Heller R., 2022, A\&A, 662, A37. doi:10.1051/0004-6361/202243129

\bibitem[\protect\citeauthoryear{Heller \& Hippke}{2023}]{Heller:2023} Heller R., Hippke M., 2023, NatAs.tmp. doi:10.1038/s41550-023-02148-w

\bibitem[\protect\citeauthoryear{Hunter}{2007}]{matplotlib} Hunter J.~D., 2007, CSE, 9, 90. doi:10.1109/MCSE.2007.55

\bibitem[\protect\citeauthoryear{Inderbitzi et al.}{2020}]{Inderbitzi:2020} Inderbitzi C., Szul{\'a}gyi J., Cilibrasi M., Mayer L., 2020, MNRAS, 499, 1023. doi:10.1093/mnras/staa2796


\bibitem[\protect\citeauthoryear{Kass \& Raftery}{1995}]{Bayes_Factors} Kass R.E., Raftery, A.E., 1995, Journal of the American Statistical Association, 90:430, 773-795, DOI: 10.1080/01621459.1995.10476572

\bibitem[\protect\citeauthoryear{Kipping}{2009a}]{timingI} Kipping D.~M., 2009, MNRAS, 392, 181. doi:10.1111/j.1365-2966.2008.13999.x

\bibitem[\protect\citeauthoryear{Kipping}{2009b}]{timingII} Kipping D.~M., 2009, MNRAS, 396, 1797. doi:10.1111/j.1365-2966.2009.14869.x

\bibitem[\protect\citeauthoryear{Kipping}{2008}]{photoeccentric} Kipping D.~M., 2008, MNRAS, 389, 1383. doi:10.1111/j.1365-2966.2008.13658.x

\bibitem[\protect\citeauthoryear{Kipping}{2011}]{LUNA} Kipping D.~M., 2011, MNRAS, 416, 689. doi:10.1111/j.1365-2966.2011.19086.x

\bibitem[Kipping(2013)]{LDCs} Kipping, D.~M.\ 2013, \mnras, 435, 2152. doi:10.1093/mnras/stt1435

\bibitem[\protect\citeauthoryear{Kipping et al.}{2015}]{HEKV} Kipping D.~M., Schmitt A.~R., Huang X., Torres G., Nesvorn{\'y} D., Buchhave L.~A., Hartman J., et al., 2015, ApJ, 813, 14. doi:10.1088/0004-637X/813/1/14

\bibitem[Kipping(2021)]{transit_origami} Kipping, D.\ 2021, \mnras, 507, 4120. doi:10.1093/mnras/stab2013

\bibitem[\protect\citeauthoryear{Kipping et al.}{2022}]{K1708} Kipping D., Bryson S., Burke C., Christiansen J., Hardegree-Ullman K., Quarles B., Hansen B., et al., 2022, NatAs, 6, 367. doi:10.1038/s41550-021-01539-1

\bibitem[\protect\citeauthoryear{Kipping et al.}{2024}]{Kipping:2024} Kipping D., Teachey A., Yahalomi D.~A., Cassese B., Quarles B., Bryson S., Hansen B., et al., 2024, arXiv, arXiv:2401.10333. doi:10.48550/arXiv.2401.10333


\bibitem[\protect\citeauthoryear{Kreidberg, Luger, \& Bedell}{2019}]{Kreidberg:2019} Kreidberg L., Luger R., Bedell M., 2019, ApJL, 877, L15. doi:10.3847/2041-8213/ab20c8

\bibitem[\protect\citeauthoryear{Martin, Fabrycky, \& Montet}{2019}]{Martin:2019} Martin D.~V., Fabrycky D.~C., Montet B.~T., 2019, ApJL, 875, L25. doi:10.3847/2041-8213/ab0aea

\bibitem[\protect\citeauthoryear{The Pandas Development Team}{2020}]{pandas} Pandas developement team, 2020, Zenodo, doi:10.5281/zenodo.3509134 

\bibitem[\protect\citeauthoryear{Rein \& Liu}{2012}]{Rebound} Rein H., Liu S.-F., 2012, A\&A, 537, A128. doi:10.1051/0004-6361/201118085

\bibitem[\protect\citeauthoryear{Teachey \& Kipping}{2018}]{K1625} Teachey A., Kipping D.~M., 2018, SciA, 4, eaav1784. doi:10.1126/sciadv.aav1784

\bibitem[\protect\citeauthoryear{Teachey et al.}{2020}]{Loose_Ends} Teachey A., Kipping D., Burke C.~J., Angus R., Howard A.~W., 2020, AJ, 159, 142. doi:10.3847/1538-3881/ab7001


\bibitem[\protect\citeauthoryear{Teachey}{2021}]{Teachey:2021} Teachey A., 2021, MNRAS, 506, 2104. doi:10.1093/mnras/stab1840

\bibitem[van der Walt et al.(2011)]{numpy1} van der Walt, S., Colbert, S.~C., \& Varoquaux, G.\ 2011, Computing in Science and Engineering, 13, 22. doi:10.1109/MCSE.2011.37

\bibitem[\protect\citeauthoryear{Virtanen et al.}{2020}]{scipy} Virtanen P., Gommers R., Oliphant T.~E., Haberland M., Reddy T., Cournapeau D., Burovski E., et al., 2020, NatMe, 17, 261. doi:10.1038/s41592-019-0686-2


\end{thebibliography}

\begin{thebibliography}{99}
\bibitem[\protect\citeauthoryear{Agnor \& Hamilton}{2006}]{agnor:2006} Agnor C.~B., Hamilton D.~P., 2006, Nature, 441, 192. doi:10.1038/nature04792

\bibitem[\protect\citeauthoryear{Agol et al.}{2005}]{Agol:2005} Agol E., Steffen J., Sari R., Clarkson W., 2005, MNRAS, 359, 567. doi:10.1111/j.1365-2966.2005.08922.x

\bibitem[\protect\citeauthoryear{Agol \& Fabrycky}{2018}]{agol:2018} Agol E., Fabrycky D.~C., 2018, haex.book, 7. doi:10.1007/978-3-319-55333-7\_7

\bibitem[\protect\citeauthoryear{Akaike}{1974}]{Akaike:1974} Akaike H., 1974, ITAC, 19, 716

\bibitem[\protect\citeauthoryear{Alshehhi et al.}{2020}]{Alshehhi:2020} Alshehhi R., Rodenbeck K., Gizon L., Sreenivasan K.~R., 2020, A\&A, 640, A41. doi:10.1051/0004-6361/201937059

\bibitem[\protect\citeauthoryear{Astropy Collaboration et al.}{2013}]{astropy1} Astropy Collaboration, Robitaille T.~P., Tollerud E.~J., Greenfield P., Droettboom M., Bray E., Aldcroft T., et al., 2013, A\&A, 558, A33. doi:10.1051/0004-6361/201322068

\bibitem[\protect\citeauthoryear{Astropy Collaboration et al.}{2018}]{astropy2} Astropy Collaboration, Price-Whelan A.~M., Sip{\H{o}}cz B.~M., G{\"u}nther H.~M., Lim P.~L., Crawford S.~M., Conseil S., et al., 2018, AJ, 156, 123. doi:10.3847/1538-3881/aa

\bibitem[\protect\citeauthoryear{Banfield \& Murray}{1992}]{banfield:1992} Banfield, D. \& Murray, N.\ 1992, \icarus, 99, 390. doi:10.1016/0019-1035(92)90155-Z

\bibitem[\protect\citeauthoryear{Borucki et al.}{2010}]{Kepler} Borucki W.~J., Koch D., Basri G., Batalha N., Brown T., Caldwell D., Caldwell J., et al., 2010, Sci, 327, 977. doi:10.1126/science.1185402


\bibitem[\protect\citeauthoryear{Cabrera \& Schneider}{2007}]{Cabrera:2007} Cabrera J., Schneider J., 2007, A\&A, 464, 1133. doi:10.1051/0004-6361:20066111

\bibitem[\protect\citeauthoryear{Canup \& Asphaug}{2001}]{canup:2001} Canup, R.~M. \& Asphaug, E.\ 2001, \nat, 412, 708

\bibitem[\protect\citeauthoryear{Canup}{2005}]{Canup:2005} Canup R.~M., 2005, Sci, 307, 546. doi:10.1126/science.1106818

\bibitem[Canup \& Ward(2006)]{canup:2006} Canup, R.~M. \& Ward, W.~R.\ 2006, \nat, 441, 834. doi:10.1038/nature04860

\bibitem[\protect\citeauthoryear{Canup}{2011}]{canup:2011} Canup R.~M., 2011, AJ, 141, 35. doi:10.1088/0004-6256/141/2/35

\bibitem[\protect\citeauthoryear{Carter et al.}{2008}]{carter:2008} Carter J.~A., Yee J.~C., Eastman J., Gaudi B.~S., Winn J.~N., 2008, ApJ, 689, 499. doi:10.1086/592321

\bibitem[\protect\citeauthoryear{Chen \& Kipping}{2017}]{forecaster} Chen J., Kipping D., 2017, ApJ, 834, 17. doi:10.3847/1538-4357/834/1/17

\bibitem[\protect\citeauthoryear{Cilibrasi et al.}{2020}]{cilibrasi:2020} Cilibrasi M., Szul{\'a}gyi J., Grimm S.~L., Mayer L., 2020, arXiv, arXiv:2011.11513

\bibitem[\protect\citeauthoryear{Cincotta, Giordano, \& Sim{\'o}}{2003}]{MEGNO} Cincotta P.~M., Giordano C.~M., Sim{\'o} C., 2003, PhyD, 182, 151. doi:10.1016/S0167-2789(03)00103-9

\bibitem[\protect\citeauthoryear{Dawson \& Fabrycky}{2010}]{Dawson:2010} Dawson, R.~I. \& Fabrycky, D.~C.\ 2010, \apj, 722, 937. doi:10.1088/0004-637X/722/1/937

\bibitem[\protect\citeauthoryear{Dobos et al.}{2021}]{Dobos:2021} Dobos V., Charnoz S., P{\'a}l A., Roque-Bernard A., Szab{\'o} G.~M., 2021, arXiv, arXiv:2105.12040


\bibitem[\protect\citeauthoryear{Domingos, Winter, \& Yokoyama}{2006}]{Domingos:2006} Domingos R.~C., Winter O.~C., Yokoyama T., 2006, MNRAS, 373, 1227. doi:10.1111/j.1365-2966.2006.11104.x

\bibitem[\protect\citeauthoryear{Ester et al.}{1996}]{DBSCAN} Ester, M., H. P. Kriegel, J. Sander, and X. Xu “A Density-Based Algorithm for Discovering Clusters in Large Spatial Databases with Noise”. In: Proceedings of the 2nd International Conference on Knowledge Discovery and Data Mining, Portland, OR, AAAI Press, pp. 226-231. 1996 doi:10.5555/3001460.3001507

\bibitem[\protect\citeauthoryear{Fabrycky et al.}{2014}]{Fabrycky:2014} Fabrycky D.~C., Lissauer J.~J., Ragozzine D., Rowe J.~F., Steffen J.~H., Agol E., Barclay T., et al., 2014, ApJ, 790, 146. doi:10.1088/0004-637X/790/2/146

\bibitem[\protect\citeauthoryear{Fox \& Wiegert}{2021}]{Fox:2021} Fox C., Wiegert P., 2021, MNRAS, 501, 2378. doi:10.1093/mnras/staa3743

\bibitem[\protect\citeauthoryear{Fujii \& Ogihara}{2020}]{Fujii:2020} Fujii Y.~I., Ogihara M., 2020, A\&A, 635, L4. doi:10.1051/0004-6361/201937192

\bibitem[\protect\citeauthoryear{Hadden \& Lithwick}{2017}]{HL2017} Hadden S., Lithwick Y., 2017, AJ, 154, 5. doi:10.3847/1538-3881/aa71ef

\bibitem[\protect\citeauthoryear{Hamers \& Portegies Zwart}{2018}]{Hamers:2018} Hamers A.~S., Portegies Zwart S.~F., 2018, ApJL, 869, L27. doi:10.3847/2041-8213/aaf3a7

\bibitem[\protect\citeauthoryear{Hansen}{2019}]{Hansen:2019} Hansen B.~M.~S., 2019, SciA, 5, eaaw8665. doi:10.1126/sciadv.aaw8665

\bibitem[\protect\citeauthoryear{Harris et al.}{2020}]{numpy2} Harris C.~R., Jarrod Millman K., van der Walt S.~J., Gommers R., Virtanen P., Cournapeau D., Wieser E., et al., 2020, arXiv, arXiv:2006.10256

\bibitem[\protect\citeauthoryear{Heller et al.}{2016}]{Heller_MMR} Heller R., Hippke M., Placek B., Angerhausen D., Agol E., 2016, A\&A, 591, A67. doi:10.1051/0004-6361/201628573

\bibitem[\protect\citeauthoryear{Holczer et al.}{2016}]{holczer:2016} Holczer, T., Mazeh, T., Nachmani, G., et al.\ 2016, \apjs, 225, 9

\bibitem[\protect\citeauthoryear{Holman \& Murray}{2005}]{Holman:2005} Holman M.~J., Murray N.~W., 2005, Sci, 307, 1288. doi:10.1126/science.1107822

\bibitem[\protect\citeauthoryear{Howard et al.}{2012}]{howard:2012} Howard A.~W., Marcy G.~W., Bryson S.~T., Jenkins J.~M., Rowe J.~F., Batalha N.~M., Borucki W.~J., et al., 2012, ApJS, 201, 15. doi:10.1088/0067-0049/201/2/15

\bibitem[\protect\citeauthoryear{Hunter}{2007}]{matplotlib} Hunter J.~D., 2007, CSE, 9, 90. doi:10.1109/MCSE.2007.55

\bibitem[\protect\citeauthoryear{Ida et al.}{2020}]{Ida:2020} Ida S., Ueta S., Sasaki T., Ishizawa Y., 2020, NatAs, 4, 880. doi:10.1038/s41550-020-1049-8

\bibitem[\protect\citeauthoryear{Inderbitzi et al.}{2020}]{inderbitzi:2020} Inderbitzi C., Szul{\'a}gyi J., Cilibrasi M., Mayer L., 2020, MNRAS, 499, 1023. doi:10.1093/mnras/staa2796

\bibitem[\protect\citeauthoryear{Kane et al.}{2019}]{Kane:2019} Kane M., Ragozzine D., Flowers X., Holczer T., Mazeh T., Relles H.~M., 2019, AJ, 157, 171. doi:10.3847/1538-3881/ab0d91

\bibitem[\protect\citeauthoryear{Kass \& Wasserman}{1993}]{kass:1993} Kass R.~E., Wasserman L., 1993, Journal of the American Statistical Association, 90:431, 928. doi:10.1080/01621459.1995.10476592

\bibitem[\protect\citeauthoryear{Kipping}{2009a}]{Kipping:2009} Kipping D.~M., 2009, MNRAS, 392, 181. doi:10.1111/j.1365-2966.2008.13999.x

\bibitem[\protect\citeauthoryear{Kipping}{2009b}]{TDV-TIP} Kipping D.~M., 2009, MNRAS, 396, 1797. doi:10.1111/j.1365-2966.2009.14869.x

\bibitem[\protect\citeauthoryear{Kipping et al.}{2012}]{HEKI} Kipping, D.~M., Bakos, G. {\'A}., Buchhave, L., et al.\ 2012, \apj, 750, 115. doi:10.1088/0004-637X/750/2/115

\bibitem[\protect\citeauthoryear{Kipping \& Teachey}{2020}]{impossible_moons} Kipping D., Teachey A., 2020, SerAJ, 201, 25. doi:10.2298/SAJ2001025K

\bibitem[\protect\citeauthoryear{Kipping}{2021}]{moon_corridor} Kipping D., 2021, MNRAS, 500, 1851. doi:10.1093/mnras/staa3398

\bibitem[\protect\citeauthoryear{Laskar}{2000}]{laskar:2000} Laskar J., 2000, PhRvL, 84, 3240. doi:10.1103/PhysRevLett.84.3240

\bibitem[\protect\citeauthoryear{Lithwick, Xie, \& Wu}{2012}]{Lithwick:2012} Lithwick Y., Xie J., Wu Y., 2012, ApJ, 761, 122. doi:10.1088/0004-637X/761/2/122

\bibitem[\protect\citeauthoryear{Lomb}{1976}]{lomb:1976} Lomb N.~R., 1976, Ap\&SS, 39, 447. doi:10.1007/BF00648343

\bibitem[\protect\citeauthoryear{Nesvorn{\'y}, Vokrouhlick{\'y}, \& Morbidelli}{2007}]{nesvorny:2007} Nesvorn{\'y} D., Vokrouhlick{\'y} D., Morbidelli A., 2007, AJ, 133, 1962. doi:10.1086/512850

\bibitem[\protect\citeauthoryear{The Pandas Development Team}{2020}]{pandas} Pandas developement team, 2020, Zenodo, doi:10.5281/zenodo.3509134 

\bibitem[\protect\citeauthoryear{Quarles, Li, \& Rosario-Franco}{2020}]{Quarles:2020} Quarles B., Li G., Rosario-Franco M., 2020, ApJL, 902, L20. doi:10.3847/2041-8213/abba36

\bibitem[\protect\citeauthoryear{Rein \& Liu}{2012}]{REBOUND} Rein H., Liu S.-F., 2012, A\&A, 537, A128. doi:10.1051/0004-6361/201118085

\bibitem[\protect\citeauthoryear{Rein \& Tamayo}{2015}]{WHFAST} Rein H., Tamayo D., 2015, MNRAS, 452, 376. doi:10.1093/mnras/stv1257

\bibitem[\protect\citeauthoryear{Ricker et al.}{2015}]{TESS} Ricker G.~R., Winn J.~N., Vanderspek R., Latham D.~W., Bakos G. {\'A}., Bean J.~L., Berta-Thompson Z.~K., et al., 2015, JATIS, 1, 014003. doi:10.1117/1.JATIS.1.1.014003

\bibitem[\protect\citeauthoryear{Rosario-Franco et al.}{2020}]{Rosario-Franco:2020} Rosario-Franco M., Quarles B., Musielak Z.~E., Cuntz M., 2020, AJ, 159, 260. doi:10.3847/1538-3881/ab89a7

\bibitem[\protect\citeauthoryear{Sartoretti \& Schneider}{1999}]{Sartoretti:1999} Sartoretti P., Schneider J., 1999, A\&AS, 134, 553. doi:10.1051/aas:1999148

\bibitem[\protect\citeauthoryear{Scargle}{1982}]{scargle:1982} Scargle J.~D., 1982, ApJ, 263, 835. doi:10.1086/160554

\bibitem[\protect\citeauthoryear{Schwarz}{1978}]{Schwarz:1978} Schwarz, G., 1978, The Annals of Statistics 6(2), 461-464 doi:10.1214/aos/1176344136

\bibitem[\protect\citeauthoryear{Szab{\'o} et al.}{2006}]{Szabo:2006} Szab{\'o} G.~M., Szatm{\'a}ry K., Div{\'e}ki Z., Simon A., 2006, A\&A, 450, 395. doi:10.1051/0004-6361:20054555

\bibitem[\protect\citeauthoryear{Szul{\'a}gyi, Cilibrasi, \& Mayer}{2018}]{szulagyi:2018} Szul{\'a}gyi J., Cilibrasi M., Mayer L., 2018, ApJL, 868, L13. doi:10.3847/2041-8213/aaeed6

\bibitem[\protect\citeauthoryear{Tamayo et al.}{2020}]{REBOUNDx} Tamayo D., Rein H., Shi P., Hernandez D.~M., 2020, MNRAS, 491, 2885. doi:10.1093/mnras/stz2870


\bibitem[\protect\citeauthoryear{Tamayo et al.}{2020}]{SPOCK} Tamayo D., Cranmer M., Hadden S., Rein H., Battaglia P., Obertas A., Armitage P.~J., et al., 2020, PNAS, 117, 18194. doi:10.1073/pnas.2001258117

\bibitem[\protect\citeauthoryear{Teachey, Kipping, \& Schmitt}{2018}]{HEKVI} Teachey A., Kipping D.~M., Schmitt A.~R., 2018, AJ, 155, 36. doi:10.3847/1538-3881/aa93f2

\bibitem[\protect\citeauthoryear{Teachey \& Kipping}{2018}]{TK18} Teachey A., Kipping D.~M., 2018, SciA, 4, eaav1784. doi:10.1126/sciadv.aav1784


\bibitem[\protect\citeauthoryear{VanderPlas}{2018}]{VP_LS} VanderPlas J.~T., 2018, ApJS, 236, 16. doi:10.3847/1538-4365/aab766

\bibitem[van der Walt et al.(2011)]{numpy1} van der Walt, S., Colbert, S.~C., \& Varoquaux, G.\ 2011, Computing in Science and Engineering, 13, 22. doi:10.1109/MCSE.2011.37

\bibitem[\protect\citeauthoryear{Virtanen et al.}{2020}]{scipy} Virtanen P., Gommers R., Oliphant T.~E., Haberland M., Reddy T., Cournapeau D., Burovski E., et al., 2020, NatMe, 17, 261. doi:10.1038/s41592-019-0686-2


\bibitem[\protect\citeauthoryear{Wright et al.}{2012}]{wright:2012} Wright, J.~T., Marcy, G.~W., Howard, A.~W., et al.\ 2012, \apj, 753, 160. doi:10.1088/0004-637X/753/2/160


\end{thebibliography}

\bsp	
\label{lastpage}
\end{document}